\documentclass[11pt]{article}

\usepackage{jheppub}
\usepackage{wasysym}

\usepackage{tabularx}
\usepackage{physics}
\usepackage{tikz}

\usepackage{graphicx} 
\usepackage{caption}
\usepackage{subcaption}

\usepackage{ae}

\usepackage{tabularx}

\usepackage{graphicx}

\usepackage{amsfonts,amsmath,amsthm,amssymb,amsbsy}

\usepackage{mathtools}

\title{Prescriptive Unitarity from Positive Geometries}

\author[]{Livia Ferro,}\emailAdd{l.ferro@herts.ac.uk}
\author[]{Ross Glew,}\emailAdd{r.glew@herts.ac.uk}
\author[]{Tomasz \L ukowski,}\emailAdd{t.lukowski@herts.ac.uk}
\author[]{and Jonah Stalknecht}\emailAdd{j.stalknecht@herts.ac.uk}

\affiliation[]{Department of Physics, Astronomy and Mathematics, \\ University of Hertfordshire, \\  Hatfield, Hertfordshire, AL10 9AB, United Kingdom}

\abstract{
In this paper, we define the momentum amplituhedron in the four-dimensional split-signature space of dual momenta. It encodes scattering amplitudes  at tree level and loop integrands for $\mathcal{N}=4$ super Yang-Mills in the planar sector.  In this description, every point in the tree-level geometry is specified by a null polygon. Using the null structure of this kinematic space, we find a geometry whose canonical differential form produces loop-amplitude integrands. Remarkably, at one loop it is a curvy version of a simple polytope, whose  vertices are specified by maximal cuts of the amplitude.  This construction allows us to find novel formulae for the one-loop integrands for amplitudes with any multiplicity and helicity. The formulae obtained in this way agree with the ones derived via prescriptive unitarity. It makes prescriptive unitarity naturally emerge from this geometric description. }

\begin{document}

\maketitle


\section{Introduction}

In the past decade we have seen substantial progress in our fundamental understanding of scattering amplitudes. Amongst the most recent advances, the advent of a new geometric framework for scattering amplitudes, that of positive geometries \cite{Arkani-Hamed:2017tmz}, has led to new and important results in various theories, see \cite{Ferro:2020ygk, Herrmann:2022nkh} for comprehensive reviews. This new interpretation has revealed a deeper understanding of the mathematical structures which govern scattering amplitudes, including the relation to various topics in combinatorics, to positivity and to the theory of cluster algebras, as well as new methods of computation. In this approach, as opposed to classical methods based on Feynman diagrams, focus is placed on the kinematic space encoding the momenta of the scattering particles, and the amplitudes are rather obtained by studying particular algebraic data associated to subsets of the kinematic space.

The main focus of this paper will be on scattering amplitudes in $\mathcal{N}=4$ super Yang-Mills (sYM) for which there exist two positive geometry constructions to date: the amplituhedron \cite{Arkani-Hamed:2013jha}, and the momentum amplituhedron \cite{Damgaard:2019ztj,Ferro:2022abq}, both of which have their advantages and drawbacks. 
On the one hand, being defined in momentum twistor space, the amplituhedron encodes polygonal Wilson loops, rather than scattering amplitudes, in the planar sector.  
But, since the momentum twistors automatically encode momentum conservation, one works with a set of unconstrained variables which in turn makes many calculations easier.
On the other hand, the momentum amplituhedron is defined directly in spinor helicity space and can therefore {\it in principle} be extended beyond the planar sector. Moreover, it encodes directly the physics of scattering amplitudes and it allows for easier access to information on the singularity structure of amplitudes \cite{Ferro:2020lgp}, on the relation of gauge theory and scalar theory amplitudes \cite{Damgaard:2020eox}, and on the relation between various color-ordered amplitudes \cite{Damgaard:2021qbi}.

In this paper we introduce a third positive geometry describing scattering processes in planar $\mathcal{N}=4$ sYM which will be defined in the space of dual momenta. We claim this description combines the advantages of both the amplituhedron and momentum amplituhedron, and will allow us to write down novel formulae for the integrands of scattering amplitudes. 
More precisely, we will consider a close cousin of four-dimensional Minkowski space, the split-signature space $\mathbb{R}^{2,2}$, into which we will translate the momentum amplituhedron. Our construction will consist of two steps: first we will translate the tree-level momentum amplituhedron into dual space encoding for us the momenta of the scattering particles as a null polygon with particular positivity conditions. Then, for each such null polygon, we will identify a region in dual space corresponding to off-shell loop momenta, which will capture the one-loop problem and allow us to write down new formulae for one-loop integrands at any multiplicity and helicity. While here we will present only the one-loop case, this idea can be naturally generalised to the higher-loop problem by considering a collection of points in the one-loop region which satisfy additional mutual positivity conditions.

In this way of thinking about amplitudes in dual space we obtain a distinct loop geometry for every fixed tree-level null polygon. However, if one is interested in finding the canonical differential form for these geometries, that is in finding the scattering amplitude integrand, many of these geometries are identical. It is therefore reasonable to classify only the distinct loop geometries. This approach was pioneered by \cite{He:2023rou} where the loop amplituhedron geometry for ABJM theory \cite{Aharony:2008ug} was described as a fibration over the tree-level geometry. This naturally led to the notion of tree-level {\it chambers} as subsets of the tree-level amplituhedron, or momentum amplituhedron, which produce the same fibers. In particular within a given chamber the loop geometry has the same shape and combinatorial structure and therefore the same loop canonical form. This idea was further explored in \cite{Lukowski:2023nnf}, where the ABJM amplituhedron was translated to its corresponding dual momentum space, and led to a classification of chambers relevant for one-loop geometries for all particle multiplicities as well as new formulae for one-loop integrands in ABJM theory. The $\mathcal{N}=4$ sYM case is similar: various tree-level null polygons produce combinatorially inequivalent loop geometries, but there is a finite number of chamber geometries that we will be able to classify and study. Having found all chambers $\mathcal{C}_{n,k}$ for a given number of particles $n$ and given helicity sector $k$, the $L$-loop integrand can be written as  
\begin{align}
\Omega_{n,k,L} = \sum_{\mathfrak{c} \in \mathcal{C}_{n,k}} \Omega^{\mathrm{tree}}_{\mathfrak{c}  } \wedge \Omega^{\mathrm{loop}}_{\mathfrak{c}}\,,
\end{align}
where the canonical differential form in each chamber naturally factorises as a wedge product of tree and loop canonical forms. 

One important and unexpected property of the chamber geometries is that they are curvy versions of four-dimensional simple polytopes, i.e. polytopes for which every vertex is incident to exactly four edges. Moreover, all vertices of the chamber geometries are either vertices of the dual polygon or correspond to maximal (quadruple) cuts of amplitudes. This observation will allow us to write down an explicit differential form $\Omega_{\mathfrak{c}}^{\mathrm{loop}}$ at one loop as the sum of contributions over vertices of the chamber geometry. Importantly, the terms coming from null polygon vertices vanish and the final answer is written purely in terms of contributions coming from maximal cuts, making the answer manifestly compatible with the prescriptive unitarity approach of \cite{Bourjaily:2017wjl}.
In generalised unitarity the integrand, being a rational function, may be expanded in some basis of rational functions multiplied by free coefficients expressed in terms of on-shell functions \cite{Arkani-Hamed:2012zlh}. The free coefficients of the ansatz are then fixed by evaluating  all possible cuts and ensuring they match the results from field theory. The prescriptive unitarity approach of \cite{Bourjaily:2017wjl} takes this a step further by selecting a particularly nice {\it prescriptive} basis. This prescriptive basis can be thought of as a diagonalization with respect to taking residues on maximal cut solutions, i.e. each basis element contributes to a single maximal cut solution whilst vanishes on all others. As such the coefficient of each basis element is simply given by the maximal cut that defines it. Note that a prescriptive basis is highly desirable as it forgoes the need of solving cumbersome linear algebra problems on the coefficients. As we shall see through the course of this paper the new formulae we provide for the one-loop integrand satisfy this prescriptive property.
Therefore, prescriptive unitarity is not assumed but rather emerges naturally from the underlying geometric description. 

This paper is organized as follows. We start by recalling basic facts about the four-dimensional split-signature kinematic space in section \ref{sec:kinematic_space}. In section \ref{sec:mom_amp} we provide the definition of the momentum amplituhedron and translate it to the space of dual momenta. Then in section \ref{sec:chambers} we discuss the tree-level geometry and introduce the notion of chambers. In section \ref{sec:nullcone} we take a different approach on the loop geometry and describe it in terms of the null structure of the dual space. Section \ref{sec:combi} introduces a set of combinatorial labels that will allow us to describe explicit examples of chamber geometries in section \ref{sec:examples}. We will also conjecture there a general formula for one-loop integrands. In section \ref{sec:cluster_algebras} we provide a glimpse at the relation of our results to cluster algebras. We end the paper with Conclusions and Outlook.


\section{Four-dimensional Positive Kinematic Space}\label{sec:kinematic_space}
The momenta of scattered particles for physically relevant amplitudes are given by elements of Minkowski space $\mathbb{R}^{1,3}$. However, in order to use the positive geometries framework, we instead consider the scattering data to be in the split-signature space $\mathbb{R}^{2,2}$ where we take the signature to be $(+,+,-,-)$. Let us begin by reviewing some basic notions about this kinematic space.

The scattering data for $n$-particle massless scattering is encoded by a set of $n$ four-dimensional on-shell momenta $p_i^\mu$, where $i=1,\ldots,n$ and $\mu=1,\ldots,4$, subject to the on-shell condition $p^2=0$ and momentum conservation 
\begin{equation}
\sum_{i=1}^np_i^\mu=0\,.
\end{equation} 
In the planar theory, this data can be equivalently encoded using dual momentum coordinates $x_i^\mu$ defined as 
\begin{equation}\label{eq.dual}
p_i^\mu = x_{i+1}^\mu-x_{i}^\mu\,,
\end{equation}
with the $x_i$ subject to the periodic boundary condition $x_{n+1}\equiv x_1$. In analogy to Minkowski space, two points $x^\mu, y^\mu\in\mathbb{R}^{2,2}$ are said to be {\it null-separated} if
\begin{equation}
(x-y)^2:=(x^1 -y^1)^2+(x^2- y^2)^2-(x^3- y^3)^2-(x^4-y^4)^2=0\,.
\end{equation} 
The collection of dual momenta $x_i$ defines a null polygon in $\mathbb{R}^{2,2}$, where consecutive points $x_i$ and $x_{i+1}$ are null-separated. Note that the definition of the dual coordinates \eqref{eq.dual} is invariant under shifts of the $x_i$ by an arbitrary constant vector and for convenience we can choose $x_1=0$. This allows us to invert relation \eqref{eq.dual} to get
\begin{equation}\label{eq:p-to-dual}
x_j^\mu=\sum_{i=1}^{j-1}p_i^\mu\,.
\end{equation}

Moreover, the on-shell condition $p^2=0$ can be resolved by introducing spinor helicity variables and writing 
\begin{equation}\label{eq:lambda-to-p}
p^{a\dot a}=\begin{pmatrix}
p^0+p^2&p^1+p^3\\-p^1+p^3&p^0-p^2
\end{pmatrix}\:=\lambda^a\tilde\lambda^{\dot a}\,,
\end{equation}
where $a=1,2$, $\dot a=1,2$, and $\lambda$, $\tilde\lambda$ are real variables defined up to little group rescaling $\lambda \to t \lambda$, $\tilde\lambda\to t^{-1}\tilde\lambda$. In the following we will make use of the familiar spinor brackets
\begin{align}
\langle ij\rangle=\lambda^1_i\lambda^2_j-\lambda^2_i\lambda_j^1\,,\qquad \qquad[ ij]=\tilde\lambda^1_i\tilde\lambda^2_j-\tilde\lambda^2_i\tilde\lambda_j^1\,,
\end{align}
and  Mandelstam variables
\begin{equation}
s_{i_1,i_2,\ldots,i_r}=(p_{i_1}+p_{i_2}+\ldots p_{i_r})^2\,.
\end{equation}

An alternative way to encode the kinematic data of the planar theory is via momentum twistors \cite{Hodges:2009hk} defined as 
\begin{equation}
\label{eq:momtw}
z^A_i= (\lambda_{i}^{a}, \tilde\mu_i^{\dot a}) \equiv (\lambda_{i}^{ a}, x_i^{a\dot a}\lambda_{ia})\,,
\end{equation}
where the spinor indices have been lowered using the two-dimensional Levi-Civita symbol.  The spinor helicity variables $\widetilde\lambda$ are determined from momentum twistors via
\begin{equation}
\widetilde\lambda_i^{\dot a} =\frac{\langle i-1\,i\rangle \tilde\mu^{\dot a}_{i+1}+\langle i+1\,i-1\rangle \tilde\mu^{\dot a}_{i}+\langle i\,i+1\rangle \tilde\mu^{\dot a}_{i-1}}{\langle i-1\,i\rangle\langle i\,i+1\rangle}\,,
\end{equation}
allowing one to easily translate between momentum twistors $z$ and spinor helicity variables $(\lambda,\tilde\lambda)$.
Furthermore, one can write the following relation between invariant brackets in momentum twistor space and distances in the dual space:
\begin{equation}
x_{ij}^2 := (x_i-x_j)^2=\frac{\langle i-1\, i\, j-1\, j\rangle}{\langle i-1\, i\rangle \langle j-1\, j\rangle} \,,
\end{equation}
where $\langle ijkl\rangle=\epsilon_{IJKL}z_i^Iz_j^Jz_k^Kz_l^L$.
%


\section{Momentum amplituhedron in dual space}
\label{sec:mom_amp}
Having established the necessary facts about the kinematic space, in this section we recall the definition of the momentum amplituhedron $\mathcal{M}_{n,k,L}$ \cite{Damgaard:2019ztj,Ferro:2022abq}, a positive geometry \cite{Arkani-Hamed:2017tmz} which encodes N$^{(k-2)}$MHV$_n$ amplitudes in $\mathcal{N}=4$ sYM via its canonical differential form. Here we will also translate the definition of the momentum amplituhedron into the space of dual momenta in $\mathbb{R}^{2,2}$, which will allow us in the following sections to identify the loop level geometry by studying the null-cone structure of a null polygon in the split-signature kinematic space.


\subsection{Tree level}
Let us denote by $M(k,n)$ the space of all $k\times n$ matrices and by $M_+(k,n)$ the space of all positive matrices in $M(k,n)$, i.e. the subset of matrices with all ordered maximal minors positive. We denote by $G(k,n)$ the Grassmannian space of $k$-planes in $n$-dimensional Euclidean space and by $G_+(k,n)$ the positive part of $G(k,n)$ which consists of points that can be parametrized by positive matrices.
The tree-level momentum amplituhedron $\mathcal{M}_{n,k,0}$ \cite{Damgaard:2019ztj}  is defined as the image of the positive Grassmannian $G_+(k,n)$ through the map
\begin{align}
\Phi_{\overline{\Lambda},\widetilde \Lambda}: G_+(k,n) \to G(2,n) \times G(2,n), \quad  C \mapsto (\lambda, \tilde\lambda),
\label{eq:map_amp}
\end{align}
where $\overline{\Lambda} \in M_+(k-2,n)$ and $\widetilde \Lambda \in M_+(k+2,n)$ are two fixed positive matrices. Given a matrix $C=(c_{\alpha i})\in G_+(k,n)$, the map is defined as
\begin{equation}
\lambda_i^a = \sum_{\alpha=1}^k(X^\perp)_\alpha^a c_{\alpha i}, \quad \quad \tilde \lambda_i^{\dot a} = \sum_{\dot A=1}^{k+2} (\widetilde{Y}^\perp)_{\dot A}^{\dot a} \widetilde \Lambda_{i}^{\dot A},
\end{equation}
where we have  introduced 
\begin{equation}
X_\alpha^{\bar A}= \sum_{i=1}^n (\overline{\Lambda})_i^{\bar A} c_{\alpha i} , \quad \quad \widetilde Y_{ \alpha}^{\dot A} = \sum_{i=1}^n c_{\alpha i} \widetilde \Lambda_i^{\dot A}.
\end{equation}
Here $\bar A=1,\ldots, k-2$, $\dot A=1,\ldots,k+2$, and  $X^\perp\in M(2,k)$ (resp. $\widetilde{Y}^\perp\in M(2,k+2)$) is the orthogonal complement of $X\in M(k-2,k)$ (resp. $\widetilde Y\in M(k,k+2)$).  Importantly, after specifying particular constraints on the $(\overline{\Lambda},\widetilde \Lambda)$, see \cite{Damgaard:2019ztj} for further details, the following positivity conditions for elements $(\lambda,\tilde\lambda)\in\mathcal{M}_{n,k,0}$ are satisfied
\begin{itemize}
\item $\langle i i+1 \rangle>0$, $[ii+1]>0$ and $ s_{i,i+1,\ldots,i+r}>0$ for all $i,r=1,\ldots,n$,
\item the sequences $\{ \langle 12 \rangle,\langle 13\rangle,\ldots, \langle 1n \rangle\}$, $\ldots$, $\{ \langle n1 \rangle,\langle n2\rangle,\ldots, \langle n\,n-1 \rangle\}$  have $k-2$ sign flips,
\item the sequences $\{ [12],[13],\ldots, [1n]\}$, $\dots$, $\{ [n1],[n2],\ldots, [n\,n-1]\}$ have $k$ sign flips.
\end{itemize}
Note that every point $(\lambda,\tilde \lambda)$ generated by the map \eqref{eq:map_amp}, that is to say every point inside the tree-level momentum amplituhedron $\mathcal{M}_{n,k,0}$, can be translated to a configuration of points $x_i$ in dual space using \eqref{eq:p-to-dual} and \eqref{eq:lambda-to-p}. Therefore, every point $(\lambda,\tilde\lambda)\in \mathcal{M}_{n,k,0}$ defines the vertices of a null polygon whose edges encode the momenta of the scattering particles. The configuration of dual points generated by this procedure satisfy the following conditions that are direct translations of the conditions in momentum twistor space \cite{Arkani-Hamed:2017vfh}:
\begin{itemize}
\item $(x_i-x_j)^2\geq 0$ for all $|i-j|>1$,
\item the sequences 
$$\{\langle i+1\,i+2\rangle(x_i-\ell^*_{i+1\,i+2})^2,\langle i+1\,i+3\rangle(x_i-\ell^*_{i+1\,i+3})^2,\ldots,\langle i+1\,i-2\rangle(x_i-\ell^*_{i+1\,i-2})^2\}$$
 have $k-2$ sign flips for all $i=1,\ldots,n$,
\end{itemize}
where we have defined
\begin{equation}\label{eq:lij}
\ell_{ij}^*=\frac{1}{\langle ij\rangle}\left(\sum_{l=1}^{j-1}\langle lj\rangle \lambda_i\tilde\lambda_l-\sum_{l=1}^{i-1}\langle li\rangle \lambda_j\tilde\lambda_l\right).
\end{equation}
In particular we have $\ell^*_{ii+1}=x_{i+1}$. 
This sign flip prescription can be easily derived from the one in \cite{Arkani-Hamed:2017vfh} by noticing that
\begin{equation}
(x_i-\ell^*_{i+1j})^2=\frac{\langle i-1ii+1j\rangle}{\langle i-1i\rangle\langle i+1j\rangle},
\end{equation}
and the fact that the brackets $\langle i-1i\rangle$ are always positive. In the following we will mostly use the sign flip definition of the momentum amplituhedron in the dual space, however, the definition as the image through the function $\Phi_{\overline{\Lambda},\widetilde \Lambda}$ is useful when generating particular configurations of points satisfying the correct sign flip patterns.


\subsection{Loop level}\label{sec:loop_mom_amp}
For a given point in the tree-level momentum amplituhedron $(\lambda,\tilde\lambda)\in\mathcal{M}_{n,k,0}$ the loop momentum amplituhedron map \cite{Ferro:2022abq} is defined as
\begin{equation}
\begin{tabular}{cccc}
$\phi_{\lambda,\tilde\lambda}:$&$G(2,n)$&$\to$&$ \mathbb{R}^{2,2}$\\
&$D$&$\mapsto$& $y$
\end{tabular}
\end{equation}
given by
\begin{equation}
y^{\mu}=\ell^{\mu}+x_1^{\mu}=\frac{\sum_{i<j}(ij)_D \langle ij\rangle\,\ell_{ij}^{*\,\mu}}{\sum_{i<j}(ij)_D\langle ij \rangle}\,,
\end{equation}
where $(ij)_D$ are $2\times 2$ minors of the matrix $D$ and the bi-spinor version of $\ell_{ij}^*$ is defined in \eqref{eq:lij}. The one-loop momentum amplituhedron $\mathcal{M}_{n,k,1}$ is then defined as the image of a particular subset of matrices $D$, as explained in \cite{Ferro:2022abq}. Importantly, this translates into the following sign flip pattern: the point $y\in \mathbb{R}^{2,2}$ is  inside $\mathcal{M}_{n,k,1}$ if and only if
\begin{itemize}
\item $(y-x_i)^2\geq 0$ for all $i=1,\ldots,n$,
\item the sequences
 \begin{equation}\label{eq:signflipsforloops}
 \{\langle i\,i+1\rangle(y-\ell^*_{i\,i+1})^2,\langle i\,i+2\rangle(y-\ell^*_{i\,i+2})^2,\ldots,\langle i\,i+n-1\rangle(y-\ell_{i\,i+n-1})^2\}
\end{equation} 
 have $k$ sign flips for all $i=1,\ldots,n$. We pick up a factor of $(-1)^{k-1}$ for $\langle i\, a\rangle$ when $a>n$ due to the \emph{twisted cyclic symmetry}, see \cite{Arkani-Hamed:2017vfh} for details.
\end{itemize}
Therefore, the one-loop momentum amplituhedron $\mathcal{M}_{n,k,1}$ is a particular subset of $\mathbb{R}^{2,2}$ determined by the tree-level configuration $(\lambda,\tilde\lambda)$.

Moving to higher loops, the $L$-loop momentum amplituhedron for $L>1$ consists of points $(y_1,\ldots,y_L)\in (\mathcal{M}_{n,k,1})^L$, i.e all loop variables $y_l$ are inside the one-loop momentum amplituhedron, that additionally satisfy the following mutual positivity conditions
\begin{equation}
(y_l-y_m)^2>0\,,\text{ for } l\neq m=1,\ldots,L\,.
\end{equation}
In this paper we will solely focus on the one-loop geometry, however its structure will also have significant implications for understanding higher loops as well, a direction that we plan to explore in the future. 


\section{Chambers}\label{sec:chambers}

Ultimately, we are interested in finding explicit forms of tree-level scattering amplitudes and integrands for loop-level amplitudes from positive geometries. For planar $\mathcal{N}=4$ sYM, these are encoded as the canonical differential form of the momentum amplituhedron $\mathcal{M}_{n,k,L}$. At tree level, there exist various approaches to find the canonical differential form of the momentum amplituhedron $\mathcal{M}_{n,k,0}$, here we will mostly focus on triangulating it using the BCFW recursion relations \cite{Britto:2005fq}. In the context of positive geometries, BCFW triangulations mean a subdivision of $\mathcal{M}_{n,k,0}$ into smaller pieces, each of which is the image through the map \eqref{eq:map_amp} of a particular boundary component of the positive Grassmannian $G_+(k,n).$ Therefore, it is useful to first recall some facts about the boundary stratification of the positive Grassmannian. Following the work of Postnikov \cite{Postnikov:2006kva}, the boundary stratification of the positive Grassmannian $G_+(k,n)$ consists of the so-called positroid cells that are in one-to-one correspondence with affine permutations of $n$ elements. We will denote the positroid cell corresponding to a permutation $\sigma$ by $S_\sigma\subset G_+(k,n)$, and for its image through the map \eqref{eq:map_amp} we use $\Gamma_\sigma=\Phi_{\overline{\Lambda},\widetilde{\Lambda}}(S_\sigma)$.

We will be particularly interested in positroid cells in $G_+(k,n)$ of dimension $2n-4$ for which the dimension\footnote{There are two publicly available packages: \texttt{positroids} \cite{Bourjaily:2012gy} and \texttt{amplituhedronBoundaries} \cite{Lukowski:2020bya} that allow for an efficient calculations of the dimensions of BCFW cells and BCFW triangles.} of $\Gamma_\sigma$ is also $2n-4$. We call such cells {\it BCFW cells} and their images {\it BCFW triangles}\footnote{In literature one can also find the notion of BCFW tiles used instead of BCFW triangles.}. It is conjectured that the momentum amplituhedron $\mathcal{M}_{n,k,0}$ can be triangulated by collections of BCFW triangles. By triangulation $\mathcal{T}$ we mean a collection of BCFW cells $\{S_\sigma\}_{\sigma\in\mathcal{T}}$ such that their images $\{\Gamma_\sigma\}_{\sigma\in\mathcal{T}}$ are pairwise disjoint and their union covers $\mathcal{M}_{n,k,0}$. Generically, one can find multiple different BCFW triangulations for a given $n$ and $k$.

\begin{figure}[t]
\center
\begin{tikzpicture}[scale=2.5]
\coordinate (p11) at   (0.5-0.15,0.866-0.0866);
\coordinate (p12) at (1-0.15, 0-0.0866);
\coordinate (p21) at (1-0.15, 0+0.0866);
\coordinate (p22) at  (0.5-0.15, -0.866+0.0866);
\coordinate (p31) at (0.5, -0.866 +0.1732);
\coordinate (p32) at (-0.5, -0.866 +0.1732);
\coordinate (p41) at (-0.5+0.15, -0.866+0.0866);
\coordinate (p42) at (-1+0.15, 0+0.0866);
\coordinate (p51) at (-1+0.15, 0-0.0866);
\coordinate (p52) at (-0.5+0.15, 0.866-0.0866);
\coordinate (p61) at (-0.5, 0.866-0.1732);
\coordinate (p62) at (0.5, 0.866-0.1732);
\coordinate(p13) at (0.6525,0.11691);
\coordinate(p23) at (0.4275,-0.50661);
\coordinate(p33) at (-0.225,-0.62352);
\coordinate(p43) at (-0.6525,-0.11691);
\coordinate(p53) at (-0.4275,0.50661);
\coordinate(p63) at (0.225,0.62352);
\coordinate(p14) at (0.4275,0.50661);
\coordinate(p24) at (0.6525,-0.11691);
\coordinate(p34) at (0.225,-0.62352);
\coordinate(p44) at (-0.4275,-0.50661);
\coordinate(p54) at (-0.6525,0.11691);
\coordinate(p64) at (-0.225,0.62352);
\draw [red,line width = 0.3mm] plot [smooth,tension=0.8] coordinates  {(p53) (0.09, 0.249408) (p13)};
\draw [red,line width = 0.3mm] plot [smooth,tension=0.8] coordinates  {(p54) (-0.09, -0.15588) (p23)};
\draw [black,line width = 0.3mm] plot [smooth,tension=0.8] coordinates { (p11)  (p14) (p13) (p12)};
\draw [black,line width = 0.3mm] plot [smooth,tension=0.8] coordinates { (p21) (p24) (p23) (p22)};
\draw [black,line width = 0.3mm] plot [smooth,tension=0.8] coordinates  {(p31)(p34) (p33)  (p32)};
\draw [black,line width = 0.3mm] plot [smooth,tension=0.8] coordinates {(p41) (p44) (p43) (p42)};
\draw [black,line width = 0.3mm] plot [smooth,tension=0.8] coordinates  {(p51) (p54) (p53) (p52)};
\draw [black,line width = 0.3mm] plot [smooth,tension=0.8] coordinates  {(p61)(p64) (p63)  (p62)};
\fill[black] (0.16,0.43) circle (0 pt) node[]{{$6$}};
\fill[black] (0.0,0.03) circle (0 pt) node[]{{$4$}};
\fill[black] (-0.2,-0.35) circle (0 pt) node[]{{$2$}};
\end{tikzpicture}
\includegraphics[height=1.5cm]{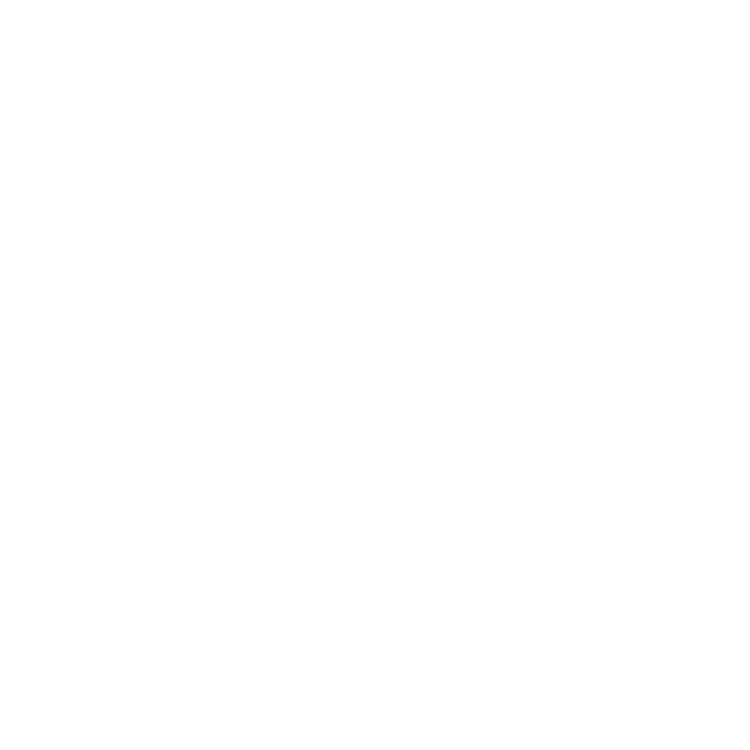}
\begin{tikzpicture}[scale=2.5]
\coordinate (p11) at   (0.5-0.15,0.866-0.0866);
\coordinate (p12) at (1-0.15, 0-0.0866);
\coordinate (p21) at (1-0.15, 0+0.0866);
\coordinate (p22) at  (0.5-0.15, -0.866+0.0866);
\coordinate (p31) at (0.5, -0.866 +0.1732);
\coordinate (p32) at (-0.5, -0.866 +0.1732);
\coordinate (p41) at (-0.5+0.15, -0.866+0.0866);
\coordinate (p42) at (-1+0.15, 0+0.0866);
\coordinate (p51) at (-1+0.15, 0-0.0866);
\coordinate (p52) at (-0.5+0.15, 0.866-0.0866);
\coordinate (p61) at (-0.5, 0.866-0.1732);
\coordinate (p62) at (0.5, 0.866-0.1732);
\coordinate(p13) at (0.6525,0.11691);
\coordinate(p23) at (0.4275,-0.50661);
\coordinate(p33) at (-0.225,-0.62352);
\coordinate(p43) at (-0.6525,-0.11691);
\coordinate(p53) at (-0.4275,0.50661);
\coordinate(p63) at (0.225,0.62352);
\coordinate(p14) at (0.4275,0.50661);
\coordinate(p24) at (0.6525,-0.11691);
\coordinate(p34) at (0.225,-0.62352);
\coordinate(p44) at (-0.4275,-0.50661);
\coordinate(p54) at (-0.6525,0.11691);
\coordinate(p64) at (-0.225,0.62352);
\draw [red,line width = 0.3mm] plot [smooth,tension=0.8] coordinates  {(p44) (-0.2025, 0.11691) (p63)};
\draw [red,line width = 0.3mm] plot [smooth,tension=0.8] coordinates  {(p33) (0.2025, -0.11691) (p14)};
\draw [black,line width = 0.3mm] plot [smooth,tension=0.8] coordinates { (p11)  (p14) (p13) (p12)};
\draw [black,line width = 0.3mm] plot [smooth,tension=0.8] coordinates { (p21) (p24) (p23) (p22)};
\draw [black,line width = 0.3mm] plot [smooth,tension=0.8] coordinates  {(p31)(p34) (p33)  (p32)};
\draw [black,line width = 0.3mm] plot [smooth,tension=0.8] coordinates {(p41) (p44) (p43) (p42)};
\draw [black,line width = 0.3mm] plot [smooth,tension=0.8] coordinates  {(p51) (p54) (p53) (p52)};
\draw [black,line width = 0.3mm] plot [smooth,tension=0.8] coordinates  {(p61)(p64) (p63)  (p62)};
\fill[black] (0.38,-0.2) circle (0 pt) node[]{{$5$}};
\fill[black] (0.0,0.0) circle (0 pt) node[]{{$3$}};
\fill[black] (-0.38,0.22) circle (0 pt) node[]{{$1$}};
\end{tikzpicture}
\caption{A sketch of the two BCFW triangulations of $\mathcal{M}_{6,3,0}$.}
\label{fig:BCFW63}
\end{figure}
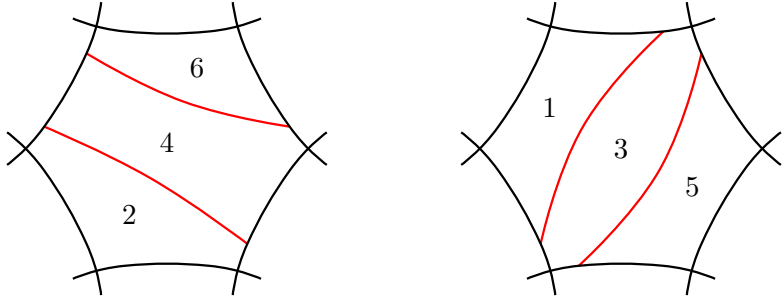

As an example, we recall the simplest case of non-trivial BCFW triangulations, relevant for the NMHV$_6$ amplitude. There are six BCFW cells in $G_+(3,6)$ defined as
\begin{equation}
S_{6,3}^{\{i\}}=\{C\in G_+(3,6):(i-1ii+1)_C=0\},
\end{equation}
where $(ijk)_C$ denotes the minors of the matrix $C\in G_+(3,6)$. 
The canonical form for $\mathcal{M}_{6,3,0}$ can be expressed in two ways 
\begin{align}
\Omega_{6,3,0} = \mathcal{I}_{6,3}^{\{1\}}+ \mathcal{I}_{6,3}^{\{3\}}+\mathcal{I}_{6,3}^{\{5\}}=  \mathcal{I}_{6,3}^{\{2\}}+ \mathcal{I}_{6,3}^{\{4\}}+ \mathcal{I}_{6,3}^{\{6\}},
\label{eq:BCFW_NMHV6}
\end{align}
where we have defined $\mathcal{I}_{6,3}^{\{ i\}}$ as the canonical form of the BCFW triangle $\Gamma_{6,3}^{\{ i \}}$. Geometrically, formula \eqref{eq:BCFW_NMHV6} translates into the fact that the momentum amplituhedron $\mathcal{M}_{6,3,0}$ can be divided into unions of BCFW triangles in two different ways:
\begin{equation}
\mathcal{M}_{6,3,0}=\Gamma_{6,3}^{\{1\}}\cup\Gamma_{6,3}^{\{3\}}\cup\Gamma_{6,3}^{\{5\}}=\Gamma_{6,3}^{\{2\}}\cup\Gamma_{6,3}^{\{4\}}\cup\Gamma_{6,3}^{\{6\}}\,,
\end{equation}
 as schematically illustrated in Fig. \ref{fig:BCFW63}.
At higher $n$ and $k$ one finds many BCFW triangulations, each of which provides an equivalent formula for the canonical differential form of the momentum amplituhedron.

\begin{figure}
\center
\begin{tikzpicture}[scale=3.5]
\coordinate (p11) at   (0.5-0.15,0.866-0.0866);
\coordinate (p12) at (1-0.15, 0-0.0866);
\coordinate (p21) at (1-0.15, 0+0.0866);
\coordinate (p22) at  (0.5-0.15, -0.866+0.0866);
\coordinate (p31) at (0.5, -0.866 +0.1732);
\coordinate (p32) at (-0.5, -0.866 +0.1732);
\coordinate (p41) at (-0.5+0.15, -0.866+0.0866);
\coordinate (p42) at (-1+0.15, 0+0.0866);
\coordinate (p51) at (-1+0.15, 0-0.0866);
\coordinate (p52) at (-0.5+0.15, 0.866-0.0866);
\coordinate (p61) at (-0.5, 0.866-0.1732);
\coordinate (p62) at (0.5, 0.866-0.1732);
\coordinate(p13) at (0.6525,0.11691);
\coordinate(p23) at (0.4275,-0.50661);
\coordinate(p33) at (-0.225,-0.62352);
\coordinate(p43) at (-0.6525,-0.11691);
\coordinate(p53) at (-0.4275,0.50661);
\coordinate(p63) at (0.225,0.62352);
\coordinate(p14) at (0.4275,0.50661);
\coordinate(p24) at (0.6525,-0.11691);
\coordinate(p34) at (0.225,-0.62352);
\coordinate(p44) at (-0.4275,-0.50661);
\coordinate(p54) at (-0.6525,0.11691);
\coordinate(p64) at (-0.225,0.62352);
\draw [red,line width = 0.3mm] plot [smooth,tension=0.8] coordinates  {(p53) (0.084375, 0.23382) (p13)};
\draw [red,line width = 0.3mm] plot [smooth,tension=0.8] coordinates  {(p54) (-0.084375, -0.146138) (p23)};
\draw [red,line width = 0.3mm] plot [smooth,tension=0.8] coordinates  {(p44) (-0.2025, 0.11691) (p63)};
\draw [red,line width = 0.3mm] plot [smooth,tension=0.8] coordinates  {(p33) (0.2025, -0.11691) (p14)};
\draw [black,line width = 0.3mm] plot [smooth,tension=0.8] coordinates { (p11)  (p14) (p13) (p12)};
\draw [black,line width = 0.3mm] plot [smooth,tension=0.8] coordinates { (p21) (p24) (p23) (p22)};
\draw [black,line width = 0.3mm] plot [smooth,tension=0.8] coordinates  {(p31)(p34) (p33)  (p32)};
\draw [black,line width = 0.3mm] plot [smooth,tension=0.8] coordinates {(p41) (p44) (p43) (p42)};
\draw [black,line width = 0.3mm] plot [smooth,tension=0.8] coordinates  {(p51) (p54) (p53) (p52)};
\draw [black,line width = 0.3mm] plot [smooth,tension=0.8] coordinates  {(p61)(p64) (p63)  (p62)};
\fill[black] (0.46,0.23) circle (0 pt) node[]{{\small $5 \cap 6$}};
\fill[black] (0.21,0.38) circle (0 pt) node[]{{\small $ 3 \cap 6$}};
\fill[black] (-0.15,0.5) circle (0 pt) node[]{{\small $ 1 \cap 6$}};
\fill[black] (-0.38,0.2) circle (0 pt) node[]{{\small $ 1 \cap 4$}};
\fill[black] (-0.5,-0.1) circle (0 pt) node[]{{\small $ 1 \cap 2$}};
\fill[black] (0.025,0.025) circle (0 pt) node[]{{\small $ 3 \cap 4$}};
\fill[black] (0.4,-0.1) circle (0 pt) node[]{{\small $ 5 \cap 4$}};
\fill[black] (0.15,-0.47) circle (0 pt) node[]{{\small $5 \cap 2$}};
\fill[black] (-0.18,-0.33) circle (0 pt) node[]{{\small $3 \cap 2$}};
\end{tikzpicture}
\caption{A sketch of the chamber decomposition of $\mathcal{M}_{6,3,0}$.}
\label{fig:chambers63}
\end{figure}
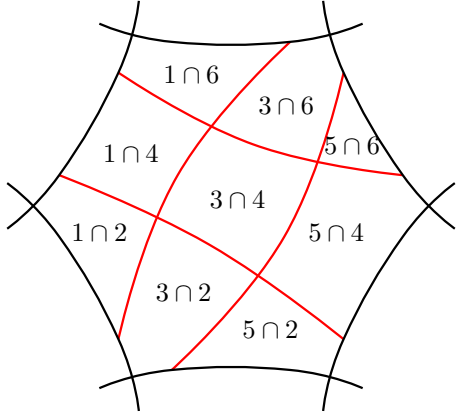

Importantly, a point in the tree momentum amplituhedron is in general an image of points in multiple BCFW cells, suggesting a natural alternative representation of the canonical form as a sum over the {\it maximal} intersection of BCFW cells. At six-points this decomposition of the tree-level geometry into maximal intersections is given explicitly by
\begin{align}
\Omega_{6,3,0}  =\ \mathcal{I}_{6,3}^{\{1\cap2\}}+  \mathcal{I}_{6,3}^{\{1\cap 4\}}+ \mathcal{I}_{6,3}^{\{1\cap6\}}
+  \mathcal{I}_{6,3}^{\{3\cap2\}}+ \mathcal{I}_{6,3}^{\{3\cap4\}}+  \mathcal{I}_{6,3}^{\{3\cap6\}} 
+  \mathcal{I}_{6,3}^{\{5\cap 2\}}+ \mathcal{I}_{6,3}^{\{5\cap4\}}+  \mathcal{I}_{6,3}^{\{5\cap6\}},
\end{align}
where we have defined $\mathcal{I}_{6,3}^{\{ i\cap j \}}$ as the canonical form associated to $\Gamma_{6,3}^{\{i\}}\cap \Gamma_{6,3}^{\{j\}}$, which is the {\it maximal} non-empty  intersection  of the images of BCFW cells for $n=6$, $k=3$.
Notice in particular the absence of the intersections of the form $\mathcal{I}_{6,3}^{\{ i\cap (i+2)\}}$ and $\mathcal{I}_{6,3}^{\{ i\cap (i+4)\}}$ which follows from non-overlapping of terms in formula \eqref{eq:BCFW_NMHV6}. The nine maximal intersections that of $\mathcal{M}_{6,3,0}$ are depicted in figure  \ref{fig:chambers63}.

We extend the notation that we introduced in the previous paragraph to all NMHV amplitudes by defining
\begin{equation}\label{eq:NMHVcells}
S_{n,3}^{\{i_1,i_2,\ldots,i_p\}}=\{C\in G_+(3,n):(i_1-1i_1i_1+1)_C=\ldots=(i_p-1i_pi_p+1)_C=0\}.
\end{equation}
Then, we can define $\Gamma_{n,3}^{\{i_1,i_2,\ldots, i_p\}}$ to be the image of $S_{n,3}^{\{i_1,i_2,\ldots, i_p\}}$ through the map \eqref{eq:map_amp}, and $\mathcal{I}_{n,3}^{\{i_1,i_2,\ldots,i_p\}}$ to be the canonical differential form of $\Gamma_{n,3}^{\{i_1,i_2,\ldots,i_p\}}$. More generally, beyond NMHV, we will use the notation $\mathcal{I}^{\sigma}_{n,k}$, where $\sigma$ labels a positroid cell in $G_+(k,n)$, to indicate the canonical differential form of the image of the cell $S_{n,k}^{\sigma}$.

At higher $n$ and $k$, the structure of maximal intersections of BCFW cells had not been previously studied, and we will partially fill this gap in the following. Importantly, it is much more intricate than that of $\mathcal{M}_{6,3,0}$. In particular, the number of BCFW cells that participate in a maximal intersection starts to vary beyond $n=6$, $k=3$ case.    
The maximal intersection of BCFW cells have been already used in the context of the momentum amplituhedron for ABJM theory \cite{He:2023rou} where they were termed {\it chambers}. They will also play a crucial role in this paper when we will study the loop geometry in the dual momentum space.

We start by introducing the  notion of compatibility of BCFW cells and using it for finding chambers for NMHV amplitudes. We will later comment on how to find chambers for $k>3$. 
We say that two BCFW cells $S^{\sigma_1}, S^{\sigma_2}\subset G_+(k,n)$ are {\it compatible}\footnote{We could equivalently consider compatibility using the amplituhedron instead of momentum amplituhedron, see App. \ref{app:chambers} for an explicit way of finding compatible BCFW cells from the NMHV amplituhedron $\mathcal{A}_{n,1,0}$.} if there exists a point $(\lambda,\tilde\lambda)$ in the momentum amplituhedron $\mathcal{M}_{n,k,0}$ such that $(\lambda,\tilde\lambda)=\Phi_{\overline{\Lambda},\widetilde{\Lambda}}(c_1)=\Phi_{\overline{\Lambda},\widetilde{\Lambda}}(c_2)$, for some $c_1\in S^{\sigma_1}$ and $c_2\in S^{\sigma_2}$. In other words two BCFW cells are compatible if their images through the tree-level momentum amplituhedron map overlap: $\Gamma^{\sigma_1}\cap\Gamma^{\sigma_2}\neq \emptyset$. Notice that if there exists a BCFW triangulation of the momentum amplituhedron that contains both cells $S^{\sigma_1}$ and $S^{\sigma_2}$, then $S^{\sigma_1}$ and $S^{\sigma_2}$ are {\it not} compatible. One can construct an adjacency graph $\mathcal{G}_{n,k}$ with vertices labelling the BCFW cells such that two vertices corresponding to cells $S^{\sigma_1}$ and $S^{\sigma_2}$ are adjacent if the cells are compatible. We conjecture that all chambers for the NMHV momentum amplituhedron $\mathcal{M}_{n,3,0}$ can be found as \emph{maximal cliques}, i.e. maximal complete subgraphs, of  $\mathcal{G}_{n,3}$. This method allows us to find all chambers for $\mathcal{M}_{n,3,0}$ up to $n=10$. The number of chambers is summarised in table \ref{tab:number_of_chambers}.
\begin{table}
\begin{center}
\begin{tabular}{c|c|c|c|c|c|c|c}
$n$&5&6&7&8&9&10\\
\hline
$\#$ chambers&1&9&71&728&15979&1144061\\
\hline
$\#$ BCFW triangulations&1&2&7&40&357&4824
\end{tabular}
\end{center}
\caption{Number of chambers and number of BCFW triangulations for the momentum amplituhedron $\mathcal{M}_{n,3,0}$.}
\label{tab:number_of_chambers}
\end{table}
Interestingly, if one takes the complement of the graph $\mathcal{G}_{n,3}$, namely the graph where vertices are connected when cells are not compatible, then the maximal cliques provide all possible BCFW triangulations of the momentum amplituhedron. This is a generalisation of the statement extensively explored in \cite{Lukowski:2020dpn} for the $m=2$ amplituhedron. The number of BCFW triangulations of $\mathcal{M}_{n,3,0}$, that can be found in table \ref{tab:number_of_chambers}, also agrees with the number of triangulations of a cyclic polytope in four dimensions and the sequence can be found on OEIS \cite{oeis}.

While we do not yet have a systematic way of finding all chambers beyond the NMHV case, it is easy to generate many chambers just by sampling points in the momentum amplituhedron $\mathcal{M}_{n,k,0}$. Given any point $(\lambda,\tilde\lambda)\in \mathcal{M}_{n,k,0}$ it is easy to find all BCFW triangles to which it belongs  by finding positive solutions, i.e. all $\alpha$'s positive, to the set of equations
\begin{equation}\label{eq:inBCFW}
C^\perp_\sigma(\alpha)\cdot\lambda=0\,,\qquad C_\sigma(\alpha)\cdot\tilde\lambda =0,
\end{equation}
where $C_\sigma(\alpha)$ is a parametrisation of the positroid cell $S_\sigma$. With this procedure we can generate a subset of chambers by performing the calculation for various points $(\lambda,\tilde\lambda)\in \mathcal{M}_{n,k,0}$. The important difference for $k>3$ is that some of the BCFW cells now have intersection number greater than one, i.e. generically there can be more than one solutions to the equations \eqref{eq:inBCFW}, see \cite{Arkani-Hamed:2012zlh} for more details. The first case where this happens is for the N$^2$MHV$_8$ amplitude, where there are two positroid cells in $G_+(4,8)$, $S_{8,4}^{\{6,5,8,7,10,9,12,11\}}$ and $S_{8,4}^{\{4,7,6,9,8,11,10,13\}}$, with intersection number equal two. To treat it properly we will split the cells into two parts, for example for $S_{8,4}^{\{6,5,8,7,10,9,12,11\}}$ we have
\begin{equation}\label{eq:split}
S_{8,4}^{\{6,5,8,7,10,9,12,11\}}=S_{8,4}^{\{6,5,8,7,10,9,12,11\},+}\cup S_{8,4}^{\{6,5,8,7,10,9,12,11\},-},
\end{equation}
where the $\pm$ indicates the sign in front of the square root in two solutions to \eqref{eq:inBCFW}. If we denote by $\Gamma_{8,4}^{\{6,5,8,7,10,9,12,11\},\pm}$ the images of this decomposition through the map $\Phi_{\overline{\Lambda},\widetilde{\Lambda}}$, then we find that $\Gamma^+\cap \Gamma^-\neq \emptyset$. This means that there can be chambers containing neither of $(\Gamma^+, \Gamma^-)$, one of them or even both. 
By sampling points in $\mathcal{M}_{8,4,0}$, we found that all these cases are realised. This means that the cells with intersection number two should be treated as unions  as in \eqref{eq:split}, with each summand treated as an independent BCFW triangle when looking for chambers.

The crucial role of chambers becomes evident at loop level, as already pointed out in \cite{He:2023rou} for the NMHV$_{6}$ amplitude based upon similar observations for ABJM theory. We find that for a fixed point inside the tree-level momentum amplituhedron, the shape of the loop momentum amplituhedron  varies from chamber to chamber but remains the same within a given chamber. This means that  the loop part of the canonical form $\Omega_{n,k,L}$ is the same within a given chamber, and the differential form of $\mathcal{M}_{n,k,L}$ factorises as the wedge product of the tree-form and loop-form for a given chamber. This leads to the following general decomposition of the full canonical form as the sum over chambers:
\begin{align}\label{eq:omegachambers}
\Omega_{n,k,L} = \sum_{\mathfrak{c} \in \mathcal{C}_{n,k}} \Omega^{\mathrm{tree}}_\mathfrak{c} \wedge \Omega^{\mathrm{loop}}_\mathfrak{c},
\end{align}
where $\mathcal{C}_{n,k}$ is the set of all chambers for $\mathcal{M}_{n,k,0}$.
As pointed out in \cite{He:2023rou}, this means that no new boundaries are introduced in the loop geometry, which is triangulated using the tree-level geometry, and the loop forms are local and give the integrands related to the prescriptive unitarity method of \cite{Bourjaily:2017wjl}.

Even though we do not have a systematic way of generating chambers, in the following sections, based on the study of examples for MHV and NMHV amplitudes, we will be able to conjecture the general form for one-loop integrands. These integrands successfully pass tests coming from generalized unitarity.

\section{Null-cone geometry}\label{sec:nullcone}
Having defined the notion of chambers, we proceed with studying the one-loop momentum amplituhedron geometry in dual space.
To understand its explicit shape, below we will take a slightly different approach compared to  section \ref{sec:loop_mom_amp} and will describe the one-loop geometry using the null structure of the kinematic space $\mathbb{R}^{2,2}$, similar to the approach pioneered in \cite{Lukowski:2023nnf} for ABJM theory. 

\subsection{Configurations of null-cones}

Let us start by defining for a given point $x\in \mathbb{R}^{2,2}$ its {\it null-cone} $\mathcal{N}_x$ as the set of all points that are null-separated from $x$:
\begin{equation}
\mathcal{N}_x=\{y\in\mathbb{R}^{2,2}:(x-y)^2=0\}\,.
\end{equation}
These are equivalent to light-cones in Minkowski space $\mathbb{R}^{1,3}$ and provide a natural generalisation of the three-dimensional light-cones studied in \cite{Lukowski:2023nnf}. In the following we will be interested in configurations of multiple null-cones and therefore we start here by a detailed study of their intersections. In the following we will use the fact that to a given point $(\lambda,\tilde\lambda)\in\mathcal{M}_{n,k,0}$ in the tree momentum amplituhedron we can associate a collection of points $x_i\in \mathbb{R}^{2,2}$, $i=1,\ldots,n$, forming a null polygon. 

\begin{figure}
\begin{center}
\begin{tikzpicture}[scale=1]
\draw[fill=blue,opacity=0.2] (-3,0,-3) -- (-3,0,3) -- (3,0,3) -- (3,0,-3) -- cycle;
\draw[fill=blue,opacity=0.1] (-3,-3,0) -- (-3,3,0) -- (3,3,0) -- (3,-3,0) -- cycle;
\draw[thick,color=red] (-3,0,0)-- (3,0,0);
\filldraw[black] (-2,0,0) circle (2pt) node[below] {$x_i$};
\filldraw[black] (1,0,0) circle (2pt) node[below] {$x_{i+1}$};
\begin{scope}[shift={(6,2)}]
    \draw (0,0) circle[radius=0.2];
    \draw[fill=gray!20] (1,0) circle[radius=0.4];
	\draw (0.18,0.07)  to [out=45,in=135] (0.67,0.2);
		\draw (0.18,-0.07)  to [out=-45,in=-135] (0.67,-0.2);
    \draw (-0.2,0)--(-0.5,0);
    \draw[color=red] (0.4,0.4)--(0.5,0.1);
    \draw[color=red] (0.4,-0.4)--(0.5,-0.1);
    \draw (1,0.4)--(1,0.8);
    \draw (1.33,0.25)--(1.7,0.6);
    \draw (1,-0.4)--(1,-0.8);
    \draw (1.33,-0.25)--(1.7,-0.6);
    \draw  (1.4,0)--(1.8,0);
\end{scope}

\begin{scope}[shift={(6,-0.5)}]
    \draw[fill=black] (0,0) circle[radius=0.2];
    \draw[fill=gray!20](1,0) circle[radius=0.4];
	\draw (0.18,0.07)  to [out=45,in=135] (0.67,0.2);
		\draw (0.18,-0.07)  to [out=-45,in=-135] (0.67,-0.2);
    \draw (-0.2,0)--(-0.5,0);
    \draw[color=red] (0.4,0.4)--(0.5,0.1);
    \draw[color=red] (0.4,-0.4)--(0.5,-0.1);
    \draw (1,0.4)--(1,0.8);
    \draw (1.33,0.25)--(1.7,0.6);
    \draw (1,-0.4)--(1,-0.8);
    \draw (1.33,-0.25)--(1.7,-0.6);
    \draw  (1.4,0)--(1.8,0);
\end{scope}

\draw[-stealth] (5,2) to [out=160,in=20] (3,2);
\draw[-stealth] (5,-0.5) to [out=160,in=20] (4,1);
\end{tikzpicture}
\end{center}
\caption{Intersection of two null-cones $\mathcal{N}_{x_i}$ and $\mathcal{N}_{x_{i+1}}$ for null-separated points $x_i$ and $x_{i+1}$. The planes are labelled by the two possible types of consecutive cuts of amplitude.}
\label{fig:Nxii1}
\end{figure}
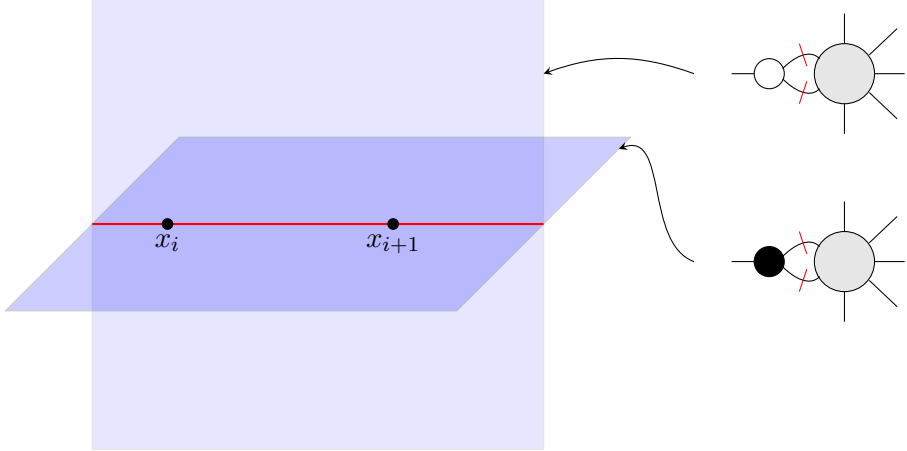

For two null-cones of generic points $x_i, x_j\in \mathbb{R}^{2,2}$, their intersection 
\begin{equation}
\mathcal{N}_{x_i}\cap \mathcal{N}_{x_j}=\{y\in \mathbb{R}^{2,2}:(y-x_i)^2=(y-x_j)^2=0\},
\end{equation}
is a union of two two-dimensional surfaces that correspond to the two solutions of the equations $(y-x_i)^2=(y-x_j)^2=0$. However, when the points $x_i$ and $x_j$ are null-separated, such as $x_i$ and $x_{i+1}$, the solutions linearize and the intersection $\mathcal{N}_{x_{i}}\cap \mathcal{N}_{x_{i+1}}$ is a union of two affine planes that intersect along a line, as illustrated in figure \ref{fig:Nxii1}. 
Importantly, further intersections of null-cones $\mathcal{N}_{x_j}$ with $\mathcal{N}_{x_i}\cap\mathcal{N}_{x_{i+1}}$ are always linear, which means that on each of these two-dimensional planes, a configuration of null-cones produces an arrangement of lines on a plane. A similar picture was already developed in twistor space in \cite{Herrmann:2020qlt}. 

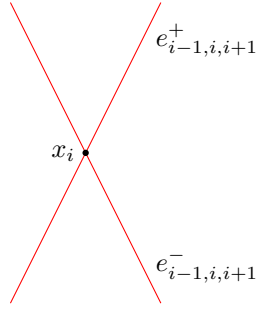
\begin{figure}
 \begin{center}
 \begin{tikzpicture}
\draw[red] (-1,-2) -- (1,2);
\draw[red] (-1,2) -- (1,-2);
\fill[black] (0.8,1.5) circle (0 pt) node[black,anchor=west]{{\small $e^+_{i-1,i,i+1}$}};
\fill[black] (0.8,-1.5) circle (0 pt) node[black,anchor=west]{{\small $e^-_{i-1,i,i+1}$}};
\fill[black] (0, 0) circle (1.2pt)node[anchor=east]{{\small $x_i$}};
\end{tikzpicture}
\end{center} 
\caption{Triple intersection of null-cones $\mathcal{N}_{x_{i-1}}\cap \mathcal{N}_{x_i}\cap \mathcal{N}_{x_{i+1}}$ for consecutive points on the null polygon.}
\label{fig:triple_intersection}
\end{figure}

The intersection of three null-cones 
 \begin{align}\label{eq:edges}
e_{ijk}^\pm =\mathcal{N}_{x_i}\cap \mathcal{N}_{x_j}\cap \mathcal{N}_{x_k}=\{y\in \mathbb{R}^{2,2}:(y-x_i)^2=(y-x_j)^2=(y-x_k)^2=0\}\,,
 \end{align}
 is generically the union of two curves in $\mathbb{R}^{2,2}$, however, when either two of the points $x_i$, $x_j$ and $x_k$ are null-separated then the intersection becomes two straight lines. Particularly interesting is the case when one takes three consecutive dual point $x_{i-1}$, $x_i$ and $x_{i+1}$ of the null polygon, in which case the two lines intersect at the point $x_i$, as illustrated in figure \ref{fig:triple_intersection}.

\begin{figure}
\center
\begin{tikzpicture}[scale=1]
\draw [line width=0.25mm ] (-2, -0.5) -- (0, -0.5) --(2, 0.5)--(0, 0.5) -- (-2,-0.5);
\draw [line width=0.25mm ] (-6, -1.5) -- (0, -1.5) --(6, 1.5)--(0, 1.5) -- (-6,-1.5);
\draw [line width=0.25mm](-2,-0.5) -- (2,0.5);
\draw [line width=0.25mm](0,-0.5) -- (0,0.5);
\draw [dashed,line width = 0.15mm] plot [smooth,tension=1] coordinates { (-2, -0.5) (-1.335+0.3, 1.035) (-0.67, 2.57)};
\draw [dashed,line width = 0.15mm] plot [smooth,tension=1] coordinates { (0, -0.5) (-0.335+0.05, 1.035) (-0.67, 2.57)};
\draw [dashed,line width = 0.15mm] plot [smooth,tension=1] coordinates { (2, 0.5) (0.665-0.4, 1.535) (-0.67, 2.57)};
\draw [dashed,line width=0.15mm] (0, 0.5) -- (-0.67, 2.57);
\draw [dashed,line width = 0.15mm] plot [smooth,tension=1] coordinates { (2, 0.5)  (1.335-0.3, -1.035) (0.67, -2.57)};
\draw [dashed,line width = 0.15mm] plot [smooth,tension=1] coordinates { (0, 0.5) (0.335-0.05, -1.035) (0.67, -2.57)};
\draw [dashed,line width = 0.15mm] plot [smooth,tension=1] coordinates { (-2,-0.5) (-0.665+0.4, -1.535) (0.67, -2.57)};
\draw [dashed,line width=0.15mm] (0, -0.5)-- (0.67, -2.57);
\fill[black] (-2, -0.5) circle (1.2pt) node[anchor=north east]{{\small $x_i$}};
\fill[black] (0, -0.5) circle (1.2pt)node[anchor=north east]{{\small $x_j$}};
\fill[black] (2, 0.5) circle (1.2pt)node[anchor=south west]{{\small $x_k$}};
\fill[black] (0, 0.5) circle (1.2pt)node[anchor=south west]{{\small $x_l$}};
\fill[black] (4.5, 1) circle (0 pt)node[anchor=south west]{{\small $3d$}};
\fill[black] (-0.67, 2.57) circle (1.2pt)node[anchor=south]{{\small $q^+_{ijkl}$}};
\fill[black] (0.67, -2.57) circle (1.2pt)node[anchor=north]{{\small $q^-_{ijkl}$}};
\end{tikzpicture}
\caption{ Quadruple intersection of four null-cones $\mathcal{N}_{x_i}\cap\mathcal{N}_{x_j}\cap\mathcal{N}_{x_k}\cap\mathcal{N}_{x_l}=\{q^+_{ijkl},q^{-}_{ijkl}\}$. The dashed lines in the figure are the intersections of three null-cones and {\it not} the null-lines connecting the points $q_{ijkl}^\pm$ with $x$'s.}
\label{fig:4point-conf}
\end{figure}
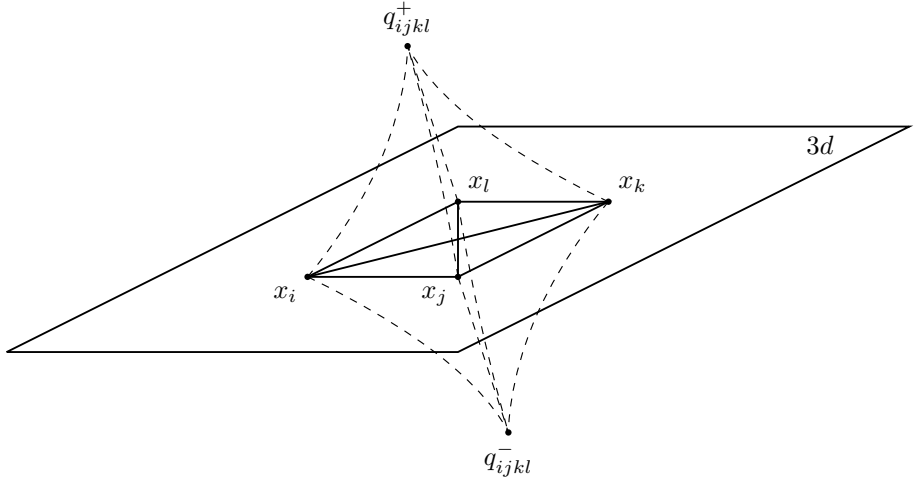

Finally, the null-cones of four generic points, $x_i,x_j,x_k,x_l\in  \mathbb{R}^{2,2}$, intersect at exactly two points  
\begin{equation}
\{q^+_{ijkl},q^{-}_{ijkl}\}=\mathcal{N}_{x_i}\cap\mathcal{N}_{x_j}\cap\mathcal{N}_{x_k}\cap\mathcal{N}_{x_l}\,,
\end{equation}
where the points are distinguished by the sign of the determinant\footnote{Notice that the assignment of $\pm$ depends on the ordering of points $x_i,x_j,x_k,x_l$. In particular, if we permute them according to a permutation $\tau$, the sign might change: $q_{\tau(i)\tau(j)\tau(k)\tau(l)}^\pm=q_{ijkl}^{sign(\tau)\pm}$.}
\begin{equation}
\mathcal{D}_{ijkl}(q)=\left|\begin{matrix}
1&1&1&1&1\\
x^\mu_i&x^\mu_j&x^\mu_k&x^\mu_l&q^\mu
\end{matrix}\right|\,.
\end{equation}
Notice that for given $x_i,x_j,x_k,x_l$, the kinematic space $\mathbb{R}^{2,2}$ is divided into two halfspaces $H_{ijkl}^+=\{q\in\mathbb{R}^{2,2}:\mathcal{D}_{ijkl}(q)>0\}$ and  $H_{ijkl}^-=\{q\in\mathbb{R}^{2,2}:\mathcal{D}_{ijkl}(q)<0\}$, with $q_{ijkl}^+\in H^+_{ijkl}$ and $q_{ijkl}^-\in H^-_{ijkl}$.
Geometrically, the points $x_i,x_j,x_k,x_l$ define a three-dimensional affine linear subspace in $\mathbb{R}^{2,2}$, and the two quadruple intersections $q_{ijkl}^\pm$ sit on opposite sides of this subspace, as illustrated in figure \ref{fig:4point-conf}.

Related to null-cones, we also define the {\it non-negative part} and {\it non-positive part} of the $\mathbb{R}^{2,2}$ with respect to point $x$:
\begin{equation}
\mathcal{N}^+_x=\{y\in\mathbb{R}^{2,2}:(x-y)^2\geq0\}\,,\qquad \mathcal{N}^-_x=\{y\in\mathbb{R}^{2,2}:(x-y)^2\leq0\}\,.
\end{equation}
Then, motivated by the construction in \cite{Lukowski:2023nnf}, we define 
\begin{equation}
\mathcal{K}_{n,k}(x):=\{y\in \mathbb{R}^{2,2}:(y-x_i)^2\geq0 \text{ for all } i=1,\ldots,n \}=\bigcap_{i=1}^n \mathcal{N}_{x_i}^+,
\end{equation}
that is the intersection of the non-negative parts $\mathcal{N}^+_{x_i}$ for all $x_i$ of the null polygon. The set $\mathcal{K}_{n,k}(x)$ is non-empty since for $i=1,\ldots,n$ we have $x_i\in \mathcal{K}_{n,k}(x)$. Moreover, it naturally decomposes into two regions: compact and non-compact
\begin{equation}
\mathcal{K}_{n,k}(x)=\Delta_{n,k}(x)\cup \overline{\Delta}_{n,k}(x).
\end{equation}
We claim that the compact region $\Delta_{n,k}(x)$ is the one-loop momentum amplituhedron $\mathcal{M}_{n,k,1}$ for the chosen tree-level configuration $(\lambda,\tilde\lambda)$!  An immediate implication of this statement is that the only vertices of $\Delta_{n,k}(x)$ are the null-polygon points $x_i$ together with quadruple intersections $q_{ijkl}^\pm$ of null-cones. In the generic case, only a subset of the quadruple intersections sit in the non-negative part of all $x_i$, and therefore not all of them are vertices of the compact region $\Delta_{n,k}(x)$. As we will see in the examples below, the explicit set of vertices of $\Delta_{n,k}(x)$ will depend on the tree-level configuration of points $x_i$ that will be compatible with our definition of tree-level chambers.
In order to determine which vertex is inside the geometry $\Delta_{n,k}(x)$ we use the sign-flip description of points \eqref{eq:signflipsforloops}. This has to be done carefully, since by the definition of points $q_{ijkl}^\pm$, some of the distances $(q_{ijkl}^\pm-x_m)^2$ are zero. To resolve it, one allows for all zero entries in the sign-flip sequences to be replaced by any combination of positive and negative entries. If there exists a resolution of zeros that gives the proper sign-flip pattern, then the point $q_{ijkl}^+$ is inside $\Delta_{n,k}(x)$. 

As an example, we will check whether $\ell^*_{13}=q_{1234}^+$ is a vertex of $\Delta_{4,2}(x)$. By direct calculation we find that
\begin{align}
	\begin{pmatrix}
		\text{sign}_{12} & \text{sign}_{13} & \text{sign}_{14} \\ 
		\text{sign}_{23} & \text{sign}_{24} & -\text{sign}_{21}\\ 
		\text{sign}_{34} & -\text{sign}_{31} & -\text{sign}_{32}\\
		-\text{sign}_{41} & -\text{sign}_{42} & -\text{sign}_{43}\\ 
	\end{pmatrix} = \begin{pmatrix}
		0 & 0 & 0\\ 0 & -1 & 0\\ 0 & 0 & 0\\ 0& -1 & 0 
	\end{pmatrix},
\end{align}
where $\text{sign}_{ab}\coloneqq\mathrm{sign}(\langle ab \rangle (\ell^*_{13}-\ell^*_{ab})^2)$, and the minus signs come from twisted cyclic symmetry. In order for $\ell^*_{13}$ to be a vertex of the geometry, we need to be able to find a resolution of this matrix such that each row has exactly two sign-flips, taking care of the fact that $\text{sign}_{ab}=-\text{sign}_{ba}$. In this case it is easy to see that the resolution 
\begin{align}
	\text{sign}_{12}\to 1,\, \text{sign}_{13}\to-1,\, \text{sign}_{14}\to 1,\, \text{sign}_{23}\to1,\,\text{sign}_{24}\to -1, \,\text{sign}_{34}\to 1\,,
\end{align}
satisfies these properties, and hence we find that $\ell^*_{13}$ is indeed one of the vertices of $\Delta_{4,2}(x)$. Similar calculations can be easily done for any $n$ and $k$ for a given configuration of points $x$ and a given quadruple intersection $q^\pm_{ijkl}$.

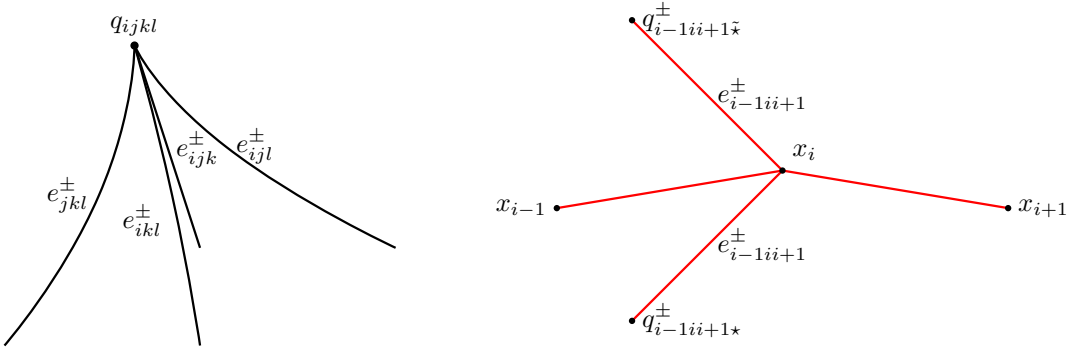
\begin{figure}
\center
\begin{tikzpicture}[scale=1.3]
\draw [black,line width = 0.3mm] plot [smooth,tension=1] coordinates { (-2, -0.5) (-1.335+0.3, 1.035) (-0.67, 2.57)};
\draw [black,line width = 0.3mm] plot [smooth,tension=1] coordinates { (0, -0.5) (-0.335+0.05, 1.035) (-0.67, 2.57)};
\draw [black,line width = 0.3mm] plot [smooth,tension=1] coordinates { (2, 0.5) (0.665-0.4, 1.535) (-0.67, 2.57)};
\draw [black,line width=0.3mm] (0, 0.5) -- (-0.67, 2.57);
\fill[black] (-1.335+0.3, 1.035) circle (0 pt) node[anchor=east]{{\small $e^\pm_{jkl}$}};
\fill[black] (-0.335+0.05, 1.035) circle (0 pt) node[anchor=north east]{{\small $e^\pm_{ikl}$}};
\fill[black] (0.665-0.4, 1.535) circle (0 pt) node[anchor=west]{{\small $e^\pm_{ijl}$}};
\fill[black] (-0.35,1.5) circle (0 pt) node[black,anchor=west]{{\small $e^\pm_{ijk}$}};
\fill[black] (-0.67, 2.57) circle (1.2pt)node[anchor=south]{{\small $q_{ijkl}$}};
\end{tikzpicture}
\begin{tikzpicture}[scale=1]
\draw [red,line width=0.3mm] (-3,-0.5) -- (0,0);
\draw [red,line width=0.3mm] (3,-0.5) -- (0,0);
\draw [red,line width=0.3mm] (-2,- 2) -- (0,0);
\draw [red,line width=0.3mm] (-2,2) -- (0,0);
\fill[black] (-5,-2.3) circle (0 pt);
\fill[black] (-1,1) circle (0 pt) node[anchor=west]{{\small $e_{i-1ii+1}^\pm$}};
\fill[black] (-1,-1) circle (0 pt) node[anchor=west]{{\small $e_{i-1ii+1}^\pm$}};
\fill[black] (0,0) circle (1.2pt)node[anchor=south west]{{\small $x_{i}$}};
\fill[black] (-3,-0.5) circle (1.2pt)node[anchor=east]{{\small $x_{i-1}$}};
\fill[black] (3,-0.5) circle (1.2pt)node[anchor=west]{{\small $x_{i+1}$}};
\fill[black] (-2,-2) circle (1.2pt)node[anchor=west]{{\small $q^\pm_{i-1ii+1\star}$}};
\fill[black] (-2,2) circle (1.2pt)node[anchor=west]{{\small $q^\pm_{i-1ii+1\tilde\star}$}};
\end{tikzpicture}
\caption{ Local geometry around points $q^{\pm}_{ijkl}$ and $x_i$ with null lines indicated in red. }
\label{fig:simple-vertices}
\end{figure}

Since the compact region $\Delta_{n,k}(x)$ coincides with the one-loop momentum amplituhedron, it is paramount to understand the properties of this region. We have done an exhaustive study of examples to arrive at the following set of properties of $\Delta_{n,k}(x)$.
First of all, we find that every vertex of $\Delta_{n,k}(x)$ has exactly four edges meeting at the vertex, see Fig. \ref{fig:simple-vertices}. This statement is rather trivial for the quadruple intersections $q^\pm_{ijkl}\in\Delta_{n,k}(x)$, for which the edges incident to $q^\pm_{ijkl}$ are $\{e_{ijk}^\pm,e_{ijl}^\pm,e_{ikl}^\pm,e_{jkl}^\pm\}$ where $e^\pm$ are defined in \eqref{eq:edges}.
Less trivially, for all $\Delta_{n,k}(x)$ that we studied, we find that for every $i=1,\ldots,n$ exactly two vertices of the form $q^{s_j}_{i-1ii+1j}$ for some $j=\star,\tilde\star$ and $s_j=\pm$, are inside $\Delta_{n,k}(x)$. Then the four edges incident to the vertex $x_i$ are: two sections of the lines $e^\pm_{i-1ii+1}$ connecting $x_i$ to vertices $q^{s_{j_1}}_{i-1ii+1\star}$ and $q^{s_{j_2}}_{i-1ii+1\tilde\star}$ respectively, together with two edges of the null polygon that connect $x_i$ to $x_{i-1}$ and $x_{i+1}$. Since all vertices  have four incident edges, we can think of  $\Delta_{n,k}(x)$ as a curvy version of a {\it simple polytope} in four dimensions. Importantly, in contrast to ordinary simple polytopes, each vertex might be incident to less than four facets of the geometry, which will have important implications for the canonical form that we will associate to $\Delta_{n,k}(x)$.

\subsection{Differential forms}

Similar to the ABJM construction developed in \cite{Lukowski:2023nnf}, the fact that $\Delta_{n,k}(x)$ is a curvy version of a simple polytope motivates us to write its canonical differential form as the sum of contributions coming from  all vertices of the geometry:
\begin{equation}\label{eq:omega-sum-over-vertices}
\Omega\left[\Delta_{n,k}(x)\right]=\sum_{v\in \mathcal{V}(\Delta_{n,k}(x))}\omega_v,
\end{equation}
where $\mathcal{V}(\Delta_{n,k}(x))$ denotes the set of all vertices of $\Delta_{n,k}(x)$. 
In particular, the set  $\mathcal{V}(\Delta_{n,k}(x))$ contains all vertices $x_i$ and a subset of the quadruple intersections of null-cones $q_{ijkl}^\pm$.
The points  $q_{ijkl}^\pm$ contribute a differential form that is a wedge product of $\dd\log$ forms associated to each facet meeting at that point:
\begin{equation}\label{eq:respm}
\omega_{ijkl}= \dd\log(x-x_i)^2\wedge  \dd\log(x-x_j)^2\wedge  \dd\log(x-x_k)^2 \wedge  \dd\log(x-x_l)^2.
\end{equation}
The forms $\omega_{ijkl}$ have unit residues at both points $q^{+}_{ijkl}$ and $q^-_{ijkl}$, and in order to remove the contribution from one of them we need to find another differential form with residues $\pm1$ on vertices $q_{ijkl}^\pm$. The natural candidate is the {\it box integrand}
\begin{align}\label{eq:box-integrand}
\omega_{ijkl}^\square&=\pm \dd\log\frac{(x-x_i)^2}{(x-q^\pm_{ijkl})^2}\wedge  \dd\log\frac{(x-x_j)^2}{(x-q^\pm_{ijkl})^2}\wedge  \dd\log\frac{(x-x_k)^2}{(x-q^\pm_{ijkl})^2} \wedge  \dd\log\frac{(x-x_l)^2}{(x-q^\pm_{ijkl})^2},\\
&=\frac{4\Delta (x_i-x_k)^2(x_j-x_l)^2 \, \dd^4x}{(x-x_i)^2(x-x_j)^2(x-x_k)^2(x-x_l)^2}\,,\qquad \Delta=\sqrt{(1-u-v)^2-4uv},
\end{align}
where $u=\frac{(x_i-x_j)^2(x_k-x_l)^2}{(x_i-x_k)^2(x_j-x_l)^2}$ and $v=\frac{(x_j-x_k)^2(x_i-x_l)^2}{(x_i-x_k)^2(x_j-x_l)^2}$ are two cross-ratios. It is a straightforward generalisation of the triangle integral used for ABJM theory in \cite{Lukowski:2023nnf}. 
We can therefore define the following combination
\begin{equation}\label{eq:combination}
\omega_{ijkl}^\pm=\frac{1}{2}(\omega_{ijkl}^\square\pm \omega_{ijkl}),
\end{equation}
which has the desired property of having non-vanishing residue only on one of the points $q^\pm_{ijkl}$. In particular, we have the following properties
\begin{equation}\label{eq:resqijkl}
\mathop{\text{Res}}_{x=q^\pm_{ijkl}}\omega_{ijkl}^\pm=1\,,\qquad\mathop{\text{Res}}_{x=q^\pm_{ijkl}}\omega_{ijkl}^\mp=0\,.
\end{equation}

Importantly, for $v=x_i$ we find $\omega_v=0$ since less than four facets meet at these vertices. This means that the dual points $x_i$ do not contribute to the sum in \eqref{eq:omega-sum-over-vertices}. At first look it is quite surprising since naively it would indicate that the form $\Omega\left[\Delta_{n,k}(x)\right]$ has vanishing residues at points $x_i$. However, we notice that the forms $\omega^\pm_{i-1ii+1j}$ have a non-zero (composite) residue on points $x_i$, coming from the box integrands, that evaluates to
\begin{equation}\label{eq:resx}
\mathop{\text{Res}}_{x=x_i}\omega_{i-1ii+1j}^\pm=\frac{1}{2}\,.
\end{equation}
Since, as we already pointed out, exactly two points of the form $q^\pm_{i-1ii+1j}$ for $i=1,\ldots,n$ are vertices of $\Delta_{n,k}(x)$, the differential form on $\Delta_{n,k}(x)$ has residues equal to 1 for all vertices of the geometry, as they should to be a positive geometry.

Given these properties of the differential forms $\omega_{ijkl}^\pm$, we arrive at a simple formula for the canonical form of $\Delta_{n,k}(x)$
\begin{align}
	\Omega\left[\Delta_{n,k}(x)\right]=\sum_{q^\pm_{ijkl}\in\mathcal{V}(\Delta_{n,k}(x))} \text{sgn}_{ijkl}\; \omega^\pm_{ijkl}\,.
\end{align}
The factor $\text{sgn}_{ijkl}$ denotes a relative sign between the terms in this sum, which can be determined by requiring that the form is \emph{projectively invariant} \cite{Arkani-Hamed:2017mur}. We say that the form $\Omega\left[\Delta_{n,k}(x)\right]$ is projectively invariant if it is invariant under local $GL(1)$ transformations 
\begin{equation}
(x-v)^2 \to \Lambda(x) (x-v)^2\,,
\end{equation} 
where $v$ is any vertex of $\Delta_{n,k}(x)$ and $\Lambda(x)$ is any function of $x$. This is essentially the same as requiring that $\Omega\left[\Delta_{n,k}(x)\right]$ can be written purely in terms of ratios of distances $\smash{\frac{(x-v_1)^2}{(x-v_2)^2}}$.  
From its explicit expression \eqref{eq:box-integrand} it is clear that $\omega^\square_{ijkl}$ is manifestly projectively invariant, we are thus only concerned with the relative signs of $\omega_{ijkl}$. We have done extensive checks for $k=2$ (for all $n$) and $k=3$ (up to $n=9$), and to a lesser extent for $k=4$, and found that the forms $\Omega\left[\Delta_{n,k}(x)\right]$ are projectively invariant if  $\text{sgn}_{ijkl}=1$ for all $i,j,k,l$! Interestingly, this is the same statement as was found for ABJM theory in \cite{Lukowski:2023nnf}. We thus arrive at the following compact result
\begin{align}\label{eq:form_final}
	\Omega\left[\Delta_{n,k}(x)\right]=\sum_{q^\pm_{ijkl}\in\mathcal{V}(\Delta_{n,k}(x))} \omega^\pm_{ijkl}\,.
\end{align}
To find an explicit form for a given $\Delta_{n,k}(x)$, it is therefore sufficient to find the vertex set $\mathcal{V}(\Delta_{n,k}(x))$, a simple task when we know the explicit positions of points $x_i$.

We conclude this section by introducing the so-called {\it parity-odd pentagon integrands} \cite{Bourjaily:2017wjl} that will be useful in the following sections:
\begin{align}
\omega^{\pentagon}_{ijklm}&= \dd\log\frac{(x-x_i)^2}{(x-x_m)^2}\wedge  \dd\log\frac{(x-x_j)^2}{(x-x_m)^2}\wedge  \dd\log\frac{(x-x_k)^2}{(x-x_m)^2} \wedge  \dd\log\frac{(x-x_l)^2}{(x-x_m)^2}\notag\\
&=\frac{\dd^4x \,\epsilon(x,i,j,k,l,m)}{(x-x_i)^2(x-x_j)^2(x-x_k)^2(x-x_l)^2(x-x_m)^2}\,,
\end{align}
where
\begin{equation}\label{eq:pentagondet}
\epsilon(x,i,j,k,l,m)=\left|\begin{matrix}
0&(x-x_i)^2&(x-x_j)^2&(x-x_k)^2&(x-x_l)^2&(x-x_m)^2\\
1&1&1&1&1&1\\
x^\mu&x^\mu_i&x^\mu_j&x^\mu_k&x^\mu_l&x^\mu_m
\end{matrix}\right|.
\end{equation}
The pentagon integrands are cyclically symmetric
\begin{equation}
\omega^{\pentagon}_{ijklm}=\omega^{\pentagon}_{jklmi}=\ldots=\omega^{\pentagon}_{mijkl}
\end{equation}
and satisfy
\begin{equation}\label{eq:perntagonformexp}
\omega^{\pentagon}_{ijklm}=\omega_{ijkl}-\omega_{ijkm}+\omega_{ijlm}-\omega_{iklm}+\omega_{jklm}\,,
\end{equation}
where the expansion \eqref{eq:perntagonformexp} can be obtained by simply expanding the determinant \eqref{eq:pentagondet} with respect to the first row.  
Importantly, the pentagon integrands integrate to zero on the Minkowski contour and therefore do not contribute to the integrals for physical amplitudes (and are therefore often neglected). We will however see in our construction that these integrands naturally arise from the null-cone geometry.


\section{Graphical representations of points in chamber geometries}\label{sec:combi}

Before proceeding with studying explicit examples of geometries $\Delta_{n,k}(x)$, in this section we introduce certain graphs encoding combinatorial information that will allow us to enumerate particular subsets of points in $\Delta_{n,k}(x)$, including its boundaries.

Let us consider an $n$-gon with boundary vertices  labelled $1,2,\ldots, n$ clockwise we call {\it corners}: these represent the vertices $x_i$ of $\Delta_{n,k}(x)$. A marked point  $y$ in the interior of the $n$-gon represents a point $y$ inside $\Delta_{n,k}(x)$. An edge internal to the $n$-gon connecting the marked point $y$ to a corner $i$ indicates that $(y-x_i)^2=0$, namely that the momentum $y-x_i$ is on-shell. Placing the difference $y-x_i$ on-shell is usually referred to as {\it cutting a propagator}, and therefore the internal edges in our labels correspond to cuts. It is important to notice that an off-shell momentum in four dimensions can only be cut four times and therefore there will be at most four internal edges in our diagrams. 
If there are two or more internal edges, the $n$-gon will be subdivided into several smaller polygons $P_a$. To each of these polygons we associate a \emph{helicity} $k_a$, which is an integer $k_a\in\{2,\ldots,m-2\}$, where $m$ is the number of vertices of $P_a$. Polygons $P_a$ with three edges, namely triangles, are an exception to this rule and they are allowed to have helicity 1 or 2. Additionally, we do not allow triangles with the same helicity to share an edge. This would imply a collinear limit, and therefore a specific kinematic configuration, while here we only consider generic kinematics. Later in this section we will explain how to assign helicities to each of the polygons $P_a$.
We further define the total helicity of the diagram to be the sum of helicities of all polygons $P_a$ minus the number of internal edges. A graph with $n$ corners and  total helicity $k$ is said to be of \emph{type} $(n,k)$. Its \emph{dimension} equals four minus the number of internal edges, and represents the number of degrees of freedom of point $y$ in this configuration. 

Additionally, we introduce two special types of graphs: a zero-dimensional diagram with the marked point on a corner of the polygon, which represents the case when $y=x_i$, and a one-dimensional diagram without a marked point, but with a marked edge connecting two adjacent corners of the polygon, which represents a configuration where $y$ is on the null-edge between $x_i$ and $x_{i+1}$. We conjecture that every boundary of $\Delta_{n,k}(x)$ can be labelled by a diagram of type $(n,k)$. For example, all boundaries of $\Delta_{4,2}(x)$ are depicted in table \ref{tab:n4-boundaries}. We note that $\Delta_{4,2}(x)$ has the same boundary structure as $G_+(2,4)$, as expected.
\begin{table}
	\begin{center}	\begin{tabular}{|c| c c c|} 
			\hline
			$d=4$ & $\begin{gathered}
				\includegraphics[ height=2.2cm]{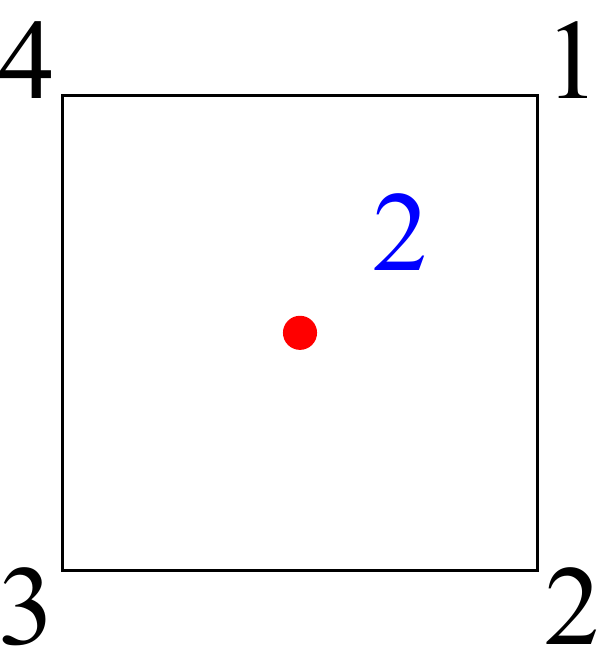} \end{gathered}\times 1$ & &  \\ 
			\hline
			$d=3$ & $\begin{gathered}
				\includegraphics[ height=2.2cm]{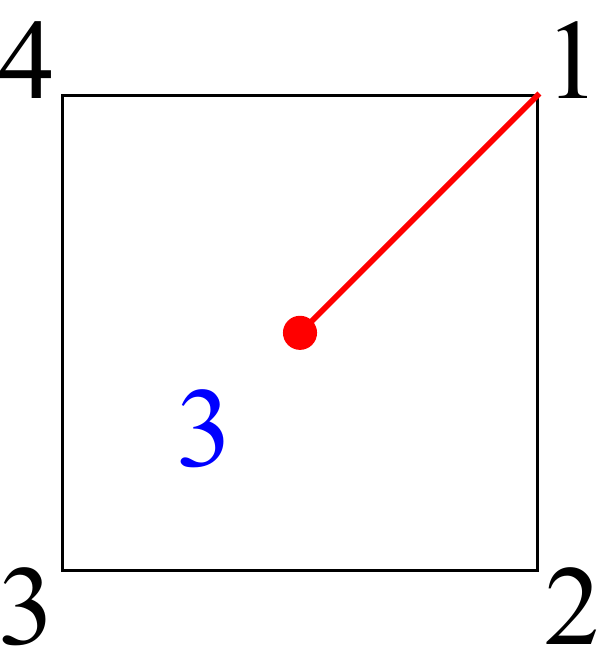} \end{gathered}\times 4$ &  &  \\
			\hline
			$d=2$ & $\begin{gathered}
				\includegraphics[ height=2.2cm]{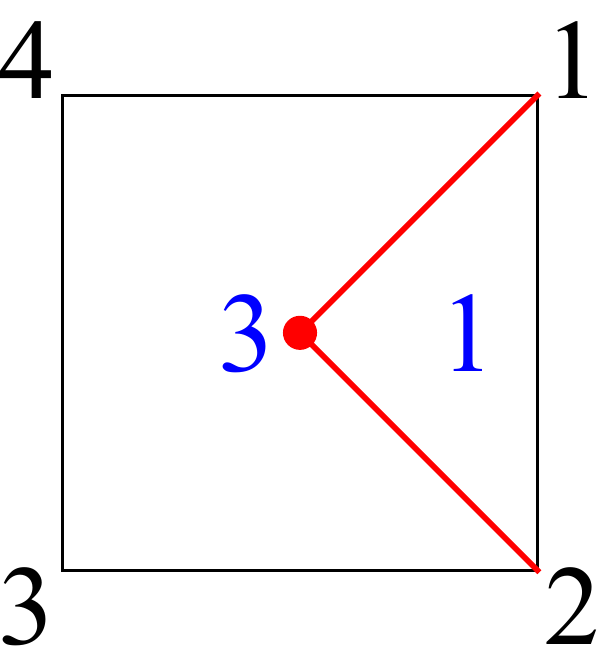} \end{gathered}\times 4$ & $\begin{gathered}
				\includegraphics[ height=2.2cm]{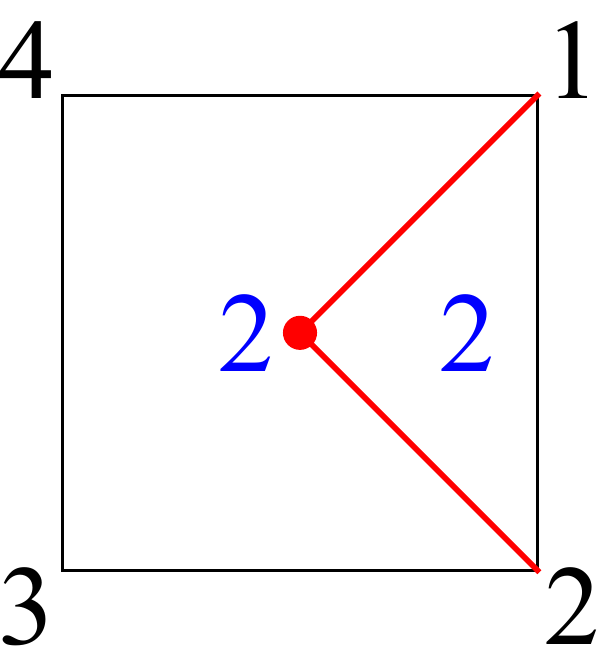} \end{gathered}\times 4$ & $\begin{gathered}
				\includegraphics[ height=2.2cm]{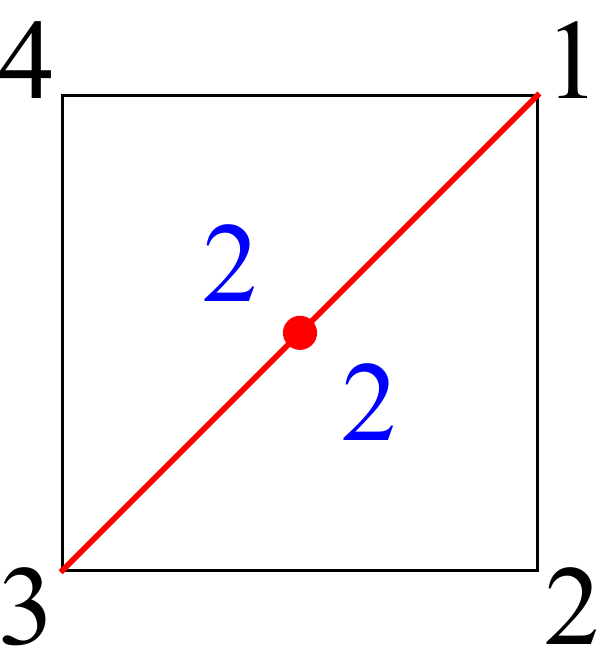} \end{gathered}\times 2$ \\
			\hline
			$d=1$ & $\begin{gathered}
				\includegraphics[ height=2.2cm]{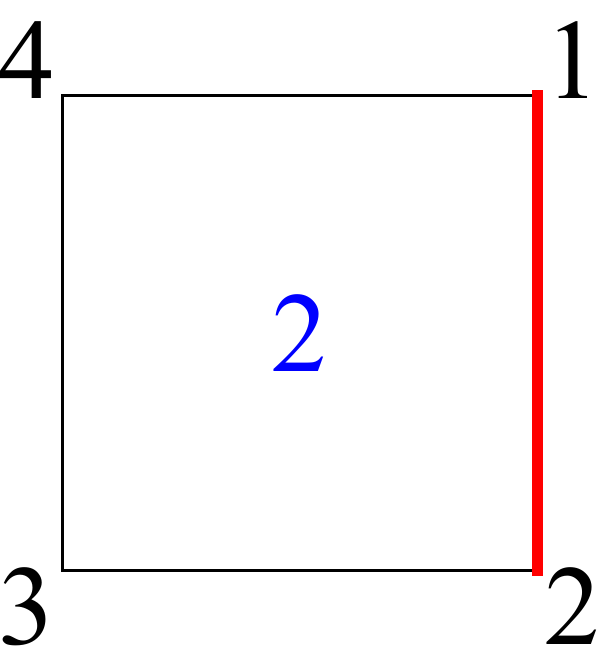} \end{gathered}\times 4$ & $\begin{gathered}
				\includegraphics[ height=2.2cm]{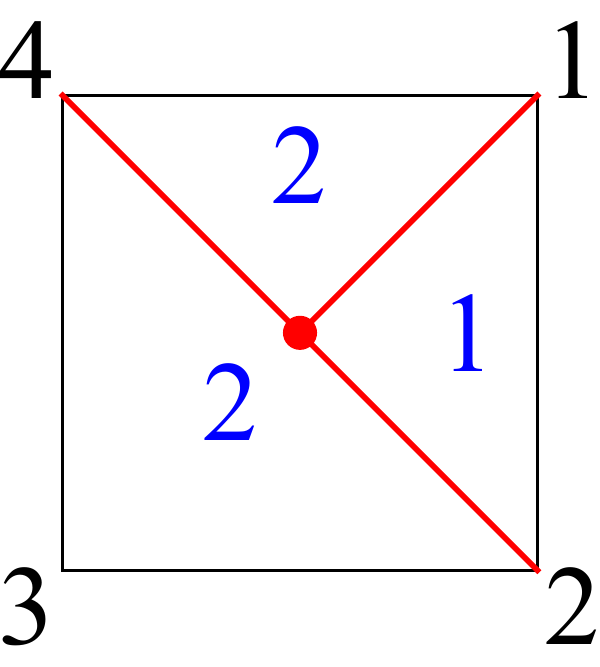} \end{gathered}\times 8$ & \\
			\hline
			$d=0$ &$\begin{gathered}
				\includegraphics[ height=2.2cm]{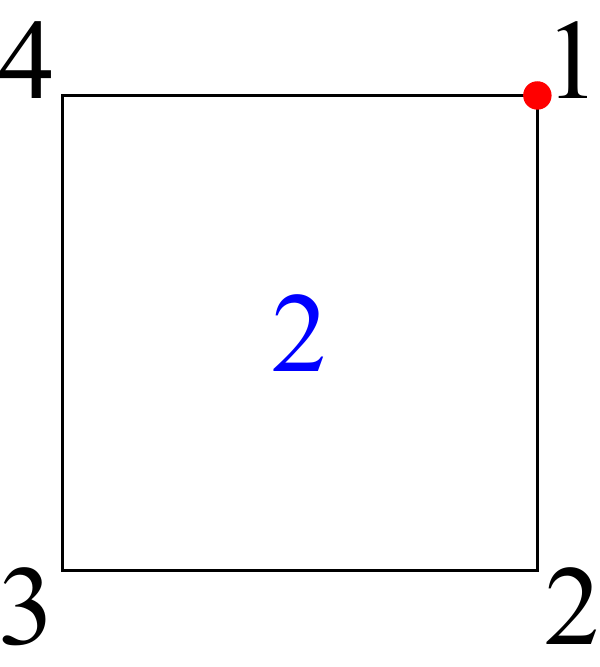} \end{gathered}\times 4$&$\begin{gathered}		\includegraphics[ height=2.2cm]{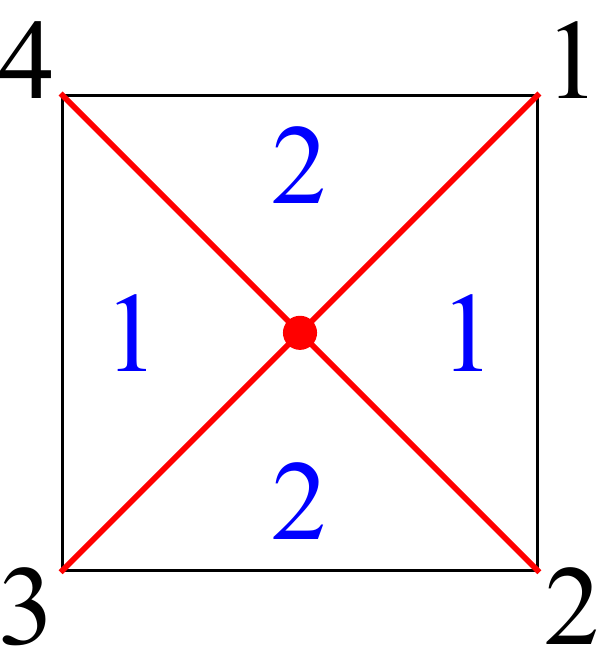} \end{gathered}\times 2$  & \\ [1ex] 
			\hline
		\end{tabular}
	\end{center}
	\caption{Graphs associated to all boundaries of $\Delta_{4,2}$.}
	\label{tab:n4-boundaries}
\end{table}

\begin{figure}
	\centering
		\includegraphics[width=0.3\textwidth]{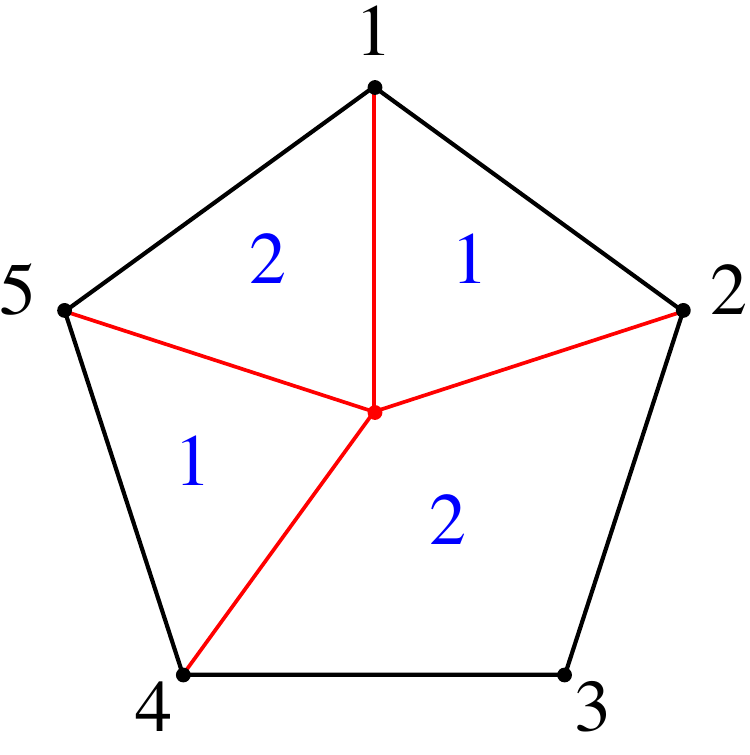}\qquad\qquad
\includegraphics[width=0.3\textwidth]{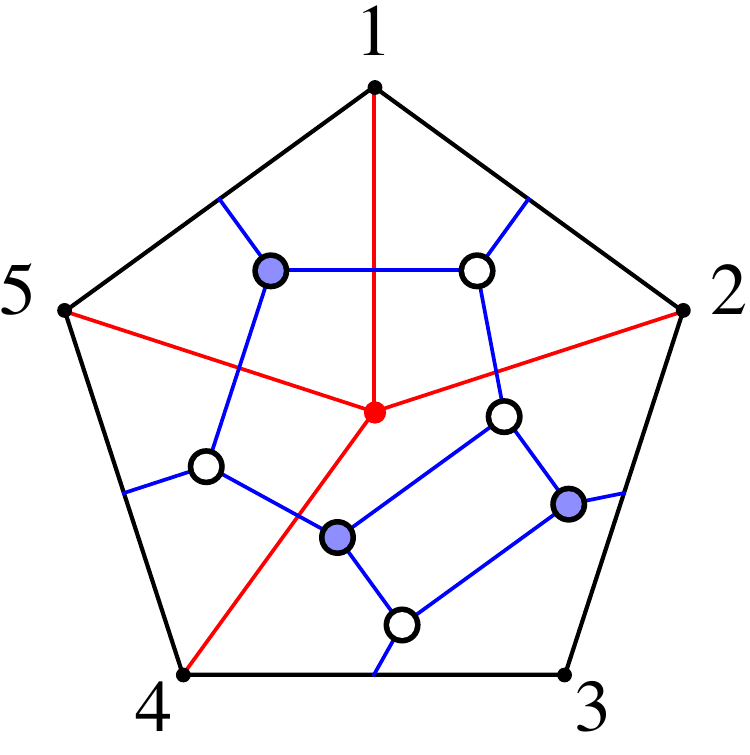}
	\caption{A simple example illustrating the relation between a quadruple intersection diagram and an on-shell graph.}
	\label{fig:diag-dual}
\end{figure}

We will make a particular use of the zero-dimensional graphs with four internal edges connected to corners $i,j,k,l$ that correspond to the position of $y$ at the quadruple intersection $y=q^\pm_{ijkl}$. First, let us observe that to each such graph, one can associate an on-shell diagram. The connection is established by replacing each internal $m$-gon by an $m$-valent vertex, and connecting all vertices whose polygons share an edge. The triangles with helicity $1 (2)$ correspond to white (black) trivalent vertices, while larger $m$-gons with helicity $h$ should be replaced by the on-shell diagram associated to the top cell of the positive Grassmannian $G_+(h,m)$. An example of this association between $n$-gon labels and on-shell diagrams can be seen in figure \ref{fig:diag-dual}. 

\begin{figure}
	\centering
	\includegraphics[height=5cm]{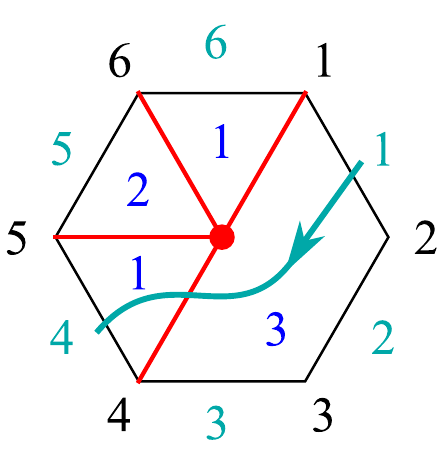}
	\caption{An example which illustrates that the permutation associated to the above diagram has $\sigma(1)=4$. The full permutation associated to this diagram is $\{4, 5, 7, 6, 8, 9\}$.}
	\label{fig:diag-path}
\end{figure}

Using the relation between quadruple-cut graphs and on-shell diagrams,  we can associate a permutation $\sigma$ to each such graph. To find $\sigma(i)$, we follow a path that starts between corners $i$ and $i+1$, and each time we enter an internal polygon we take the $h^\text{th}$ left turn, where $h$ is the helicity of the polygon. If we exit the diagram between corners $j$ and $j+1$, then $\sigma(i)=j$. This is illustrated in \ref{fig:diag-path} for a particular case.
Starting from any $n$-gon with a quadruple cut and total helicity $k$, its associated permutation $\sigma$ labels a positroid cell in $G_+(k,n)$, which we call the \emph{positroid cell associated to the graph}. In all cases we studied, we found that if we take a point $C$ in a positroid cell associated to the graph of a quadruple intersection $q^\pm_{ijkl}$, the associated configuration of points $x_i$ in dual space has the quadruple intersection point $q^\pm_{ijkl}$ as one of the vertices of $\Delta_{n,k}(x)$!

\begin{figure}[h!]
\begin{center}\raisebox{0.3cm}{\includegraphics[scale=0.4]{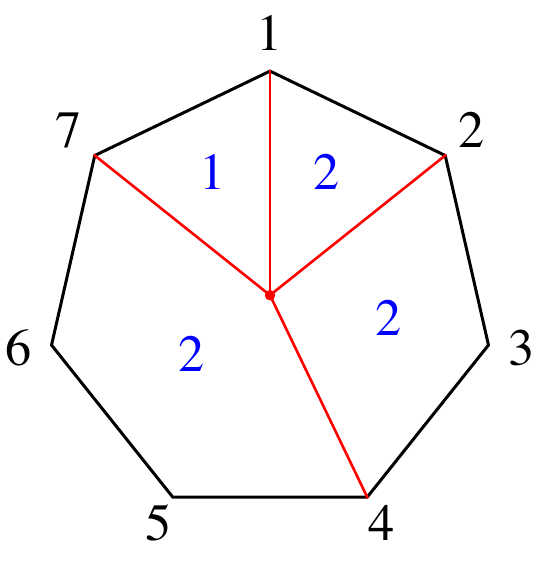}}\qquad \raisebox{2.2cm}{=}\qquad \includegraphics[scale=0.4]{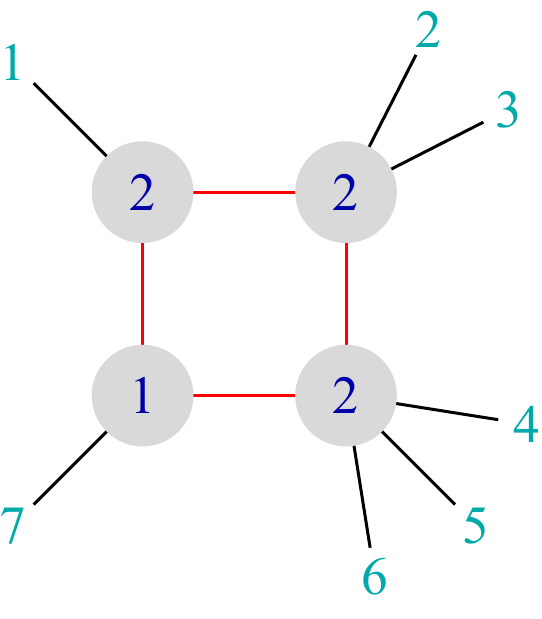}\end{center}
\caption{Relation between our graphs and the notation for quadruple cuts of box integrals.}
\label{fig:dualbox}
\end{figure}

Finally, for the quadruple-cut graphs, there is an alternative way to label them, more familiar to the usual labels used in this context. If we take any quadruple-cut graph, then its dual graph looks exactly like a quadruple cut of a box integral, see figure \ref{fig:dualbox} for a 7-point example. 

\paragraph{Refinements of graphs.} 

The graphs that we introduced above are actually very close to a true configuration of points in dual space. Let us focus on a particular quadruple cut $q_{ijkl}^\pm$ and let us assume that $1\leq i<j<k<l\leq n$. Since $q_{ijkl}^{\pm}$ is null-separated from vertices $x_i,x_j,x_k$ and $x_l$, it naturally splits the null polygon of $x$'s into four null polygons $(P_1,P_2,P_3,P_4)$ with vertices:
\begin{align}\label{eq:squashed-configuration}
&P_1:\{q_{ijkl}^\pm,x_i\ldots,x_j\}\,,\qquad P_2:\{q_{ijkl}^\pm,x_j\ldots,x_k\}\,,\nonumber\\ 
&P_3:\{q_{ijkl}^\pm,x_k\ldots,x_l\}\,,\qquad P_4: \{q_{ijkl}^\pm,x_l\ldots,x_n,x_1,\ldots x_i\}\,,
\end{align}
respectively.
Each such null polygon can be associated with a configuration of dual momenta for some amplitude with given number of particles $m<n$ and helicity $h\leq k$. Therefore, the quadruple-cut graph with $y=q_{ijkl}^\pm$ represents a two-dimensional projection of the configuration of points $x_i$ and $q_{ijkl}^\pm$, with all internal and external edges in the graph corresponding to null segments in the polygons \eqref{eq:squashed-configuration}.
In the remaining part of this section we explain how to assign 
the helicities $h_a$ to the smaller polygons $P_a$ in the case of a quadruple cut. While we consider this case as it will be relevant in the following sections, the same procedure can be used to assign helicities for graphs with a smaller number of cuts.

Let us start by considering the case where
the polygon $P_a$ is a triangle with fixed kinematic configuration  $(y,x_i,x_{i+1})$. For this case we can determine if it has helicity $1$ or $2$ by numerically checking whether the null momentum of the cut is proportional to $\lambda$ or $\tilde\lambda$, i.e.
\begin{align}\label{eq:3-particle-labels}
\epsilon_{\alpha\beta}\lambda_i^\beta (q^{\alpha\dot\alpha}_{i,i+1,j,k}-x_i^{\alpha\dot\alpha})=0 \quad \mathrm{or} \quad
\epsilon_{\dot\alpha\dot\beta}\tilde\lambda_i^{\dot\beta} (q^{\alpha\dot\alpha}_{i,i+1,j,k}-x_i^{\alpha\dot\alpha})=0.
\end{align} 
In the first case  $h_a=1$, while in the second $h_a=2$. 
For polygons $P_a$ with four vertices, the only allowed helicity is $h_a=2$, since these correspond to $\mathrm{MHV}$ (sub)amplitudes. For higher number of vertices of $P_a$, we can find their helicities by working recursively. If $P_a$ is formed of vertices $(y,x_i,\ldots,x_{i+m-2})$ then it should correspond to an $m$-particle configuration. We can determine all quadruple cuts for this configuration and consider any one of them that is inside $\Delta_{n,k}(x)$. The graph associated to this cut is divided into four polygons with number of vertices smaller than $m$. In this way we can reduce the problem of finding the helicities $h_a$ to simply checking one of the two conditions \eqref{eq:3-particle-labels}. 

Let us look at an explicit example for $n=8$ and $k=3$. First consider the point in the momentum amplituhedron $\mathcal{M}_{8,3,0}$ with 
\begin{align}
\lambda&=\left(\begin{smallmatrix}-96&-576&-960&-814&-526&150&1302&0\\152&478&652&-194&-216&-780&0&434\end{smallmatrix}\right),\notag\\
\tilde\lambda&=\left(\begin{smallmatrix}-62294&-9962&7736&6858&0&-3704&1418&17576\\56248&-5774&-11716&0&6858&2248&-4534&11714\end{smallmatrix}\right),
\end{align}
whose corresponding configuration of dual points are given by 
\begin{equation}
x=\left(\begin{smallmatrix}0& 7264960& 8754030& 
  1221334& -1569872& -2310536& -3465056&-2541938\\0& 2034440& 
  6078270& 9180014& 9845240& 8041586& 6765626& 3813992\\0& -1284736&
   2964306& 3070442& 279236& 1019900& 1618820& 
  2541938\\0& -7434248& -8152254& -6638& -671864& -2475518& 
-862358& -3813992\end{smallmatrix}\right),
\end{equation}
where the coordinates of $x_i$ can be read from the $i$-th column. This point lies in the chamber defined by the intersection of six BCFW cells in $G_+(3,8) $ associated to the following permutations:
\begin{align}\label{eq:chamber83}
\{& \{3, 5, 7, 6, 8, 9, 12, 10\}, \{3, 6, 5, 7, 8, 9, 12, 10\}, \{5, 4, 7, 6, 8, 9, 11, 10\}, \notag\\
  & \{6, 4, 5, 7, 8, 9, 11, 10\}, \{3, 4, 7, 6, 8, 9, 10, 13\}, \{3, 4, 5, 7, 8, 9, 10, 14\}\}\,.
\end{align}
By checking the sign flip patterns, we find that the vertices of the geometry $\Delta_{n,k}(x)$ associated to this chamber are
\begin{align}
&\{ q^+_{1234},q^+_{1245},q^+_{1256},q^+_{1268},q^+_{2345},q^+_{2356},q^+_{2368},q^+_{3456},\notag\\&\qquad q^+_{3468},q^+_{4567}, q^+_{1678},q^-_{1238},q^-_{1348},q^-_{1457},q^-_{1478},q^-_{1567},q^-_{4678} \}\,,
\end{align}
together with the points $x_1,\ldots,x_8$.

\begin{figure}
	\centering
	\includegraphics[height=4.5cm]{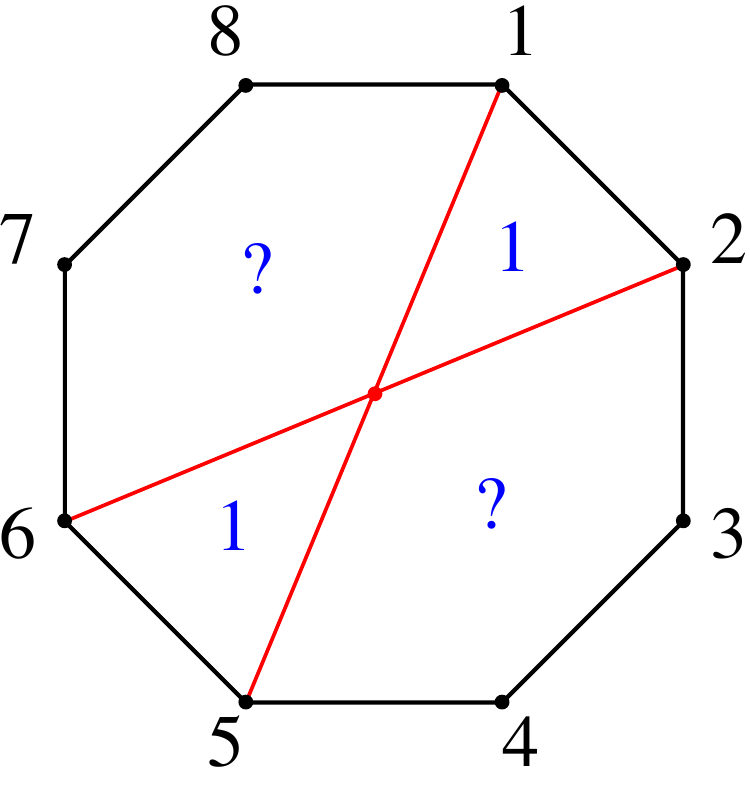}
	\includegraphics[height=1cm]{Figures/white}
	\includegraphics[height=4.5cm]{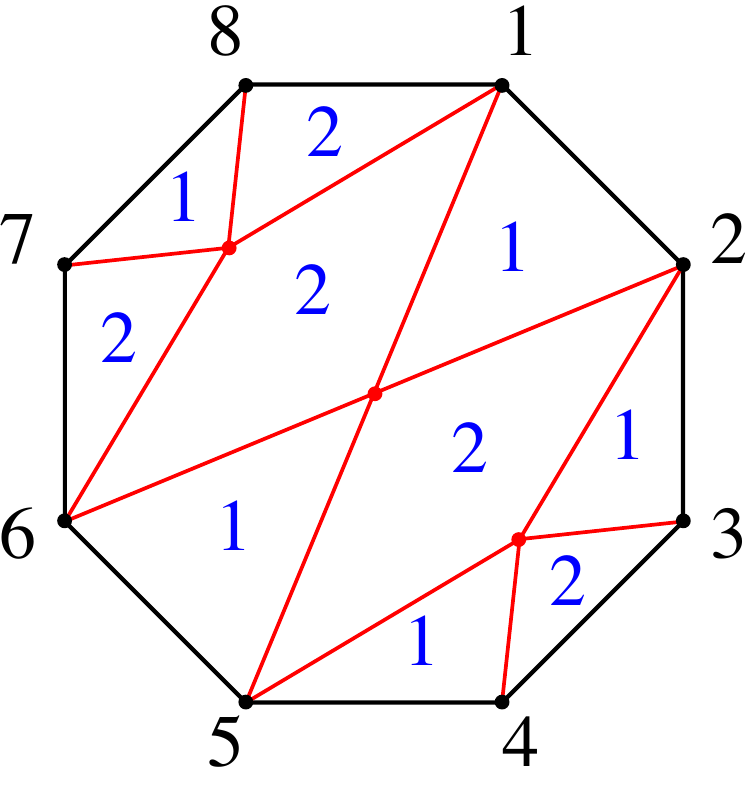}
	\includegraphics[height=1cm]{Figures/white}
	\includegraphics[height=4.5cm]{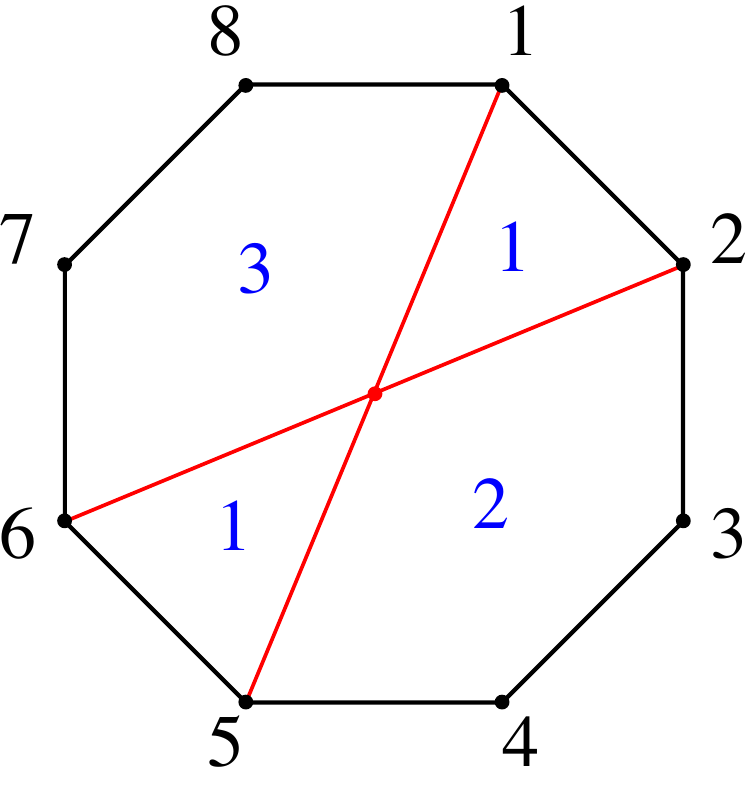}
	\caption{Illustration of the process of assigning helicity labels to the graph associated to the quadruple cut $q^+_{1256}$.}
	\label{fig:example_38}
\end{figure}

We wish to assign the correct helicity labels to the graph associated with e.g. the quadruple cut $q^+_{1256}$. The graph is an octagon with an internal marked point $y=q^+_{1256}$ and edges connecting it to points $(1,2,5,6)$. The quadruple cut splits the octagon into two triangles and two pentagons,  as depicted on the left of figure \ref{fig:example_38},  which we label by
\begin{align}
P_1 : \{ q^+_{1256}, x_1,x_2 \}, \, P_2 : \{ q^+_{1256}, x_5,x_6 \},  \, P_3 : \{ q^+_{1256},x_2,x_3,x_4,x_5 \}, \, P_4 : \{ q^+_{1256}, x_6,x_7,x_8,x_1 \}. 
\end{align}
The helicity of both triangles can be fixed by simply evaluating
\begin{align}\label{eq:3-particle-labels2}
&\epsilon_{\alpha\beta}\lambda_1^\beta (q^{+\ \alpha\dot\alpha}_{1256}-x_1^{\alpha\dot\alpha})=\epsilon_{\alpha\beta}\lambda_1^\beta (q^{+\ \alpha\dot\alpha}_{1256}-x_2^{\alpha\dot\alpha})=0\quad\rightarrow\quad h_{1}=1\,,\notag \\
&\epsilon_{\alpha\beta}\lambda_5^\beta (q^{+\ \alpha\dot\alpha}_{1256}-x_5^{\alpha\dot\alpha})=\epsilon_{\alpha\beta}\lambda_5^\beta (q^{+\ \alpha\dot\alpha}_{1256}-x_6^{\alpha\dot\alpha})=0\quad\rightarrow\quad h_2=1\,.
\end{align} 
However, the helicities of the pentagons are as of yet undetermined. To determine them we can for instance consider the quadruple cuts $\{ q^+_{1678}, q^+_{2345}\}$ which both lie in the chamber geometry $\Delta_{8,3}(x)$ for this chamber. These quadruple cuts further decompose each pentagon into three triangles and one quadrilateral, as depicted in the middle of figure \ref{fig:example_38}, labelled by
\begin{align}
P_{3,1} : \{q^+_{2345},x_2,x_3 \}, \,  P_{3,2} :\{q^+_{2345},x_3,x_4 \}, \,  P_{3,3} :\{q^+_{2345},x_4,x_5 \}, \,  P_{3,4} :\{q^+_{2345},x_5,q^+_{1256},x_2 \}, \notag \\
P_{4,1} : \{q^+_{1678},x_6,x_7 \}, \,  P_{4,2} :\{q^+_{1678},x_7,x_8 \}, \,  P_{4,3} :\{q^+_{1678},x_8,x_1 \}, \,  P_{4,4} :\{q^+_{1678},x_1,q^+_{1256},x_6 \}.
\end{align}
As mentioned above the helicity of the quadrilateral is always 2, i.e. $h_{3,4}=h_{4,3}=2$, and the helicity of the six triangles are again fixed using the conditions \eqref{eq:3-particle-labels}, to find the graph labels as displayed in the middle of figure \ref{fig:example_38}. Having assigned helicities to all subpolygons we can read of the helicity of the pentagons to be $h_{3}=2$ and $h_4=3$ respectively, as shown on the right of figure \ref{fig:example_38}.

\begin{figure}[h]
	\centering
	\includegraphics[height=5.5cm]{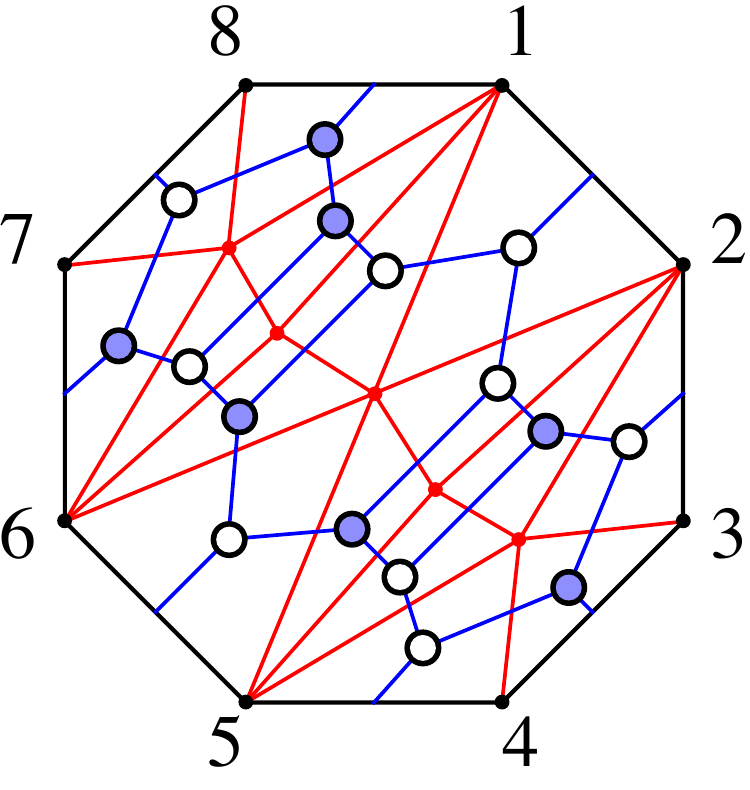}
	\caption{The plabic graph dual to the refinement of the quadruple cut $q_{1256}^+$. From top left to bottom right the red points are given respectively by the quadruple intersections $\{ q^+_{1678},q^+_{1268} ,q^+_{1256}, q^+_{1245},q^+_{2345} \}$.}
	\label{fig:dual_plab_38}
\end{figure}

Even though, at this point, we have already fixed the correct helicity configuration, it is instructive to decompose the two quadrilaterals further in order to see the connection to on-shell diagrams. This can be achieved by considering either of the points from each of the following intersections:
\begin{align}
\mathcal{N}_{q^+_{1256}} \cap \mathcal{N}_{q^+_{2345}} \cap \mathcal{N}_{x_2} \cap \mathcal{N}_{x_5} = \{ q^+_{1245}, q^+_{2356}\},\notag \\
\mathcal{N}_{q^+_{1256}} \cap \mathcal{N}_{q^+_{1678}} \cap \mathcal{N}_{x_1} \cap \mathcal{N}_{x_6} = \{ q^+_{1268}, q^-_{1567} \} ,
\end{align}
which all lie in the chamber geometry. Let us  choose $\{ q^+_{1245},q^+_{1268}$\}. This decomposes $P_{3,4}$ and $P_{4,4}$ further into
\begin{align}
&P_{3,4,1} : \{q^+_{1245},q^+_{1256},x_2 \}, \quad  P_{3,4,2} :\{q^+_{1245},x_2,q^+_{2345} \}, \, \notag\\& P_{3,4,3} :\{q^+_{1245},q^+_{2345},x_5 \}, \quad  P_{3,4,4} :\{q^+_{1245},x_5,q^+_{1256} \}, \notag \\
&P_{4,4,1} : \{q^+_{1268},q^+_{1678},x_1 \}, \quad  P_{4,4,2} :\{q^+_{1268},x_1,q^+_{1256} \}, \,  \notag\\&P_{4,4,3} :\{q^+_{1268},q^+_{1256},x_6 \}, \quad  P_{4,4,4} :\{q^+_{1268},x_6,q^+_{1678} \},
\end{align}
whose helicities can be calculated as above and are encoded in figure \ref{fig:dual_plab_38}. Since the original octagon is now fully divided into triangles with helicities 1 or 2, we can  also depict the dual trivalent graph where black (respectively white) vertex corresponds to the triangle with helicity 1 (respectively 2). Moreover, we can associate the affine permutation $\sigma$ to this on-shell diagram using the standard methods and we get $\sigma=\{3,4,5,7,8,9,10,14\}$, that was one of the permutations defining the chamber \eqref{eq:chamber83}.

\section{Examples}
\label{sec:examples}
In this section we will discuss the one-loop momentum amplituhedron geometries $\Delta_{n,k}(x)$ in more detail. In particular, we provide a detailed structure of the geometry for small $n$ and conjecture how the geometry looks for all $n$. We then conjecture a general one-loop integrand expression for any $n$ and $k$. It naturally agrees with the predictions obtained using the prescriptive unitarity methods \cite{Bourjaily:2017wjl}.

\subsection{MHV amplitudes}
 The null-cone geometries for MHV amplitudes are particularly basic because, at fixed $n$, all configurations of dual points $x_i$ give geometries that are combinatorially identical. This comes from the fact that for the MHV case there exists only one chamber, and therefore the shape of $\Delta_{n,2}\equiv \Delta_{n,2}(x)$ is independent of the positions of $x_i$. In this section we start by considering low number of points $n$, for which we provide explicit formulae for the canonical forms of $\mathcal{M}_{n,2,1}$, and then conjecture a general formula for generic $n$. Since there is only one chamber, the final formula for the differential form will be the wedge product  of the one-loop form with the tree-level one: 
\begin{equation}
\Omega_{n,2,1} = \Omega_{n,2,0}\wedge \Omega\left[\Delta_{n,2}\right]\,.
\end{equation}

\paragraph{MHV$_3$.}
Three-particle one-loop geometries are trivial since they are not fully dimensional. Indeed, in the $n=3$ case one cannot construct any quadruple intersections of null-cones since the geometry is spanned by three points $x_1,x_2$ and $x_3$. Therefore $\Delta_{3,2}$ is a triangle embedded in four-dimensional space. We can still associate a differential form to this geometry
\begin{equation}
\Omega\left[\Delta_{3,2}\right]=\dd\log \frac{\langle \ell 1 \rangle}{\langle \ell 3 \rangle} \wedge \dd\log \frac{\langle \ell 2 \rangle}{\langle \ell 3 \rangle}\,.
\end{equation}
However, it is not a top-dimensional form and therefore it integrates to 0 on any four-dimensional contour. This correctly leads to a vanishing three-point one-loop amplitude which is a well-established result.


\paragraph{MHV$_4$.}
The four-particle MHV geometry $\Delta_{4,2}$ is just an embedding of the positive Grassmannian $G_{+}(2,4)$ in the kinematic space $\mathbb{R}^{2,2}$. In particular, the vertices of  $\Delta_{4,2}$ are the four dual momenta $x_i$, $i=1,2,3,4$ and two quadruple intersections $q^\pm_{1234}$ corresponding to the quadruple cuts: 
\begin{figure}[h!]
\begin{center}\includegraphics[scale=0.3]{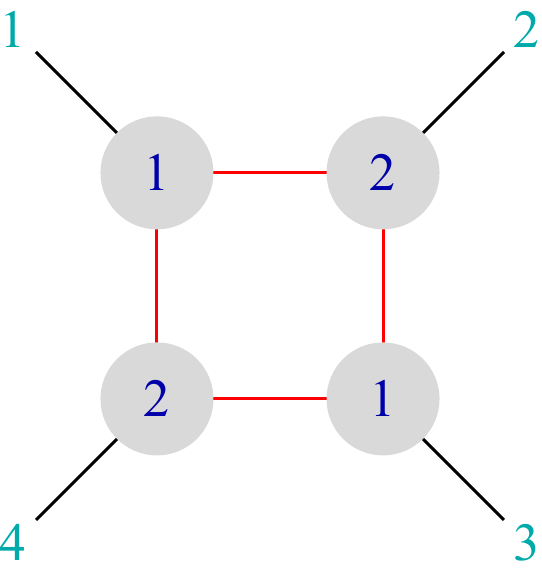}\qquad\includegraphics[scale=0.3]{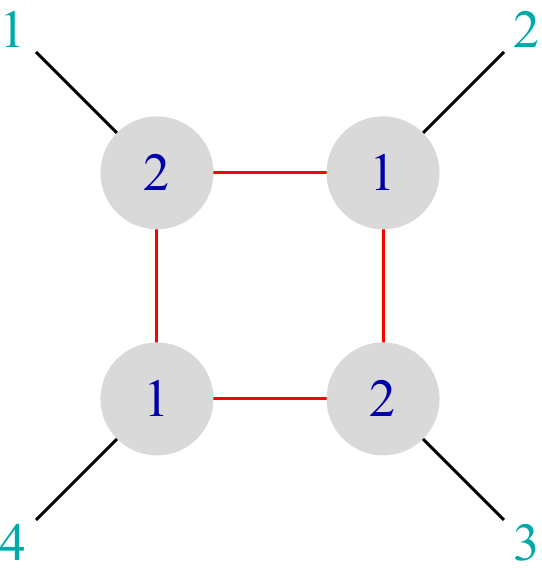}\end{center}
\caption{Quadruple cuts contributing to the geometry $\Delta_{4,2}$.}
\end{figure}

\noindent The boundaries of $\Delta_{4,2}$ of other dimensions can be naturally labelled by the set of graphs we introduced in the previous section, which in this case is in one-to-one correspondence with the set of on-shell graphs associated to $G_{+}(2,4)$, see table \ref{tab:n4-boundaries} for details.

The canonical differential form on $\Delta_{4,2}$ is easily found from \eqref{eq:form_final}:
\begin{equation}
\Omega\left[\Delta_{4,2}\right]=\omega^+_{1234}+\omega^-_{1234}=\omega^\square_{1234}\,,
\end{equation}
which is simply the massless box integrand, as expected.
Importantly, one can check using \eqref{eq:respm} and \eqref{eq:resx} that $\Omega\left[\Delta_{4,2}\right]$ has unit residues at all six vertices of $\Delta_{4,2}$. 


\paragraph{MHV$_5$.}
The  five-point case is the first time when not all quadruple intersections are inside of the loop geometry $\Delta_{5,2}$. Using the sign-flip description, see section \ref{sec:nullcone}, one finds that only the points of the form $q^+_{ii+1i+2i+3}$ are vertices of $\Delta_{5,2}$, while the remaining vertices are located outside. Therefore, the vertices of $\Delta_{5,2}$ are the five points $x_i$, $i=1,\ldots,5$ together with five quadruple intersections: 
\begin{equation}
\mathcal{V}(\Delta_{5,2})=\{x_i,q^{+}_{1234},q^{+}_{2345},q^{+}_{3451}=q^-_{1345},q^{+}_{4512}=q_{1245}^+,q^{+}_{5123}=q^-_{1235}\}\,.
\end{equation}
\begin{figure}[h!]
\begin{center}\raisebox{-0.25cm}{\includegraphics[scale=0.3]{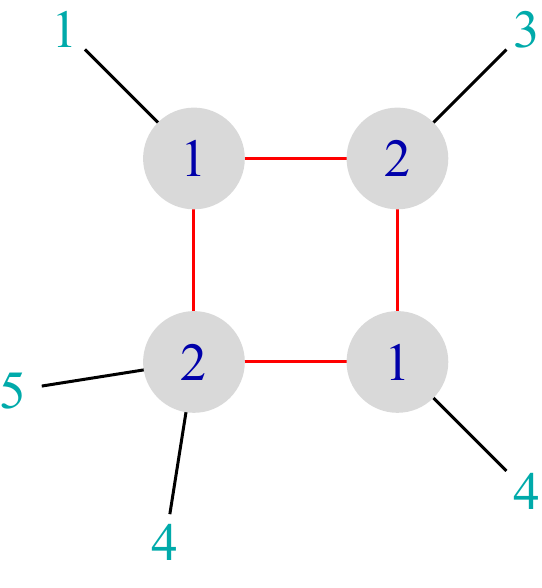}}\quad\includegraphics[scale=0.3]{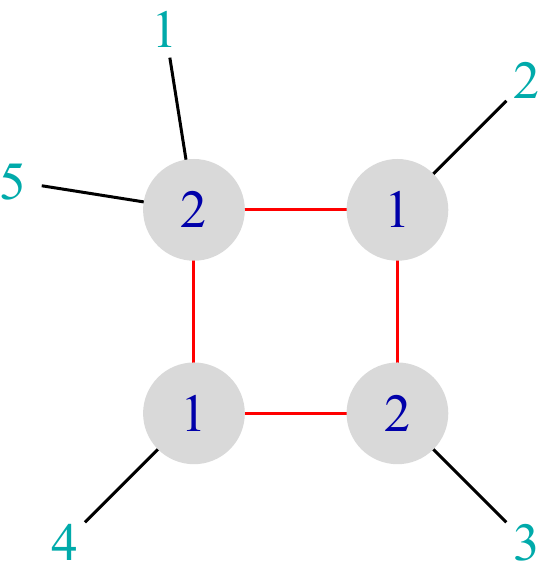}\quad\includegraphics[scale=0.3]{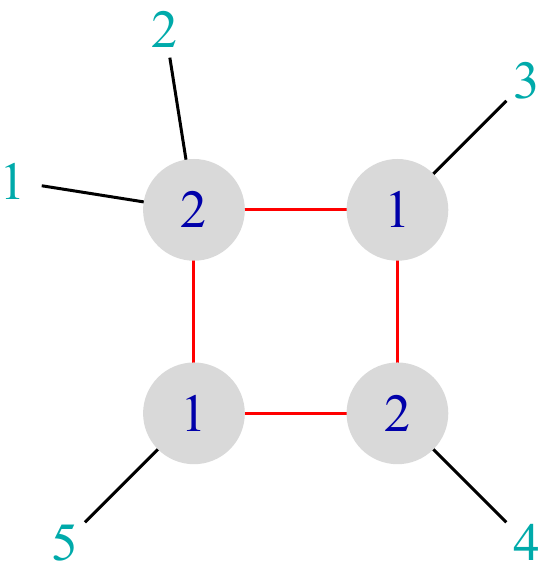}\quad\includegraphics[scale=0.3]{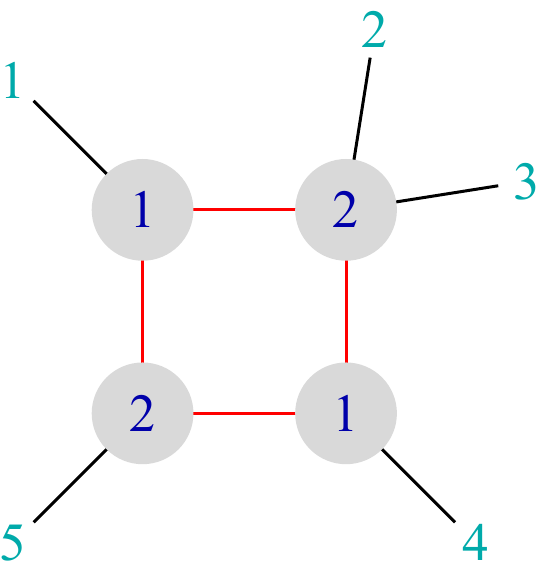}\quad\raisebox{-0.25cm}{\includegraphics[scale=0.3]{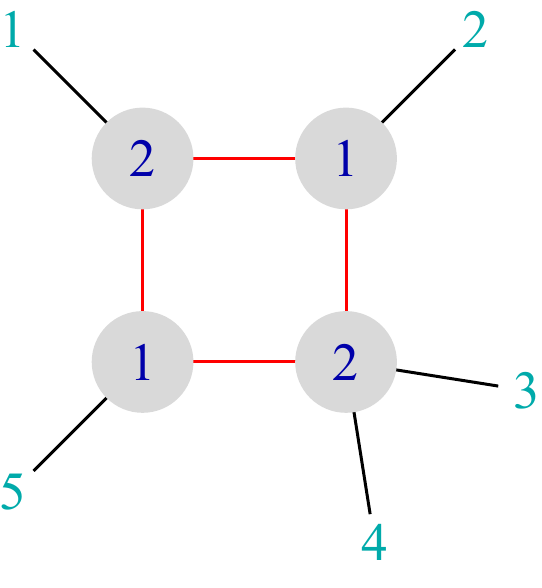}}\end{center}
\caption{Quadruple cuts corresponding to the quadruple-intersection vertices of $\Delta_{5,2}$.}
\end{figure}

The canonical differential form on $\Delta_{5,2}$ can be again found from formula \eqref{eq:form_final}:
\begin{align}
\Omega\left[\Delta_{5,2}\right]&=\omega^+_{1234}+\omega^+_{2345}+\omega_{1345}^-+\omega^+_{1245}+\omega_{1235}^- \nonumber \\
&=\frac{1}{2}\left(\omega^\square_{1234}+\omega^\square_{2345}+\omega^\square_{1345}+\omega^\square_{1245}+\omega^\square_{1235}+\omega_{12345}^{\pentagon}\right),
\end{align}
where we used \eqref{eq:combination} and \eqref{eq:perntagonformexp} to arrive at the second line.
Notice that the contributions coming from the differential forms $\omega_{ijkl}$ naturally combine into the pentagon integrand $\omega_{12345}^{\pentagon}$, making the differential form projectively invariant.


\paragraph{MHV$_6$.}
For the MHV geometry $\Delta_{6,2}$ coming from a null hexagon, we find that the set of its vertices consists of the six points $x_i$, $i=1,\ldots,6$ and nine quadruple intersections:
\begin{equation}
\mathcal{V}(\Delta_{6,2})=\{x_i,q^+_{1234},q^+_{2345},q^+_{3456},q^+_{4561}=q^-_{1456},q^+_{1256},q^+_{6123}=q^-_{1236},q^+_{1245},q^+_{2356},q^+_{6134}=q^-_{1346}\}\,.
\end{equation}
Notice that all quadruple intersections are of the form $q_{ii+1jj+1}^+$.
The explicit formula for the canonical form on $\Delta_{6,2}$ is:
\begin{align}\label{eq:six_point_MHV}
\Omega\left[\Delta_{6,2}\right]&=\omega^+_{1234}+\omega^+_{2345}+\omega^+_{3456}+\omega_{1456}^-+\omega_{1256}^++\omega_{1236}^-+\omega^+_{1245}+\omega_{2356}^++\omega_{1346}^-\\
&=\frac{1}{2}\left(\omega^\square_{1234}+\omega^\square_{2345}+\omega^\square_{3456}+\omega^\square_{1456}+\omega^\square_{1256}+\omega^\square_{1236}+\omega^\square_{1245}+\omega^\square_{2345}+\omega^\square_{1346}\right.\nonumber \\
&\left.+\omega_{12345}^{\pentagon}+\omega_{12356}^{\pentagon}+\omega_{13456}^{\pentagon}\right),
\end{align}
where as before we have  recombined all differential forms $\omega_{ijkl}$ into pentagon integrands. Notice that there is no unique way to do this, and there is an equivalent version of this answer 
\begin{equation}
\Omega\left[\Delta_{6,2}\right]=\frac{1}{2}\sum_{i<j}\omega_{ii+1jj+1}^{\square}+\frac{1}{2}\left(\omega_{12346}^{\pentagon}+\omega_{12456}^{\pentagon}+\omega_{23456}^{\pentagon}\right),
\end{equation}
that only differs by pentagon integrands.

It is easy to check that the formula \eqref{eq:six_point_MHV} possesses all required properties that we demand from the canonical form on $\Delta_{6,2}$: it is projectively invariant and the residues evaluated at {\it each} vertex of $\Delta_{6,2}$ equal one. 


\paragraph{General MHV amplitudes.}

\begin{figure}
	\centering
	\includegraphics[height=4.5cm]{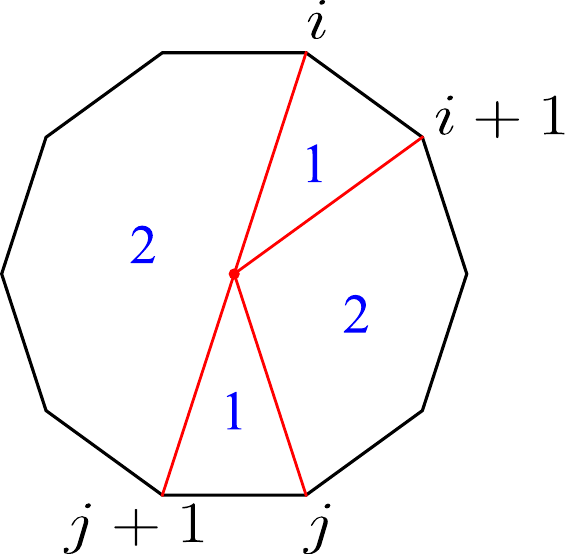}
	\caption{All quadruple-cut diagrams that contribute to MHV integrands. The associated vertex is $q^+_{i\,i+1\,j\,j+1}$.}
	\label{fig:diag-MHV}
\end{figure}

Proceeding in the same way for MHV geometries for $n>6$, we find a general pattern allowing us to conjecture the behaviour of these geometries at any $n$.
First, we find that there is only a single family of quadruple-cut diagrams that contribute to MHV\textsubscript{$n$}. These are of the type  $q^+_{i\,i+1\,j\,j+1}$,  see figure \ref{fig:diag-MHV},
and all these diagrams have the same associated permutation with $\sigma(i)=i+2$, i.e. the top cell of $G_+(2,n)$.
Therefore, using the fact that we know all vertices of $\Delta_{n,2}$, we can write the following general formula for differential forms for  $k=2$ and any $n$:
\begin{equation}
\Omega\left[\Delta_{n,2}\right]= \sum_{i<j}\omega_{ii+1jj+1}^{+} =\frac{1}{2}\sum_{i<j}\omega_{ii+1jj+1}^{\square}+\frac{1}{2}\sum_{1<i<j}\omega_{1ii+1jj+1}^{\pentagon} \,,
\end{equation}
where the answer is recast in terms of box and pentagon integrands making it manifestly projectively invariant. Moreover, notice that   the forms with indices $\omega^{\square}_{i-1ii+1j}$ for some $j$ appear exactly twice in this expression, namely we have $\omega^{\square}_{i-1ii+1i+2}$ and  $\omega^{\square}_{i-2i-1ii+1}$, and therefore, using \eqref{eq:resx}, the residues of $\Omega\left[\Delta_{n,2}\right]$ at $x_i$  are always equal to one.


\subsection{$\overline{\mathrm{MHV}}$ amplitudes}
The answers for all $\overline{\mathrm{MHV}}$ amplitudes can be obtained by parity operation acting on the MHV amplitudes, that exchanges all signs $(+\leftrightarrow -)$. In particular, only points $x_i$ and $q^-_{ii+1jj+1}$ are vertices of the geometries $\Delta_{n,n-2}\equiv \Delta_{n,n-2}(x)$. The canonical form for any $n$ becomes:
\begin{equation}
\Omega\left[\Delta_{n,n-2}\right]=\frac{1}{2}\sum_{i<j}\omega_{ii+1jj+1}^{\square}-\frac{1}{2}\sum_{1<i<j}\omega_{1ii+1jj+1}^{\pentagon} \,.
\end{equation}


\subsection{NMHV amplitudes}
NMHV amplitudes are the simplest case where the one-loop geometry {\it does} depend on the details of the tree-level configuration of points $x_i$. In particular, for distinct configurations of $x_i$ in $\mathcal{M}_{n,3,0}$ one might get a different subset of vertices $q_{ijkl}^\pm$ that are vertices of the geometry. This splits the space of possible tree-level configurations into subsets  -- the chambers. However, for any configuration of  $x_i$ within a given chamber, the loop differential form remains the same and we can use formula \eqref{eq:omegachambers} to calculate $\Omega_{n,3,1}$.


\paragraph{NMHV$_6$.}
We start this section by recalling the results found in \cite{He:2023rou} for the NMHV$_{6}$ amplitude. It was already pointed out there that the loop geometry naturally introduces the notion of chambers, and an explicit expression for the canonical form as a sum over chambers of the wedge products of the tree and loop parts was found to be
\begin{align}\label{eq:omega63}
\Omega_{6,3,1} = \sum_{\begin{smallmatrix}i \in \{2,4,6\} \\ j \in \{ 1,3,5 \} \end{smallmatrix}} \mathcal{I}_{6,3}^{\{ i\cap j \}} \wedge \widetilde\Omega_{6,3}^{\{i \cap j\}}.
\end{align}
Here, $\mathcal{I}_{6,3}^{\{ i\cap j \}}$ denotes the tree-level canonical form associated to the chamber $\Gamma_{6,3}^{\{i\}}\cap \Gamma_{6,3}^{\{j\}}$, that is the maximal intersection of the BCFW triangles $\Gamma_{6,3}^{\{ i \}}$ and $\Gamma_{6,3}^{\{ j\}}$, and  $\widetilde\Omega_{6,3}^{\{i\cap j\}}$ is the canonical form for the loop geometry associated to this chamber. 
In \cite{He:2023rou} this was found to be 
\begin{equation}
\label{eq:heboxes}
\widetilde\Omega_{6,3}^{\{ i\cap j \}} = \sum_{\begin{smallmatrix} a \in \{1,2\} \\ b \in \{ i,j \} \end{smallmatrix}}\bigg(I^{1\text{mb}}_a(b) +I_{a}^{2\text{mh}}(b) \bigg)+ \sum_{i=1}^{6} I_{\text{tri}}^{(i)},
\end{equation}
where the integrals are the familiar one-mass box, two-mass hard and triangle integrals. Importantly, every term in formula \eqref{eq:heboxes} depends on an auxiliary bi-twistor which cancels in the total sum. In the following, we will rederive $\Omega_{6,3,1}$ using the geometry $\Delta_{6,3}(x)$.

\begin{figure}[t]
\begin{center}
\includegraphics[scale=0.3]{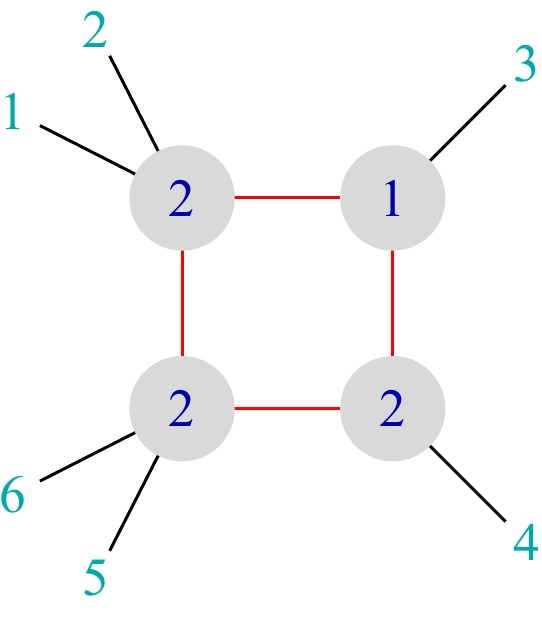}\quad\includegraphics[scale=0.3]{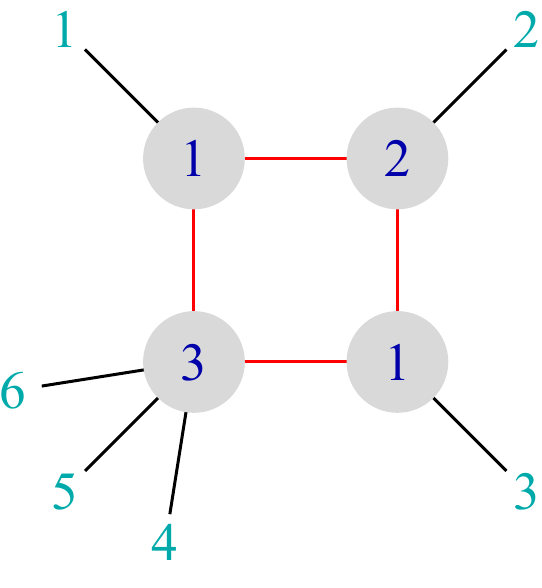}\quad\includegraphics[scale=0.3]{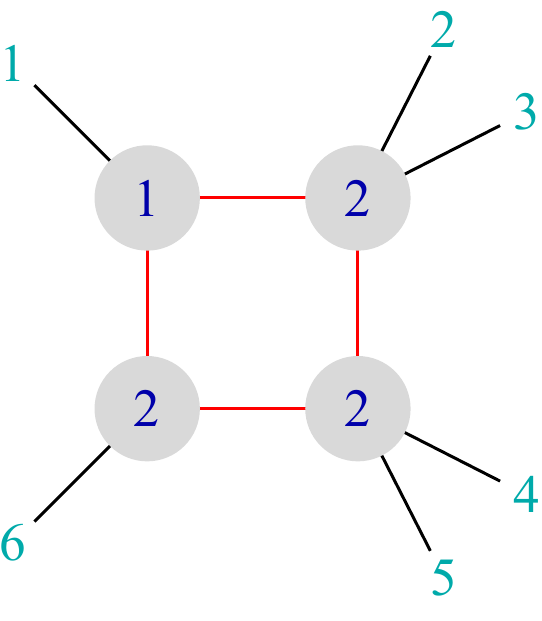}\quad\includegraphics[scale=0.3]{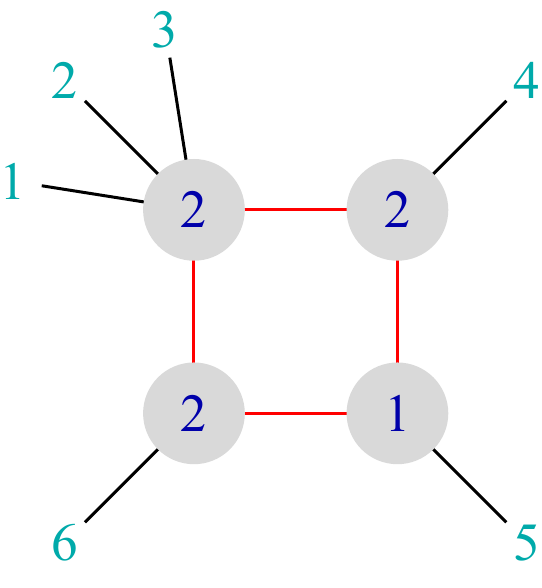}\\
\includegraphics[scale=0.3]{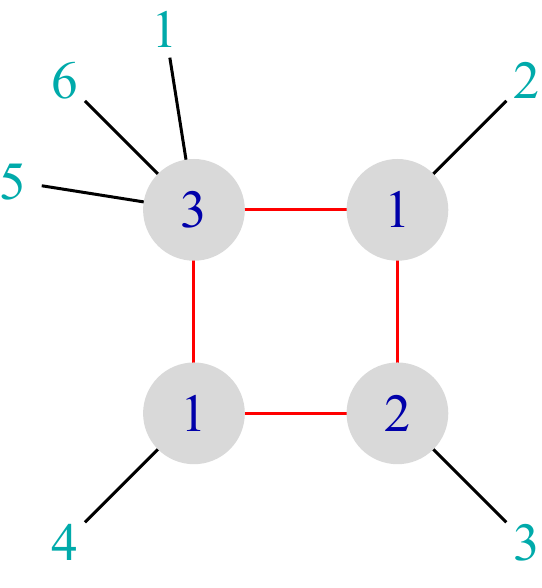}\quad\includegraphics[scale=0.3]{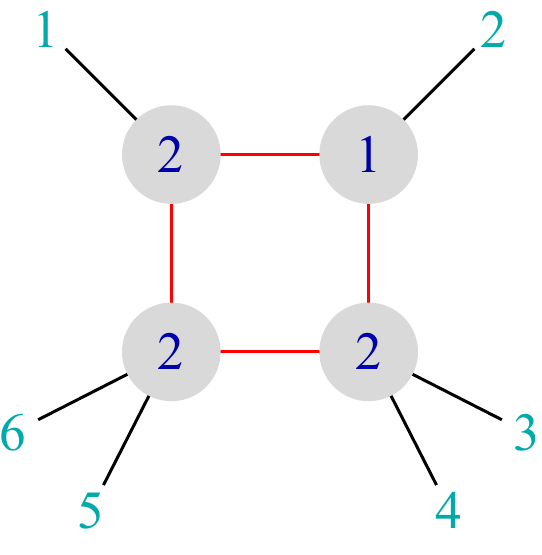}\quad\includegraphics[scale=0.3]{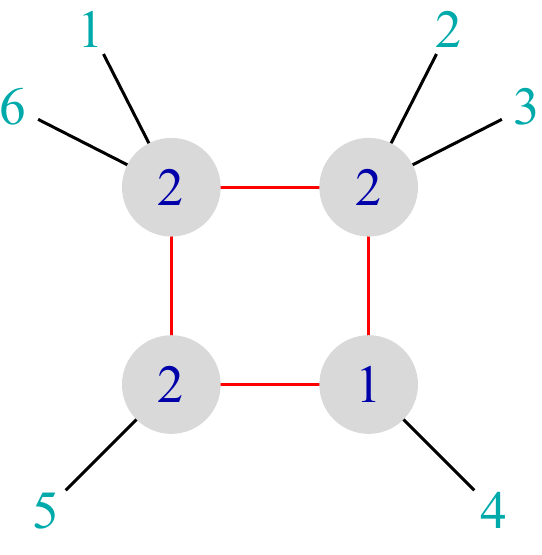}\quad\includegraphics[scale=0.3]{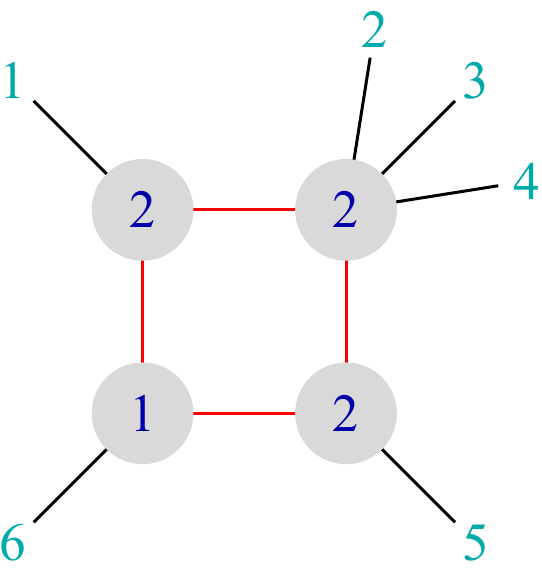}
\end{center}
\caption{Quadruple cuts corresponding to quadruple-intersection vertices in the chamber geometry $\Delta_{6,3}^{\{1\cap 2\}}$.}
\label{fig:6pointsdiagrams}
\end{figure}

By a careful study of all tree-level configurations of points $x_i$, we find that there are exactly nine distinct geometries $\Delta^{\{i\cap j\}}_{6,3}$ with $i=1,3,5$ and $j=2,4,6$, as expected from the discussion about chambers in section \ref{sec:chambers}. Moreover, if we take a configuration in a given chamber $\{i,j\}$, we find that the geometry has exactly the same shape and combinatorial structure for all configurations of $x_i$ in this chamber, as conjectured. In particular, the set of vertices of $\Delta^{\{i\cap j\}}_{6,3}$ contains the points $x_i$ together with the quadruple cuts in the set $ Q_{6,3}^i\cup Q_{6,3}^j$ where
\begin{equation}
	Q_{6,3}^i=\{q^+_{i-1ii+1i+2},q^-_{i+2i+3i+4i+5},q^+_{i+1i+2i+3i+5},q^-_{i-2i-1ii+2}\}\,.
\end{equation}
Since the sets $Q_{6,3}^i$ and $Q_{6,3}^j$ are disjoint for $i\neq j$, this immediately allows us to find the canonical form associated to $\Delta^{\{i\cap j\}}_{6,3}$ in each chamber using \eqref{eq:form_final}:
\begin{align}
	\Omega\left[\Delta^{\{i\cap j\}}_{6,3}\right]&=\omega^+_{i-1ii+1i+2}+\omega^-_{i+2i+3i+4i+5}+\omega^+_{i+1i+2i+3i+5}+\omega^-_{i-2i-1ii+2}\nonumber\\&+\omega^+_{j-1jj+1j+2}+\omega^-_{j+2j+3j+4j+5}+\omega^+_{j+1j+2j+3j+5}+\omega^-_{j-2j-1jj+2} \nonumber\\
	&= {\Omega}_{6,3}^{\{i\}}+{\Omega}_{6,3}^{\{j\}}\,,
\end{align}
where we defined
\begin{align}
	{\Omega}_{6,3}^{\{i\}}=\omega^+_{i-1ii+1i+2}+\omega^-_{i+2i+3i+4i+5}+\omega^+_{i+1i+2i+3i+5}+\omega^-_{i-2i-1ii+2}\,.
\end{align}
To illustrate this result, let us consider the chamber $\{1\cap 2\}$ as an example. The quadruple-intersection vertices in the chamber geometry $\Delta_{6,3}^{\{1\cap 2\}}$ are depicted in figure \ref{fig:6pointsdiagrams}
 and the differential form is
\begin{align}
\Omega\left[\Delta^{\{1\cap 2\}}_{6,3}\right]&=\omega^-_{6123}+\omega^+_{3456}+\omega^-_{2346}+\omega^+_{1356}+\omega^-_{1234}+\omega^+_{4561}+\omega^-_{3451}+\omega^+_{2461} \nonumber\\
&=\frac{1}{2}\big(\omega^{\square}_{1236}+\omega^{\square}_{3456}+\omega^{\square}_{2346}+\omega^{\square}_{1356}+\omega^{\square}_{1234}+\omega^{\square}_{1456}+\omega^{\square}_{1345}+\omega^{\square}_{1246}\nonumber \\
&+\omega^{\pentagon}_{13456}-\omega^{\pentagon}_{12346}\big)\,,
\end{align}
which is manifestly projectively invariant.
We can now add the results from all chambers to get the full one-loop answer. After including the tree-level contributions we find the full one-loop differential form for the NMHV$_6$ case
\begin{align}\label{eq:tildeomega63}
{\Omega}_{6,3,1}&=\sum_{\begin{smallmatrix}i \in \{2,4,6\} \\ j \in \{ 1,3,5 \} \end{smallmatrix}} \mathcal{I}_{6,3}^{\{ i\cap j \}} \wedge \Omega\left[\Delta^{\{i\cap j\}}_{6,3}\right]\notag \\
&=\sum_{\begin{smallmatrix}i \in \{2,4,6\} \\ j \in \{ 1,3,5 \} \end{smallmatrix}} \mathcal{I}_{6,3}^{\{ i\cap j \}} \wedge\left({\Omega}_{6,3}^{\{i\}}+{\Omega}_{6,3}^{\{j\}}\right)=\sum_{i=1}^6 \mathcal{I}_{6,3}^{\{i\}}\wedge {\Omega}_{6,3}^{\{i\}}\,,
\end{align}
where in the last equality $\mathcal{I}_{6,3}^{\{i\}}$ is the canonical form of the BCFW cell $\Gamma_{6,3}^{\{i\}}$, and we used the fact that chambers subdivide BCFW cells. This implies that we can directly associate loop differential forms, and therefore quadruple cuts, to positroid cells $S^{\{i\}}_{6,3}$. We notice that, while $\Omega\left[\Delta^{\{i\cap j\}}_{6,3}\right]$ is projectively invariant, the differential forms ${\Omega}_{6,3}^{\{i\}}$ are not. We have checked that formula \eqref{eq:tildeomega63} agrees up to a trivial numerical factor with $\Omega_{6,3,1 }$ given by \eqref{eq:omega63}.

Finally we mention that it is also possible to rewrite formula \eqref{eq:tildeomega63} in terms of box and pentagon integrands:
\begin{align}
	{\Omega}_{6,3,1}&=\frac{1}{2}\left(\mathcal{I}_{6,3}^{\{1\}}+\mathcal{I}_{6,3}^{\{4\}}\right)\wedge\left(\omega^{\square}_{1236}+\omega^{\square}_{3456}+\omega^{\square}_{2346}+\omega^{\square}_{1356}\right)\notag\\
	&+\frac{1}{2}\left(\mathcal{I}_{6,3}^{\{2\}}+\mathcal{I}_{6,3}^{\{5\}}\right)\wedge\left(\omega^{\square}_{1234}+\omega^{\square}_{1456}+\omega^{\square}_{1345}+\omega^{\square}_{1246}\right)\notag\\
	&+\frac{1}{2}\left(\mathcal{I}_{6,3}^{\{3\}}+\mathcal{I}_{6,3}^{\{6\}}\right)\wedge\left(\omega^{\square}_{1256}+\omega^{\square}_{2345}+\omega^{\square}_{1235}+\omega^{\square}_{2456}\right)\notag\\
	&+\frac{1}{2}\left(\mathcal{I}_{6,3}^{\{1\}}-\mathcal{I}_{6,3}^{\{4\}}\right)\wedge\left(\omega^{\pentagon}_{12356}-\omega^{\pentagon}_{23456}\right)\notag\\
	&+\frac{1}{2}\left(\mathcal{I}_{6,3}^{\{2\}}-\mathcal{I}_{6,3}^{\{5\}}\right)\wedge\left(\omega^{\pentagon}_{12345}-\omega^{\pentagon}_{23456}\right),
\end{align}
where we used the tree-level relation \eqref{eq:BCFW_NMHV6} to be able to recast the answer into a manifestly projectively invariant form.
 
\paragraph{NMHV$_7$.}

For NMHV$_7$, the chamber structure of the tree-level momentum amplituhedron $\mathcal{M}_{7,3,0}$ was not previously known. In the following we will first find all chambers and their corresponding chamber geometries that will allow us to write down a novel formula for the one-loop integrand of the NMHV$_7$ amplitude.  We will use the method explained in section \ref{sec:chambers} based on the notion of BCFW cells adjacency. Importantly, the BCFW cells adjacency graph $\mathcal{G}_{7,3}$ can be easily constructed using a variety of methods, see the Appendix \ref{app:chambers} for one possibility. Then using the software \texttt{IGraphM} \cite{Horvat2023} one can immediately find 71 maximal cliques, and therefore 71 chambers $\mathfrak{c}_a$, $a=1,\ldots 71$. They are organised in 11 families with members related by cyclic symmetry, see \eqref{eq:chambers73}. For each chamber $\mathfrak{c}_a$ we find tree-level configurations $x_i$ and construct the corresponding chamber geometry $\Delta_{7,3}^{\mathfrak{c}_a}$. Then, we can find all points in $\Delta_{7,3}^{\mathfrak{c}_a}$ and therefore find the canonical forms $\Omega\left[\Delta_{7,3}^{\mathfrak{c}_a}\right]$. After summing over all chambers we find a surprisingly simple one-loop integrand for NMHV$_7$ amplitude:
\begin{equation}
\Omega_{7,3,1}=\sum_{i_1<i_2}\mathcal{I}_{7,3}^{\{i_i,i_2\}}\wedge \Omega_{7,3}^{\{i_1,i_2\}}+\sum_{i=1}^7 \mathcal{I}_{7,3}^{\{i\}}\wedge \Omega_{7,3}^{\{i\}}\,,
\end{equation}
where
\begin{align}
\Omega_{7,3}^{\{1,2\}}&=\omega^-_{1347}+\omega^-_{1467}+\omega^{-}_{4567}+\omega^{+}_{3457}\,, \nonumber\\
\Omega_{7,3}^{\{1,3\}}&=\omega_{2357}^+\,, \nonumber\\
\Omega_{7,3}^{\{1,4\}}&=\omega^-_{1367}+\omega_{3567}^-\,, \nonumber\\
\Omega_{7,3}^{\{1\}}&=\omega^-_{1237}\,,
\end{align}
and cyclically rotated answers for other indices. Again, this implies that the quadruple intersection vertices of chamber geometries can be directly associated to positroid cells. The tree-level differential forms $\mathcal{I}_{7,3}^{\{i,j\}}$ and $\mathcal{I}_{7,3}^{\{i\}}$ are defined as canonical differential forms associated to positroid cells $S^{\{i,j\}}_{7,3}$ and $S^{\{i\}}_{7,3}$ defined in \eqref{eq:NMHVcells}, respectively. We also have the following relation
\begin{equation}
\mathcal{I}_{7,3}^{\{1\}}=\mathcal{I}_{7,3}^{\{1,2\}}+\mathcal{I}_{7,3}^{\{1,4\}}+\mathcal{I}_{7,3}^{\{1,6\}}=\mathcal{I}_{7,3}^{\{1,3\}}+\mathcal{I}_{7,3}^{\{1,5\}}+\mathcal{I}_{7,3}^{\{1,7\}}\,.
\end{equation}

\paragraph{NMHV$_8$}
We can repeat the construction from the previous case for $n=8$ and $k=3$. We find that there are 728 maximal cliques of the graph $\mathcal{G}_{8,3}$ and therefore 728 chambers for $\mathcal{M}_{8,3,0}$. For each chamber we can construct the chamber geometry $\Delta_{8,3}^{\mathfrak{c}_a}$ for $a=1,\ldots,728$ and find all their vertices. This allows us to write the explicit form of the differential form for these chamber geometries. Similar as in the case when $n=7$, after we add all chamber canonical forms, the final answer can be rewritten using only contributions coming from positroid cells as follows:
\begin{equation}
\Omega_{8,3,1}=\sum_{i_1<i_2<i_3}\mathcal{I}_{8,3}^{\{i_1,i_2,i_3\}}\wedge \Omega_{8,3}^{\{i_1,i_2,i_3\}}+\sum_{i=1}^8 \mathcal{I}_{8,3}^{\{i,i+1\}}\wedge \Omega_{8,3}^{\{i,i+1\}}+\sum_{i=1}^8 \mathcal{I}_{8,3}^{\{i\}}\wedge \Omega_{8,3}^{\{i\}}\,,
\end{equation}
where
\begin{align}
\Omega_{8,3}^{\{1,2,3\}}&=\omega^-_{1458}+\omega^-_{1578}+\omega^{+}_{4568}+\omega^{-}_{5678} \,,\nonumber\\
\Omega_{8,3}^{\{1,2,4\}}&=\omega_{3468}^+\,, \nonumber\\
\Omega_{8,3}^{\{1,3,4\}}&=\omega_{2368}^+\,, \nonumber\\
\Omega_{8,3}^{\{1,4,5\}}&=\omega^-_{1378}+\omega_{3678}^+\,, \nonumber\\
\Omega_{8,3}^{\{1,2,5\}}&=\omega^-_{1478}+\omega_{4678}^-\,, \nonumber\\
\Omega_{8,3}^{\{1,4,6\}}&=\omega_{3568}^-\,, \nonumber\\
\Omega_{8,3}^{\{1,2\}}&=\omega_{1348}^- \,,\nonumber\\
\Omega_{8,3}^{\{1\}}&=\omega^-_{1238}\,,
\end{align}
and non-zero contributions with other indices can be obtained by cyclic shifts. It is easy to check that this differential form has all correct maximal cuts.

\paragraph{General NMHV amplitudes.}
 
\begin{figure}
	\centering
	\begin{subfigure}{.5\textwidth}
		\centering
		\includegraphics[width=0.6\textwidth]{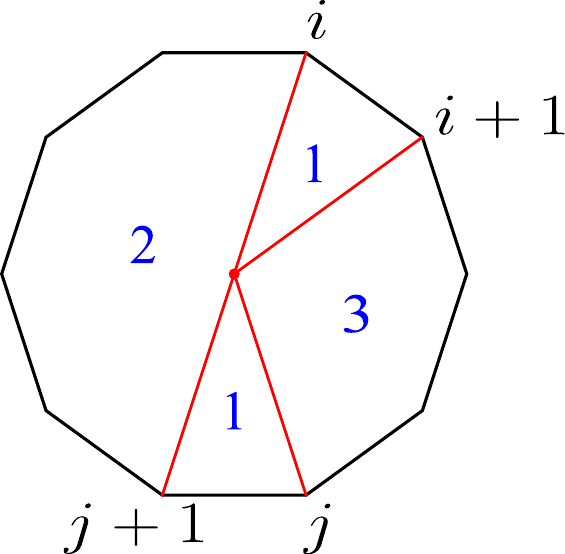}
	\end{subfigure}%
	\begin{subfigure}{.5\textwidth}
		\centering
		\includegraphics[width=0.6\textwidth]{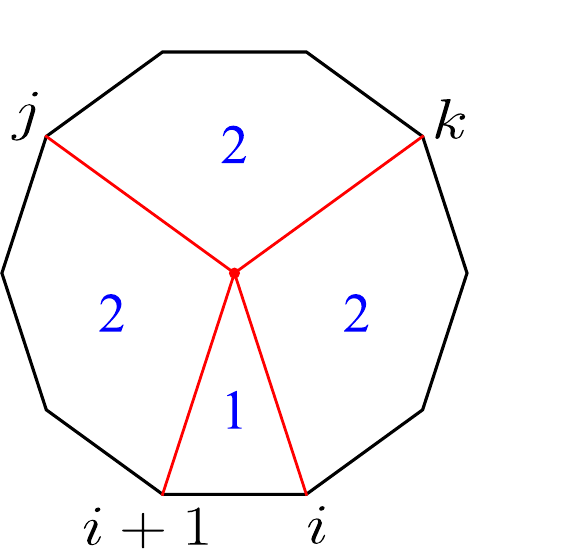}
	\end{subfigure}
	\caption{The two types of quadruple-cut diagrams that contribute to NMHV integrands. The associated vertices are $q^+_{i\,i+1\,j\,j+1}$ (left), and $q^+_{i\,i+1\,j\,k}$ (right).}
	\label{fig:diag-NMHV}
\end{figure}

For general NMHV amplitudes, we find that only two families of quadruple cuts shown  in figure \ref{fig:diag-NMHV} can be vertices of the geometries $\Delta_{n,3}(x)$. 
Then the canonical form for any $n$ can be written as the sum over positroid cells
\begin{equation}
\Omega_{n,3,1}=\sum_{i_1<\ldots<i_{n-5}}\mathcal{I}_{n,3}^{\{i_{1},\ldots,i_{n-5}\}}\wedge \Omega_{n,3}^{\{i_{1},\ldots,i_{n-5}\}}+\sum_{p=0}^{n-7}\sum_{i=1}^n \mathcal{I}_{n,3}^{\{i,i+1,\ldots,i+p\}}\wedge \Omega_{n,3}^{\{i,i+1,\ldots,i+p\}}\,,
\end{equation}
where $\Omega_{n,3}^I$ can be derived for all $n$ by finding all vertices in the geometries associated to positroid cell $S_{n,3}^I$.

\subsection{Beyond NMHV}

\paragraph{N$^2$MHV$_8$}

Before proposing a general one-loop integrand, we study one more non-trivial example: $n=8$, $k=4$. This is the first time where positroid cells with intersection numbers larger than one can be found. We do not know the complete classification of chambers in this case, however, using the methods outlined in the previous sections, we were able to generate many of them to confirm that the chamber geometries possess the same properties as for MHV and NMHV amplitudes. In particular, the canonical forms of $\Delta_{8,4}(x)$ are projectively invariant in all cases we investigated. 

The only case which needs special attention is when the positroid cells have intersection number equal two. As we already mentioned in section \ref{sec:chambers}  there are two such cells and their images intersect inside the momentum amplituhedron $\mathcal{M}_{8,4,0}$. Therefore, we need to treat them as two independent cells when looking for chambers. 
After taking this into account, we find the following one-loop answer 
\begin{equation}\label{eq:n84}
\Omega_{8,4,1}=\sum_{\sigma\in \mathcal{Q}_{8,4}^{(1)}}\mathcal{I}_{8,4}^\sigma\wedge \Omega_{8,4}^\sigma +\sum_{\sigma\in \mathcal{Q}_{8,4}^{(2)}}\left(\mathcal{I}_{8,4}^{\sigma,+}\wedge \Omega_{8,4}^{\sigma,+}+\Omega_{8,4}^{\sigma,-}\wedge \Omega_{8,4}^{\sigma,-}\right),
\end{equation}
where $\mathcal{I}_{8,4}^\sigma$ are the tree-level canonical forms for $\Gamma_{8,4}^\sigma$. Here, $\mathcal{Q}_{8,4}^{(1)}$ is the set of all permutations for positroid cells with intersection number equal one that contribute to quadruple intersections, and  $\mathcal{Q}_{8,4}^{(2)}=\{\{6,5,8,7,10,9,12,11\},\{4,7,6,9,8,11,10,13\}\}$ are the two permutations labelling cells with intersection number equal two. We indicated by $\mathcal{I}_{8,4}^{\sigma,\pm}$ and $\Omega_{8,4}^{\sigma,\pm}$ the two contributions coming from two quadruple cut solutions associated to intersection number two cells. In particular
\begin{equation}
\Omega_{8,4}^{\{6,5,8,7,10,9,12,11\},\pm}=\omega_{2468}^\pm\,,\qquad\qquad \Omega_{8,4}^{\{4,7,6,9,8,11,10,13\},\pm}=\omega_{1357}^\pm \,.
\end{equation}

\subsection{General one-loop integrand and prescriptive unitarity}
We now conjecture how this construction generalizes to any $n$ and $k$. First of all, for a given point $(\lambda,\tilde\lambda)\in \mathcal{M}_{n,k,0}$ in a chamber $\mathfrak{c}=\Gamma_{n,k}^{\sigma_1}\cap\ldots\cap\Gamma_{n,k}^{\sigma_p}$, we can obtain a complete list of all vertices that appear in the chamber geometry $\Delta_{n,k}^{\mathfrak{c}}$ by listing all allowed quadruple cut diagrams of type $(n,k)$, and selecting those diagrams whose associated positroid cells are one of those participating in the intersection: $S_{n,k}^{\sigma_1},\ldots ,S_{n,k}^{\sigma_p}$. This allows us to associate vertices directly to positroid cells  $S_{n,k}^{\sigma}$ and write the one-loop contribution from each of them as 
\begin{equation}
\Omega_{n,k}^{\sigma}=\sum_{q_{ijkl}^\pm}\omega^\pm_{ijkl} \,.
\end{equation}
The general one-loop integrand is then
\begin{equation}\label{eq:main}
\Omega_{n,k,1}=\sum_{\sigma\in\mathcal{Q}_{n,k}}\mathcal{I}_{n,k}^{\sigma} \wedge \Omega_{n,k}^{\sigma}\,,
\end{equation}
where the sum runs over all permutations $\sigma$ that can be associated to quadruple cuts and  $\mathcal{I}_{n,k}^{\sigma}$ is the canonical form of $\Gamma_{n,k}^\sigma$. 
 If a permutation $\sigma$ in \eqref{eq:main} corresponds to a positroid cell with intersection number two, we get two contributions, as in \eqref{eq:n84}. We conjecture that cells with intersection number larger than two do not contribute to the sum in \eqref{eq:main}\footnote{For example, we checked that the cells $S_{12,6}^{\{10,8,12,7,11,9,16,14,18,13,17,15\}}$ and $S_{16,8}^{\{11,5,16,10,15,9,20,14,19,13,24,18,23,17,28,22\}}$ with intersection number equal to 4 do not correspond to quadruple cuts \cite{Arkani-Hamed:2012zlh}.}. 
 We have verified that this formula agrees with the ones provided in the Mathematica package attached to \cite{Bourjaily:2013mma}.
Formula \eqref{eq:main} is the main result of this paper.


\section{Cluster structure in dual space}
\label{sec:cluster_algebras}
The combinatorial labels introduced in section \ref{sec:combi} for the quadruple cuts bare a close resemblance to the pre-existing notion of the {\it dual quiver} associated to an on-shell diagram. As we will now explain this connection to quivers will allow us to identify all points in dual space contained in the one-loop MHV$_n$ geometry $\Delta_{n,2}$ in a cluster-like fashion i.e., starting from an `initial seed' and performing `mutations' on the dual space points, we will generate all points in the geometry. This analysis is far from complete and should be treated as an invitation to explore the connection to cluster algebras further. In the following we assume that the reader is familiar with basic notions related to cluster algebras, see \cite{2012arXiv1212.6263W} for a review.

\begin{figure}
	\centering
	\includegraphics[height=4.5cm]{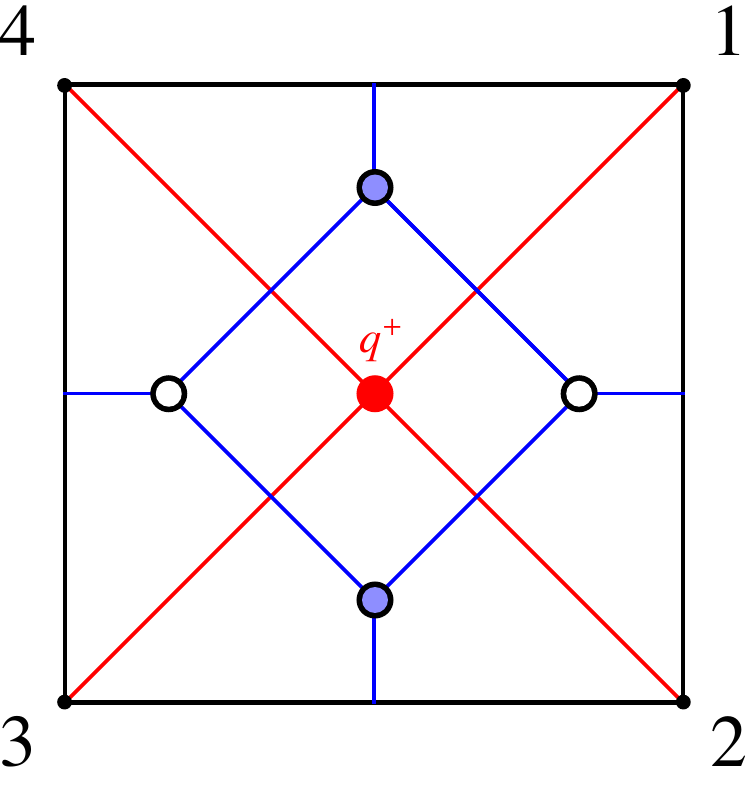}
	\includegraphics[height=3cm]{Figures/white}
	\includegraphics[height=4.5cm]{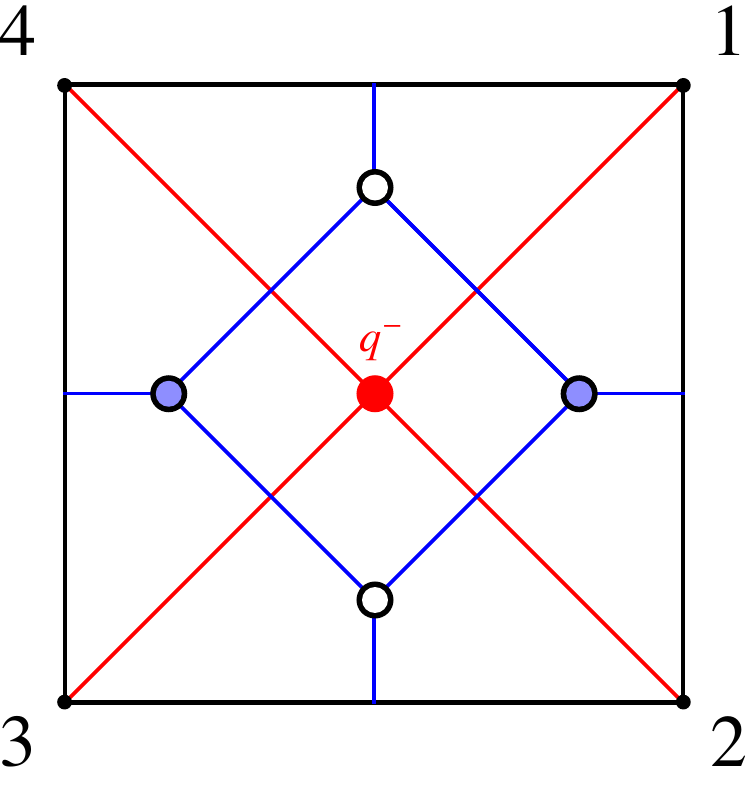}
	\caption{On-shell diagrams associated to the quadruple cuts $q^+_{1234}$ and $q^-_{1234}$.}
	\label{fig:example_MHV_4}
\end{figure}

Let us begin with the simplest case of the MHV$_4$ geometry $\Delta_{4,2}$. Recall that this geometry consists of the four points $\{ x_1,x_2,x_3,x_4\}$ together with the two points $\{ q^+_{1234},q^-_{1234}\}$. As described in section \ref{sec:combi}, these quadruple intersections can be assigned a  graph related to an on-shell diagram; for example the on-shell diagrams associated to the two points $\{ q^+_{1234},q^-_{1234}\}$  are depicted in figure \ref{fig:example_MHV_4}. In particular, this allows us to assign to each internal region of an on-shell diagram a point in dual space $q^{+}_{ijkl}$ or $q^{-}_{ijkl}$. 

\begin{figure}
	\centering
	\includegraphics[height=4.5cm]{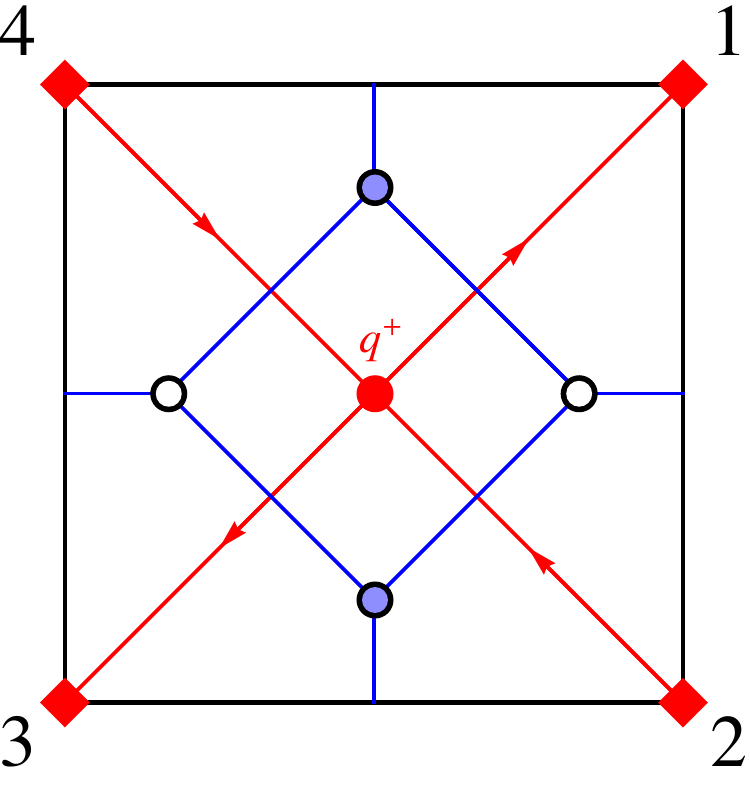}
	\includegraphics[height=3cm]{Figures/white}
	\includegraphics[height=4.5cm]{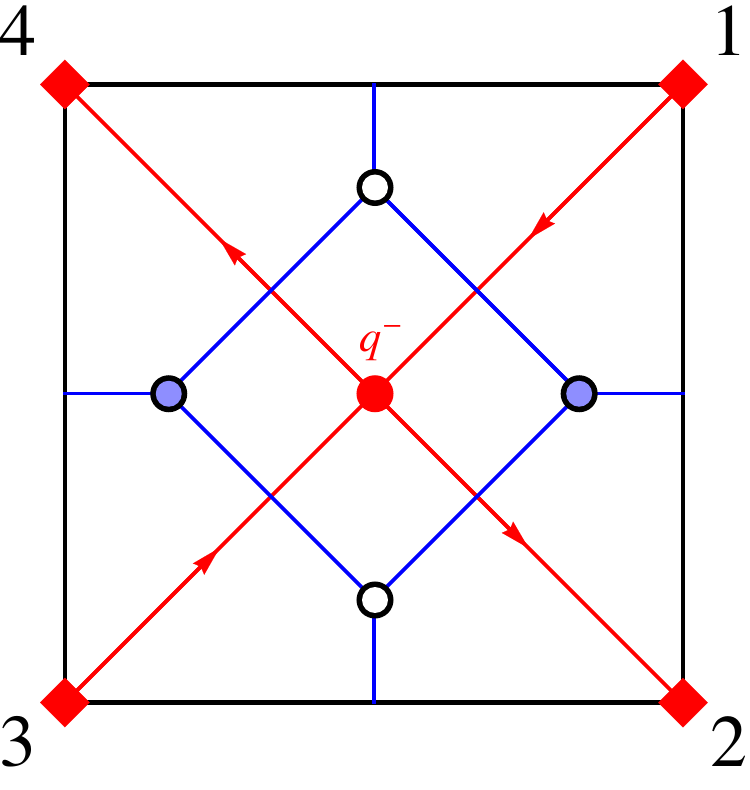}
	\caption{The dual quiver of the plabic graph. }
	\label{fig:dual_cluster}
\end{figure}

As alluded to these graphs have a very close connection to the dual quivers associated to on-shell diagrams. Given an on-shell diagram to obtain its dual quiver one first collapses all white-white and black-black vertices to produce a bipartite graph, assigns a (frozen) node to each (external) internal region together with a cluster coordinate, and connects neighbouring nodes by an arrow such that the black vertex is to the left of the arrow. As an example the dual quiver for the MHV$_4$ on-shell diagrams are depicted in figure \ref{fig:dual_cluster}. In this language performing a square move in the on-shell diagram corresponds to performing a mutation on the associated node in the cluster quiver. Notice that this relates the two cluster coordinates $\{ \tilde q^+_{1234},\tilde q^-_{1234} \}$ via the {\it exchange relation}
\begin{align}
\tilde q^+_{1234} \tilde q^-_{1234} =\tilde x_1 \tilde x_3 + \tilde x_2 \tilde x_4.
\end{align}
Here we use the notation $\tilde x$ and $\tilde q$ to emphasise that we are considering the cluster coordinates of the dual quiver rather than four-dimensional vectors in the dual space geometry. Importantly, this allows one to generate new cluster coordinates from old.

Remarkably, when studying points in the dual space geometry we find an {\it exchange-like} relation between various quadruple intersection points
\begin{equation}\label{eq:cluster_exchange}
q^+_{1234} + q^-_{1234}  =\frac{(x_2-x_4)^2(x_1+x_3)+(x_1-x_3)^2(x_2+x_4)}{(x_1-x_3)^2+(x_2-x_4)^2}.
\end{equation}
This relation allows us to express new dual vectors $q^-_{1234}$ in terms of the $\{ q^+_{1234},x_1,x_2,x_3,x_4 \}$. As this is a relation linking vectors associated to nodes of a quiver, it is very tempting to draw analogy between \eqref{eq:cluster_exchange} and the mutation rules acting on $g$-vectors in the cluster algebra.

\begin{figure}
	\centering
	\includegraphics[height=3.2cm]{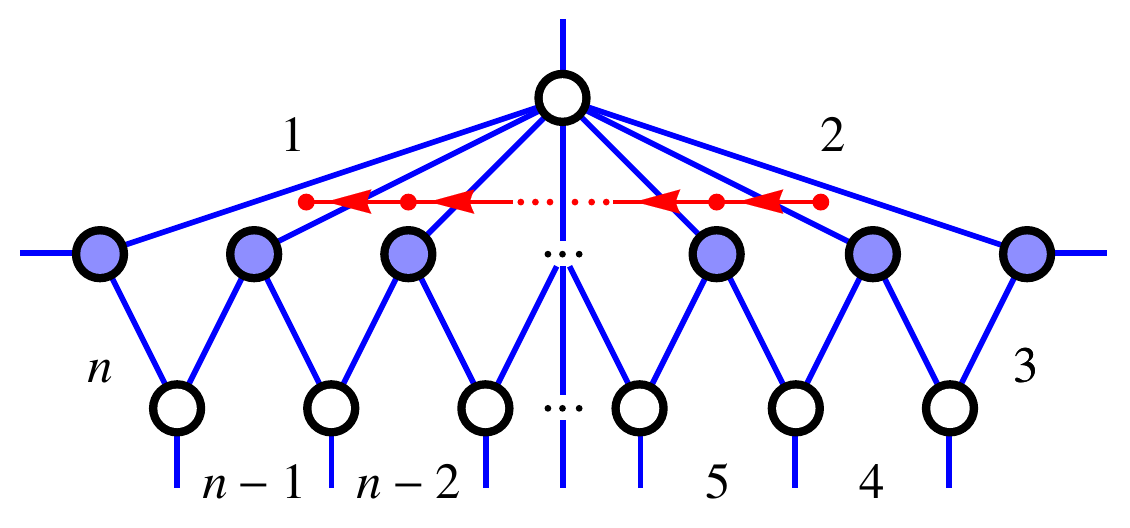}
		\caption{The on-shell diagram for MHV$_n$. Its associated cluster is $A_{n-3}$. The coordinates appearing in this cluster from left to right on this diagram are given by $\{q^+_{12n-1n}, q^+_{12n-2n-1} ,\ldots, q^+_{1245},q^+_{1234} \}$.}
	\label{fig:mhv_init}
\end{figure}

\begin{figure}
	\centering
	\includegraphics[height=5cm]{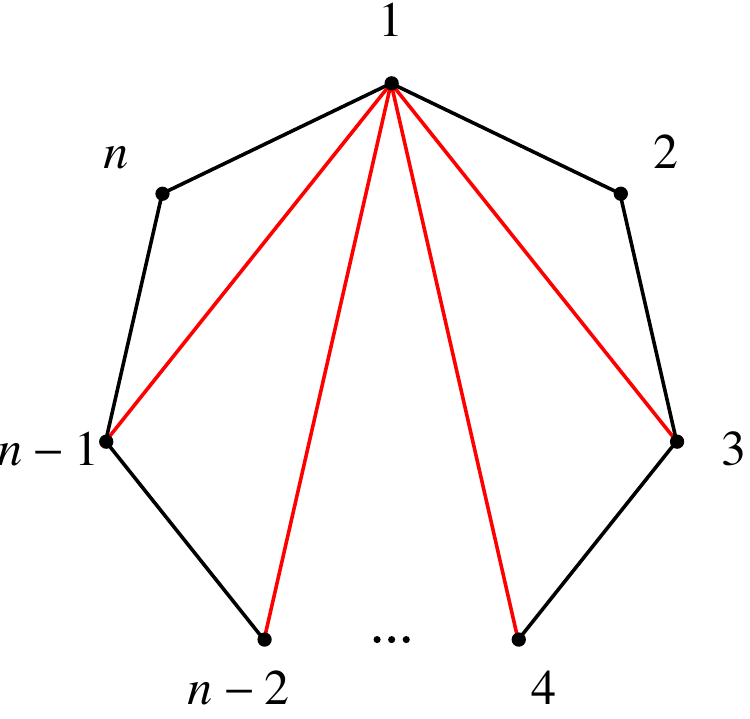}
		\caption{The triangulation of the $n$-gon associated to the general MHV$_n$ on-shell diagram. The labels for the chord $(i,j)$ for $i<j$ is given by $q^+_{ii+1jj+1}$.}
	\label{fig:mhv_triang}
\end{figure}

In the example given above the cluster algebra associated to the MHV$_4$ on-shell diagram consists of a single node i.e. the $A_1$ cluster algebra. It is straightforward to extend this analysis to all MHV$_n$ on-shell diagrams. A representative for the MHV$_n$ on-shell diagram, together with its dual quiver $A_{n-3}$, is depicted in figure \ref{fig:mhv_init}. Again we label each internal facet (node of the dual quiver) by a point in dual space which from left to right are given by
\begin{equation}
\{q^+_{12n-1n}, q^+_{12n-2n-1} ,\ldots, q^+_{1245},q^+_{1234} \}.
\label{eq:cluster_init}
\end{equation}
Notice that, as is well established for the $A_{n-3}$ cluster algebras, each of the above dual points can alternatively be labelled as the chord of an $n$-gon, where we assign to the chord $(ij)$ with $i<j$ the dual point $q^+_{ii+1jj+1}$. With this labelling the dual points of \eqref{eq:cluster_init} are given by the chords 
\begin{equation}
\{(1n-1), (1n-2) ,\ldots, (14),(13) \},
\end{equation}
which has the nice interpretation of the triangulation of the $n$-gon depicted in figure \ref{fig:mhv_triang}. Given this initial triangulation, one can perform all flips, corresponding to square moves in the on-shell diagram, and use \eqref{eq:cluster_exchange} to generate all possible dual points appearing in the geometry $\Delta_{n,2}$.

\begin{figure}
	\centering
	\includegraphics[height=5cm]{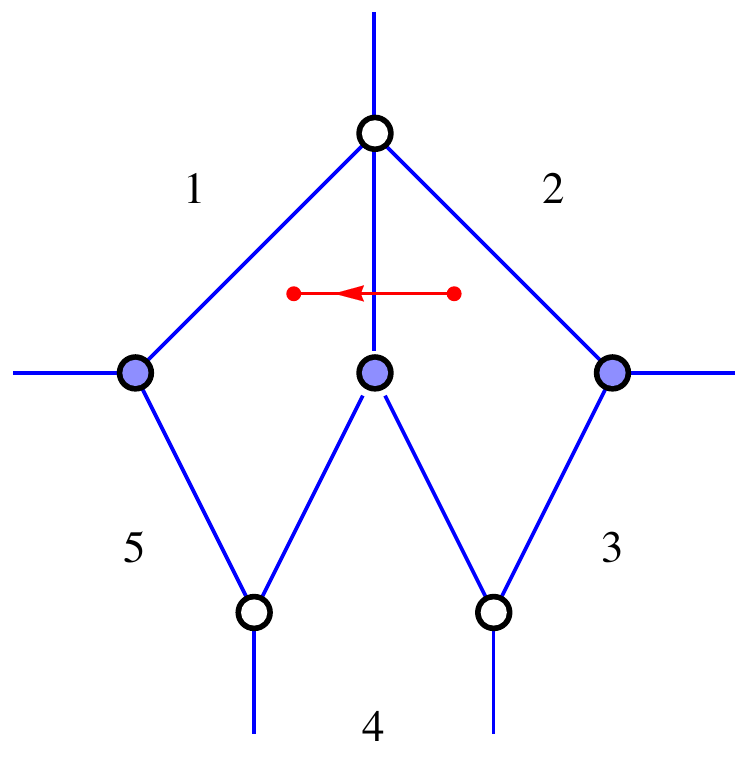}
	\includegraphics[height=1cm]{Figures/white}
	\includegraphics[height=5cm]{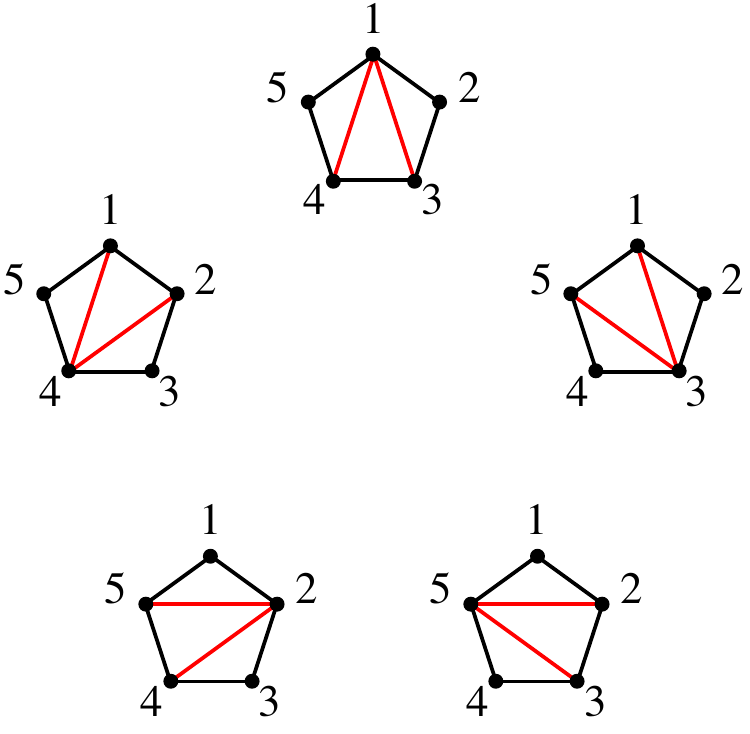}
		\caption{(Left) A representative on-shell diagram for MHV$_5$ along with the dual $A_2$ quiver. The nodes on the dual quiver are from left to right are $q^+_{1245}$ and $q^+_{1234}$ corresponding to the triangulation with all chords emanating from $1$. (Right) The five triangulations of the pentagon whose corresponding dual points are given by $\{ \{q^+_{1245},q^+_{1234}\} , \{q^+_{1234},q^-_{1345}\} , \{q^-_{1345},q^-_{1235}\} , \{q^-_{1235},q^+_{2345}\} , \{q^+_{2345},q^+_{1245}\} \}$.}
	\label{fig:MHV_cluster_5}
\end{figure}

To see this in action let us focus on the explicit example of MHV$_5$. Its associated on-shell diagram and dual quiver is given on the left of figure \ref{fig:MHV_cluster_5} and corresponds to the triangulation of the $n$-gon with all chords emanating from $1$. The dual points/chords associated to this triangulation are 
\begin{align}
\{(14),(13)\} \equiv \{q^+_{1245},q^+_{1234}\}.
\end{align}
By performing flips on the chords, and reading off the labels, the five triangulations are given by
\begin{equation}
\{ \{q^+_{1245},q^+_{1234}\} , \{q^+_{1234},q^-_{1345}\} , \{q^-_{1345},q^-_{1235}\} , \{q^-_{1235},q^+_{2345}\} , \{q^+_{2345},q^+_{1245}\} \},
\end{equation}
as depicted on the right of figure \ref{fig:MHV_cluster_5} starting at the top and going clockwise. Note that the minus signs arise from reordering the labels i.e.~using our prescription the chords in the bottom right triangulation are given by
\begin{align}
\{ (35),(25) \} \equiv \{q^+_{3451},q^+_{2351}\}= \{q^-_{1345},q^-_{1235}\}.
\end{align}
This procedure generates the five points in dual space given by
\begin{equation}
\{q^+_{1234},q^-_{1345},q^-_{1235},q^+_{2345},q^+_{1245} \},
\end{equation}
exactly those appearing in the geometry $\Delta_{5,2}$.


%


\section{Conclusions and outlook}
In this paper we have presented a new perspective on the tree- and loop-level positive geometries for planar $\mathcal{N}=4$ sYM. In particular, we translated the momentum amplituhedron positive geometry to the space of dual momenta. This allowed us to describe the loop geometries based on the null structure of the $\mathbb{R}^{2,2}$ kinematic space built from the tree-level configuration of points specified by a null polygon. As a result, we proposed a novel formula for the one-loop integrand $\Omega_{n,k,1}$ for any $n$ and $k$.

Our construction of the positive geometry in dual momentum space relevant for planar $\mathcal{N}=4$ sYM opens many possible avenues of further research. The most pressing one is to explore this construction beyond one loop. As we already mentioned in the main text, to generalise it to more loops we simply need to study collections of $L$ points inside $\mathcal{M}_{n,k,1}$ (or equivalently $\Delta_{n,k}(x)$) that are additionally positively separated from each other. Since the prescriptive unitarity method can be applied beyond one loop, we expect that the geometry for higher loops will have similar properties to the ones we studied at one loop, with the maximal cut contributions coming from the vertices of higher loop geometry. It has also been  suggested in \cite{Arkani-Hamed:2021iya} that additional simplifications come from considering negative geometries and we hope that it will be possible in our constructions as well.    

Another very interesting problem is to understand the dual momentum amplituhedron, following the lines of \cite{Herrmann:2020qlt} where the dual geometry was explored for one-loop MHV amplitudes. In particular, the authors of \cite{Herrmann:2020qlt} identified regions in the amplituhedron space corresponding to the contributions from box integrands and chiral pentagon integrands, and used projective duality to explore the dual geometry. A similar analysis can be done in our construction for the box integrands and parity-odd pentagons for any helicity. We leave the details of this to future work.

Various other questions remain open and deserve further investigation.
The first question is how to find  a complete classification of chambers beyond NMHV geometries. In particular, it would be interesting to check whether our prescription using maximal cliques of adjacent BCFW cells works beyond the NMHV sector. Moreover, while integrating the differential forms on the Minkowski contour gives the amplitude, a natural question to address is whether it would be meaningful to integrate it over the chamber geometry $\Delta_{n,k}(x)$, and what would be the relation of such obtained result to the amplitude. 
Finally, we have only hinted at the possible cluster algebra structures which emerges from our construction, and it would be very interesting to see whether it can be generalised to all helicity sectors. Another related direction would be to study whether our construction relates to the symbol alphabet and cluster adjacencies at the level of the integrated answer, see for instance \cite{Golden:2013xva,Golden:2014xqa}. A prime candidate being the case of N$^2$MHV$_8$, the first example where chambers containing cells of intersection number two appear, it would be interesting to see whether a full analysis of the chamber structure in this case would shed light on the cluster adjacency properties of square root letters appearing in the symbol of eight point amplitudes \cite{Drummond:2019cxm,Henke:2019hve,Arkani-Hamed:2019rds}.


\section{Acknowledgements}

We would like to thank Andrew McLeod for useful discussions. We would also like to thank CERN for their hospitality during the final stages of writing this paper.
This work was partially supported by the Deutsche Forschungsgemeinschaft (DFG, German Research Foundation) -- Projektnummer 404362017.

\appendix

\section{NMHV chambers}\label{app:chambers}
In this appendix we provide an efficient strategy for determining the compatibility of cells in the case of NMHV amplitudes which is most easily described using the original definition of the amplituhedron \cite{Arkani-Hamed:2013jha}. At tree level the amplituhedron for NMHV amplitudes, $\mathcal{A}_{n,1,0}$, is the positive geometry in $G(1,5)$ defined by the map
\begin{equation}
Y^{I} = \sum_{a=1}^n c_{a} Z^I_a, \quad \quad \quad a=1, \ldots, n, \quad I =1,\ldots, 5,
\end{equation}
where $c_a$ are the matrix elements of $C = (c_1,\ldots,c_n) \in G_+(1,n)$, and $Z_a^I$ are the matrix elements of the positive matrix $Z \in M_+(5,n)$. 
The BCFW cells relevant for triangulating $\mathcal{A}_{n,1,0}$ are the four-dimensional cells of $G_+(1,n)$, we will label these cells by $( a_1\ldots a_{5})$ where the indices specify the location of the non zero entries. Furthermore, we shall denote the collection of all BCFW cells as
\begin{equation}
\Sigma_{\text{NMHV}_n} = \left\{ ( a_1\ldots a_{5}) | ( a_1\ldots a_{5}) \in \left(\begin{smallmatrix}[n] \\ 5 \end{smallmatrix}\right) \right\}\,,
\end{equation}
where $[n] = \{ 1,2, \ldots, n \}$ and  $\left(\begin{smallmatrix}[n] \\ k \end{smallmatrix}\right)$ denotes $k$ element subsets of $[n]$.
As an example at six-points we have six BCFW cells
\begin{equation}
\Sigma_{\text{NMHV}_6} = \{ (23456),(13456),(12456),(12356),(12346),(12345) \},
\end{equation}
where for instance cell $(23456)$ is parameterised as
\begin{equation}
C_{(23456)} = \left(\begin{matrix}0&\alpha_2&\alpha_3&\alpha_4&\alpha_5&\alpha_6\end{matrix}\right).
\end{equation}
The boundaries of the BCFW cells take the form $\langle Y ijkl \rangle=0$ and are split into two types: the physical boundaries of $\mathcal{A}_{n,1,0}$ which take the form $\langle Y i i+1 j j+1 \rangle=0$; and all other boundaries not of this form which are referred to as {\it spurious} boundaries. As an example for $n=6$ the six spurious boundaries are given by
\begin{equation}
\left\{\langle Y1235\rangle=0,\langle  Y1245\rangle=0,\langle Y1345\rangle=0,\langle Y1356\rangle=0,\langle Y2346\rangle=0,\langle Y2456 \rangle =0\right\}.
\end{equation}
We say two BCFW cells $( a_1\ldots a_{5})$ and $( b_1\ldots b_{5})$ are {\it compatible} if their images through the $C \cdot Z$ map intersect\footnote{Note that depending on the external data some of the chambers may in fact be empty.}. We will be interested in finding the maximal sets of compatible BCFW cells referred to as {\it chambers} in the main text. As we shall now explain the compatibility of BCFW cells, or rather the incompatibility, is succinctly encoded in the sign patterns of the invariants $\langle Yijkl\rangle$ associated to {\it spurious} boundaries evaluated on each BCFW cell. 

To see why this is the case recall that each spurious boundary slices $\mathcal{A}_{n,1,0}$ into two regions depending on the sign of $\langle Y ijkl \rangle$. If two BCFW cells lie on opposite sides of this boundary they are hence incompatible. Consider again $n=6$ and the spurious boundary invariant $\langle Y 1235\rangle $ which tells us that the following BCFW cells
\begin{align}
& (12356) :  \langle Y 1235\rangle = \alpha_6 \langle 12356 \rangle \geq 0, \notag \\
& (12345) :  \langle Y 1235\rangle = -\alpha_4 \langle 12345 \rangle \leq 0,
\end{align}
are incompatible. By constructing these sets of mutually incompatible cells for all possible spurious boundaries $\langle Yijkl \rangle$ we arrive at the full list of incompatible cells.
\begin{figure}
\center
\begin{tikzpicture}[scale=0.8]
\draw[gray, thick] (0,0) -- (10,0);
\filldraw[black] (0,0) circle (2pt) node[anchor=north]{$1$};
\filldraw[black] (2,0) circle (2pt) node[anchor=north]{$i$};
\filldraw[black] (4,0) circle (2pt) node[anchor=north]{$j$};
\filldraw[black] (6,0) circle (2pt) node[anchor=north]{$k$};
\filldraw[black] (8,0) circle (2pt) node[anchor=north]{$l$};
\filldraw[black] (10,0) circle (2pt) node[anchor=north]{$n$};
\filldraw[black] (1,0) circle (0pt) node[anchor=south]{$+$};
\filldraw[black] (3,0) circle (0pt) node[anchor=south]{$-$};
\filldraw[black] (5,0) circle (0pt) node[anchor=south]{$+$};
\filldraw[black] (7,0) circle (0pt) node[anchor=south]{$-$};
\filldraw[black] (9,0) circle (0pt) node[anchor=south]{$+$};
\end{tikzpicture}
\caption{The positive and negative regions of label space for the spurious boundary $\langle Yijkl \rangle$.}
\label{fig:label_space}
\end{figure}
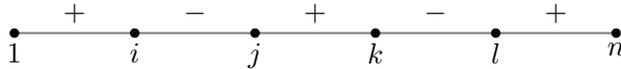

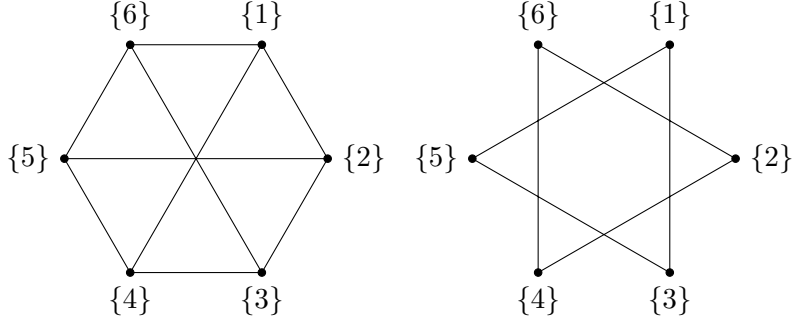
\begin{figure}[h]
\center
\begin{tikzpicture}[scale=0.7]
   \newdimen\R
   \R=2.5cm
   \draw (0:\R) \foreach \x in {60,120,...,360} {  -- (\x:\R) };
    \draw (0:\R) \foreach \x in {60,120,180} { (\x-180:\R) -- (\x:\R) };
   \foreach \x/\l/\p in
     { 60/{\{1\}}/above,
      120/{\{6\}}/above,
      180/{\{5\}}/left,
      240/{\{4\}}/below,
      300/{\{3\}}/below,
      360/{\{2\}}/right
     }
     \node[inner sep=1pt,circle,draw,fill,label={\p:\l}] at (\x:\R) {};
\end{tikzpicture}
\begin{tikzpicture}[scale=0.7]
   \newdimen\R
   \R=2.5cm
    \draw (0:\R) \foreach \x in {60,120,...,360} { (\x+120:\R) -- (\x:\R) };
   \foreach \x/\l/\p in
     { 60/{\{1\}}/above,
      120/{\{6\}}/above,
      180/{\{5\}}/left,
      240/{\{4\}}/below,
      300/{\{3\}}/below,
      360/{\{2\}}/right
     }
     \node[inner sep=1pt,circle,draw,fill,label={\p:\l}] at (\x:\R) {};
\end{tikzpicture}
\caption{(Left) The compatibility graph for BCFW cells $\Sigma_{\text{NMHV}_6}$ whose maximal cliques correspond to chambers. (Right) The incompatibility graph whose maximal cliques correspond to triangulations of $\mathcal{A}_{6,1,0}$.}
\label{fig:adjacency_NMHV_6}
\end{figure}
The incompatibility conditions for each spurious boundary $\langle Yijkl \rangle $, with $i<j<k<l$, follow a simple combinatorial rule which we now describe. We begin by defining two subsets of the set of labels
\begin{align}
[n]^+_{ijkl} = \{ 1\leq a<i \}\cup\{ j<a<k\}\cup\{l <a\leq n\}, \quad \quad  [n]^-_{ijkl} = \{ i<a<j\}\cup \{k<a<l \},
\end{align}
such that for all $a_\pm \in [n]^\pm_{ijkl}$ we have 
\begin{equation}
\pm \text{sgn}\langle a_{\pm}ijkl \rangle \geq 0 \,,
\end{equation}
see figure \ref{fig:label_space}. 
It follows that the following subsets of BCFW cells 
\begin{equation}
S_{ijkl}^\pm  = \{ (a_1 \ldots a_5) | ( a_1\ldots a_{5}) \in \left(\begin{smallmatrix}[n]^\pm_{ijkl} \\ 5 \end{smallmatrix}\right) \} \subset  \Sigma_{\text{NMHV}_n},
\end{equation}
live on opposite sides of the spurious boundary $\langle Y ijkl \rangle =0$ and are hence mutually incompatible. Specifying this to $n=6$ the mutually incompatible sets of BCFW cells for each spurious boundary are given by 
\begin{align}
&S^+_{1235} = \{ ( 12356) \},  && S^-_{1235} = \{ (12345) \}, \notag \\
& S^+_{1246} = \{ ( 12346) \}, && S^-_{1246} = \{ ( 12456) \}, \notag \\
& S^+_{1345} = \{ ( 13456) \}, && S^-_{1345} = \{ ( 12345) \}, \notag \\
& S^+_{1356} = \{ ( 13456) \}, &&S^-_{1356} = \{ ( 12356) \}, \notag \\
& S^+_{2346} = \{ ( 12346) \}, && S^-_{2346} = \{ ( 23456) \}, \notag \\
& S^+_{2456} = \{ ( 12456) \}, && S^-_{2456} = \{ ( 23456) \}.
\end{align}
These incompatibilities can be encoded in a graph as depicted on the right of figure \ref{fig:adjacency_NMHV_6} where we have introduced the notation $\{i \} \equiv ([6]\setminus \{i\})$. The maximal cliques of this incompatibility graph encode the two BCFW triangulations of $\mathcal{A}_{6,1,0}$
\begin{align}
\overline{\Omega}_{6,1,0} = \overline{\mathcal{I}}_{6,3}^{\{ 1 \}}+ \overline{\mathcal{I}}_{6,3}^{\{ 3 \}}+ \overline{\mathcal{I}}_{6,3}^{\{ 5 \}}
=  \overline{\mathcal{I}}_{6,3}^{\{ 2 \}}+ \overline{\mathcal{I}}_{6,3}^{\{ 4 \}}+ \overline{\mathcal{I}}_{6,3}^{\{ 6 \}}. 
\end{align}
Alternatively, we can look at the maximal cliques of the complement graph, depicted on the left of figure \ref{fig:adjacency_NMHV_6}, which encode the chambers of $\mathcal{A}_{6,1}$, given explicitly by
\begin{align}
\mathcal{C}_{6,1} = \{ & \{ 1 \} \cap \{ 2 \}, \{ 1 \} \cap \{ 4 \}, \{ 1 \} \cap \{ 6 \}  \notag\\ 
& \{ 3 \} \cap \{ 2 \}, \{ 3 \} \cap \{ 4 \}, \{ 3 \} \cap \{ 6 \}  \notag\\ 
& \{ 5 \} \cap \{ 2 \}, \{ 5 \} \cap \{ 4 \}, \{ 5 \} \cap \{ 6 \} \}.
\end{align}
This procedure can be carried out for any $n$ and we find 9, 71, 728, 15979, 1144061 chambers for $n=6, \ldots 10$, respectively. For example, this leads to the following explicit chamber structure for NMHV$_7$ 
\begin{align}\label{eq:chambers73}
\mathcal{C}_{7,1} = \{&\{1, 2\} \cap \{1, 3\}\cap \{2, 3\}, 
\notag \\&\{1, 2\}\cap \{1, 5\}\cap \{2, 5\},\notag 
\\&\{1, 2\}\cap \{1, 3\}\cap \{2, 5\}\cap \{3, 5\}, \notag \\
 & \{1, 2\}\cap \{1, 3\}\cap \{2, 7\}\cap \{3, 7\}, \notag \\
 &\{1, 3\}\cap \{1,4\}\cap \{3, 7\}\cap \{4, 7\} , \notag \\
   & \{1, 2\}\cap \{1, 5\}\cap \{2, 7\}\cap \{5, 7\}, \notag \\
 & \{1, 2\} \cap\{1, 3\}\cap \{2, 7\}\cap \{3, 5\}\cap \{5, 7\} , \notag 
 \\&\{1, 3\} \cap \{1, 4\}\cap \{2, 4\}\cap \{2, 5\}\cap \{3, 5\},\notag  \\
 & \{1, 3\} \cap \{1, 4\}\cap \{3,5\}\cap \{4, 7\}\cap \{5, 7\},\notag 
 \\&\{1, 3\}\cap \{1, 4\} \cap \{2, 4\}\cap \{2, 7\}\cap \{3, 5\}\cap \{5, 
   7\},\notag  \\
 & \{1, 3\}\cap \{1, 6\}\cap \{2, 4\}\cap \{2, 7\}\cap \{3, 5\}\cap \{4, 6\}\cap \{5, 7\}\},
\end{align}
where we explicitly included only cyclic representatives of each family. Again we use the notation $\{ i,j \} \equiv ([7]\setminus \{i,j\})$.

\bibliographystyle{nb}

\bibliography{N4_lightcones}

\end{document}